\documentclass[12pt]{article}
\usepackage{graphicx}
\usepackage{amsmath} 
\usepackage{amssymb}
\usepackage{epsfig}
\begin{document}
\title{\hfill \small FERMILAB-TM-2617-AD-CD-E
\\
{\bigskip}
\Large \bf  Calibration and GEANT4 simulations of the Phase~II Proton Compute Tomography (pCT) Range Stack Detector }
% \author{\large Sergey A. Uzunyan, \bigskip \\
% {\it  $^1$~Department of Physics, Northern Illinois University, DeKalb, IL 60115, USA}}
%{\small Femilab TM, FERMILAB-TM-2617-AD-CD-E}
\maketitle
\begin{center}
\author{ S.~A.~Uzunyan, G.~Blazey, S.~Boi, G.~Coutrakon, \\ A.~Dyshkant, K.~Francis, D.~Hedin,  E.~Johnson, J.~Kalnins, V.~Zutshi,\\ ~{\it Department of Physics, Northern Illinois University, DeKalb, IL 60115, USA};\\
      R.~Ford, J~.E.~Rauch, P.~Rubinov, G.~Sellberg , P.~Wilson,\\ 
      {~\it Fermi National Accelerator Laboratory, Batavia, IL 60510, USA; }\\  M.~Naimuddin,{~\it Delhi University, 110007, India}}
%% make the title area
%\maketitle
\end{center}
\section{Introduction\label{intro}}
\begin{figure}[ht]
\centering
\includegraphics[scale=.28]{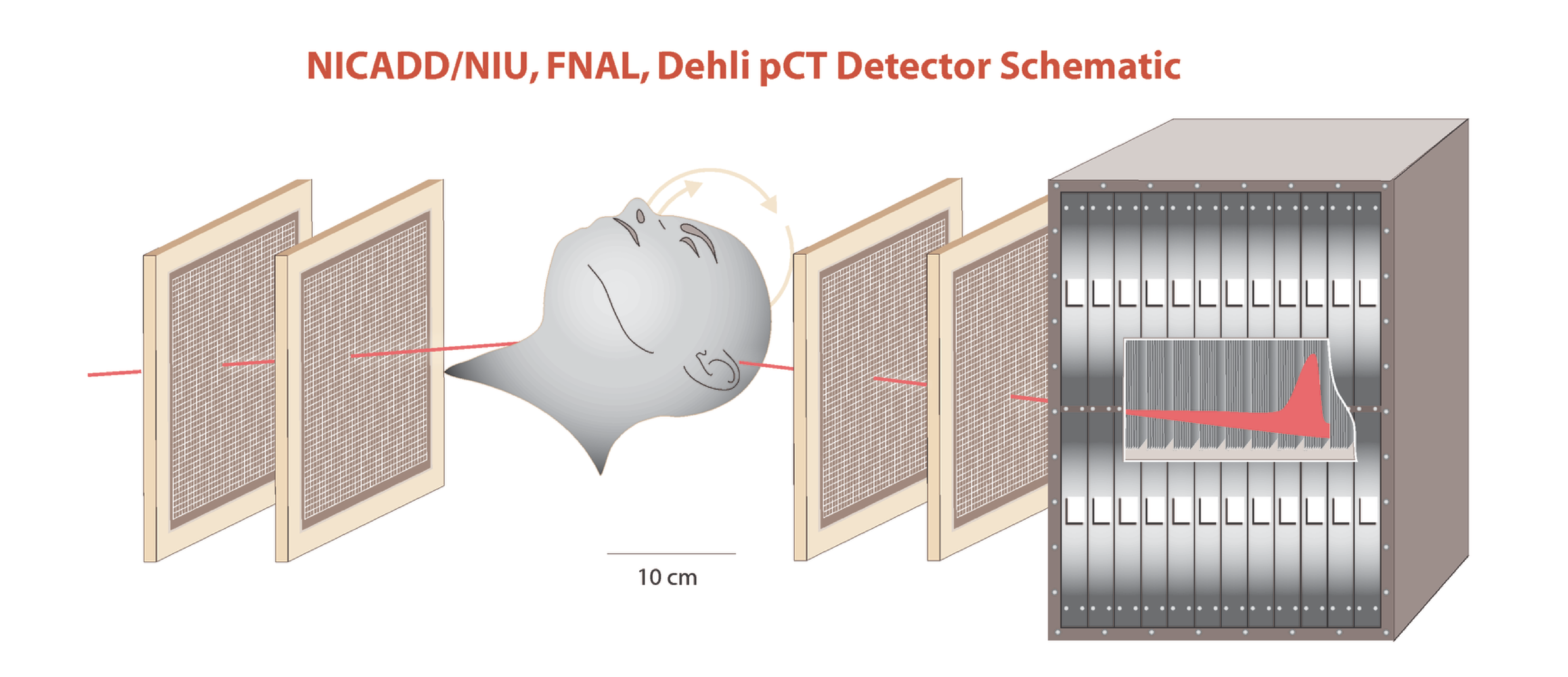}
\caption{\label{fig:pct_design} Four (X,Y) stations measure the proton trajectory before and after the patient.
A  stack of 3.2~mm thick scintillator tiles measures the residual energy or range after the patient.
}
\end{figure}

% 
%%%
Northern Illinois University in collaboration with Fermi National Accelerator Laboratory (FNAL) and Delhi University
has been designing and building a proton CT scanner~\cite{brudge_paper} for applications in proton treatment planning. 
In proton therapy, the current treatment planning systems are based on X-ray CT images that have intrinsic
limitations in terms of dose accuracy to tumor volumes and nearby critical structures. Proton CT aims to overcome
these limitations by determining more accurate relative proton stopping powers  directly as a result of imaging
with protons. 
%At present, the proton RSPs for various tissues, as derived from X-ray CT, 
%produce range uncertainties \cite{xct_uncert} of about 3 to 4\%.  We hope to reduce this  uncertainties using proton CT.
Fig.~\ref{fig:pct_design} shows a schematic proton CT scanner, which consists of eight planes of tracking detectors with
two X and two Y coordinate measurements both before and after the patient. 
%This provides the information for finding the trajectory through the head to correct, as much as possible,  for multiple coulomb scattering in the patient. 
In  addition, a calorimeter consisting of a stack of thin scintillator tiles, arranged in twelve eight-tile frames,
is used to determine the water equivalent path length (WEPL) of each track through the patient.  
The X-Y coordinates and WEPL are required input for image reconstruction software to find the relative (proton) stopping powers (RSP) value of each voxel in the patient and generate a corresponding 3D image.  In this note we describe tests conducted in 2015 at the  proton beam at the  Central DuPage Hospital in Warrenville, IL,
focusing on the range stack calibration procedure and comparisons with the GEANT~4 range stack simulation.

\section{The GEANT 4 model\label{g4}}
\begin{figure}[ht]
\centering
\includegraphics[scale=.50]{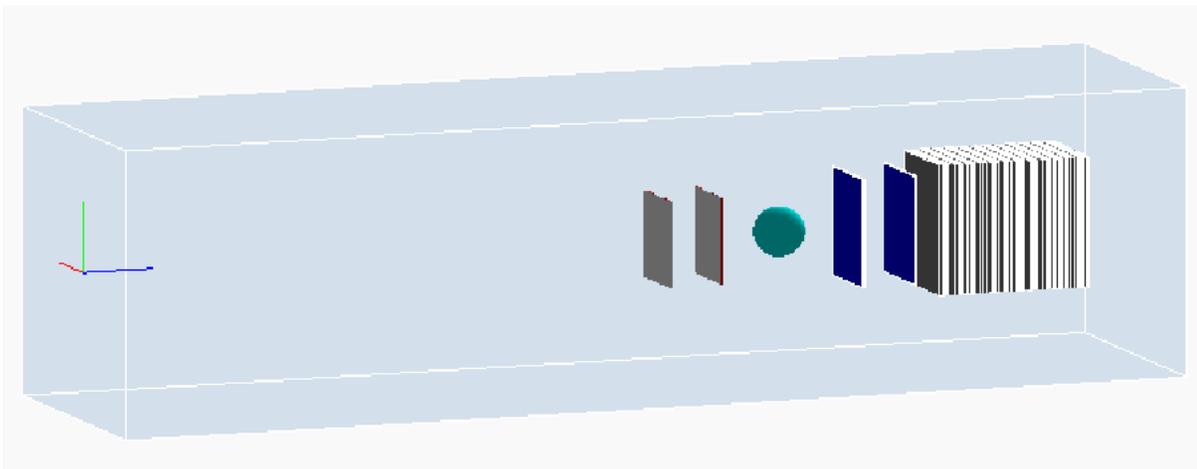}
\caption{\label{fig:pct_g4} The GEANT4 visualization of the scanner model used in the simulations.}
\end{figure}

To verify measurements obtained by the scanner at the CDH proton beam the scanner response was simulated using
a detailed model based on the GEANT-4 software.  Fig.~\ref{fig:pct_g4} shows a spherical water phantom  between 
the tracker planes of the scanner model. The simulated responses of the range stack and tracker stations
were analyzed with the same software as for the data.
\section{The CDH test beam\label{testbeam}}
\begin{figure}[ht]
\centering
\includegraphics[scale=1.90]{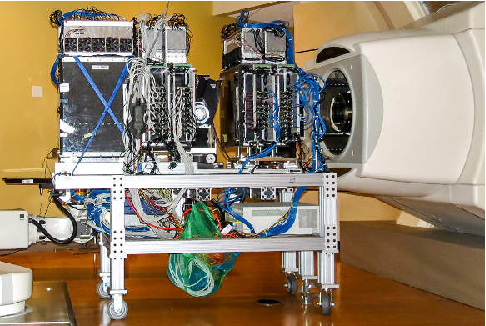}
\caption{\label{fig:assembled_scaner} Fully assembled proton CT scanner at CDH Proton center. 
From right to left, beam enters the  upstream tracker planes followed by the downstream 
tracker planes and finally the range stack. The gap in the middle is the position of the rotation 
stage for  the head phantom in the horizontal plane.}
\end{figure}

Figure~\ref{fig:assembled_scaner}  shows the  NIU scanner mounted on a cart in a treatment room at Central DuPage Hospital.
The proton beam enters the upstream tracker planes from the right followed by the downstream tracker planes and finally the range stack. 
In this note the range stack tiles are labeled from zero (the tile closest to the tracker) to 95.  Data were obtained using proton beams of energy in range from 103-225~MeV, equivalent to 8-32~cm proton stopping range in water.  
\subsection{Data acquisition (DAQ) system and event selection}
The DAQ system of the scanner is described in  ~\cite{su_daq}. The range stack
data are collected by twelve front-end boards. Each board provides the readout of 
one eight-tile range stack frame in form of time-stamped records of signal amplitudes in all 
tiles of the frame. We form the proton candidate event by combining records with close time-stamps.
We remove events candidates with duplicated frames (overlapped tracks). We then found the frame
with a Bragg peak, or stopping frame, and check that all frames before the stopping frame are 
also present in the event.
\subsection{Units of measurement}
The CDH accelerator control system is tuned to operate with proton beams with energies 
expressed in units of  the proton stopping range in water in $cm$,
$R_w (cm)$.   One can also express the proton stopping range $R_w$, and thus the beam energy $E_{beam}$,
in  density-independent units of  $g/cm^2$ :  
\begin{equation}
\label{bb0}         
E_{beam} (g/cm^2) \equiv R_w(g/cm^2) = R_{w} (cm) \times  \rho_{w} (g/cm^3)
\end{equation}

\begin{figure}[ht]
\centering
\includegraphics[scale=.35]{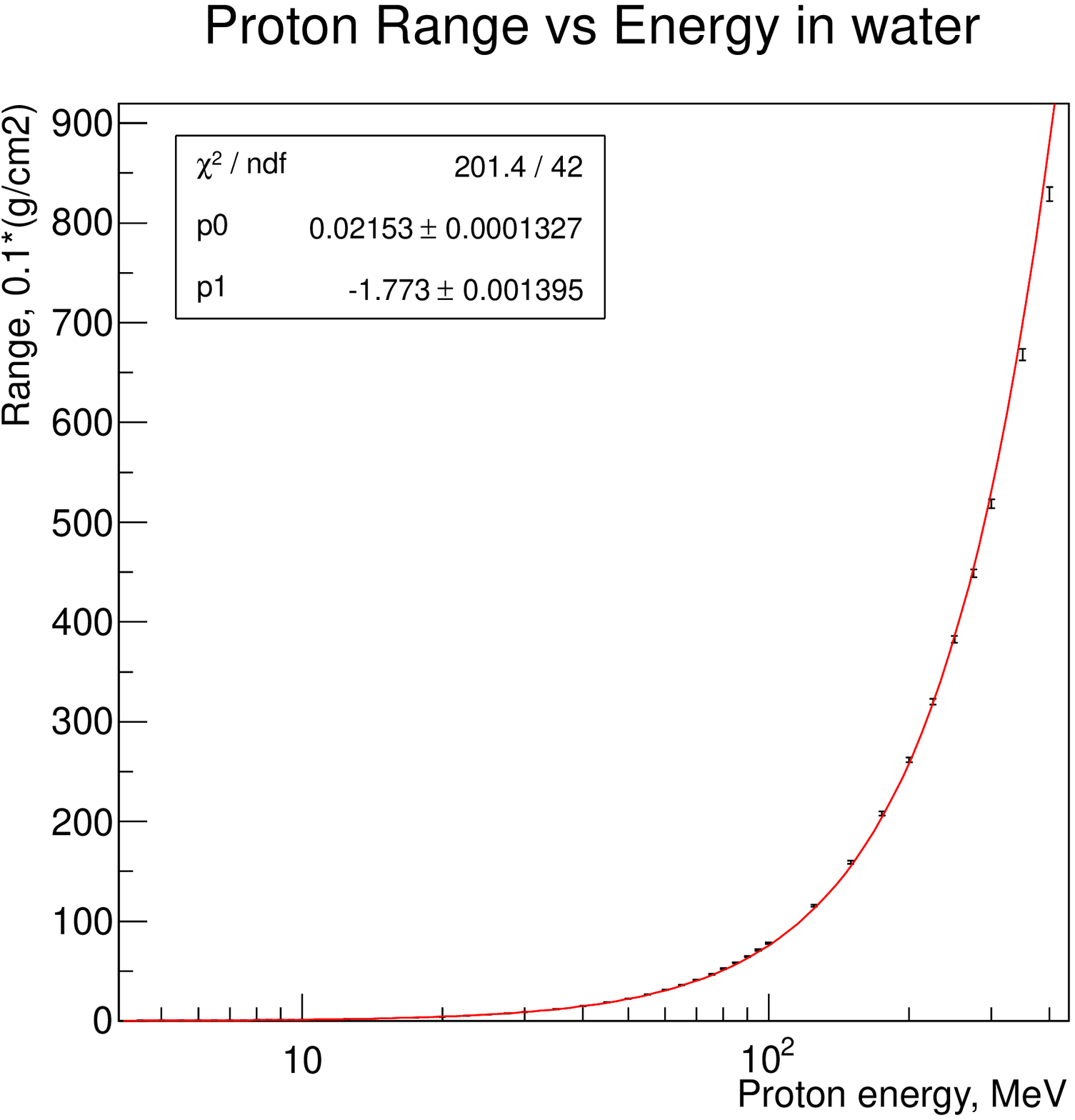}
\includegraphics[scale=.35]{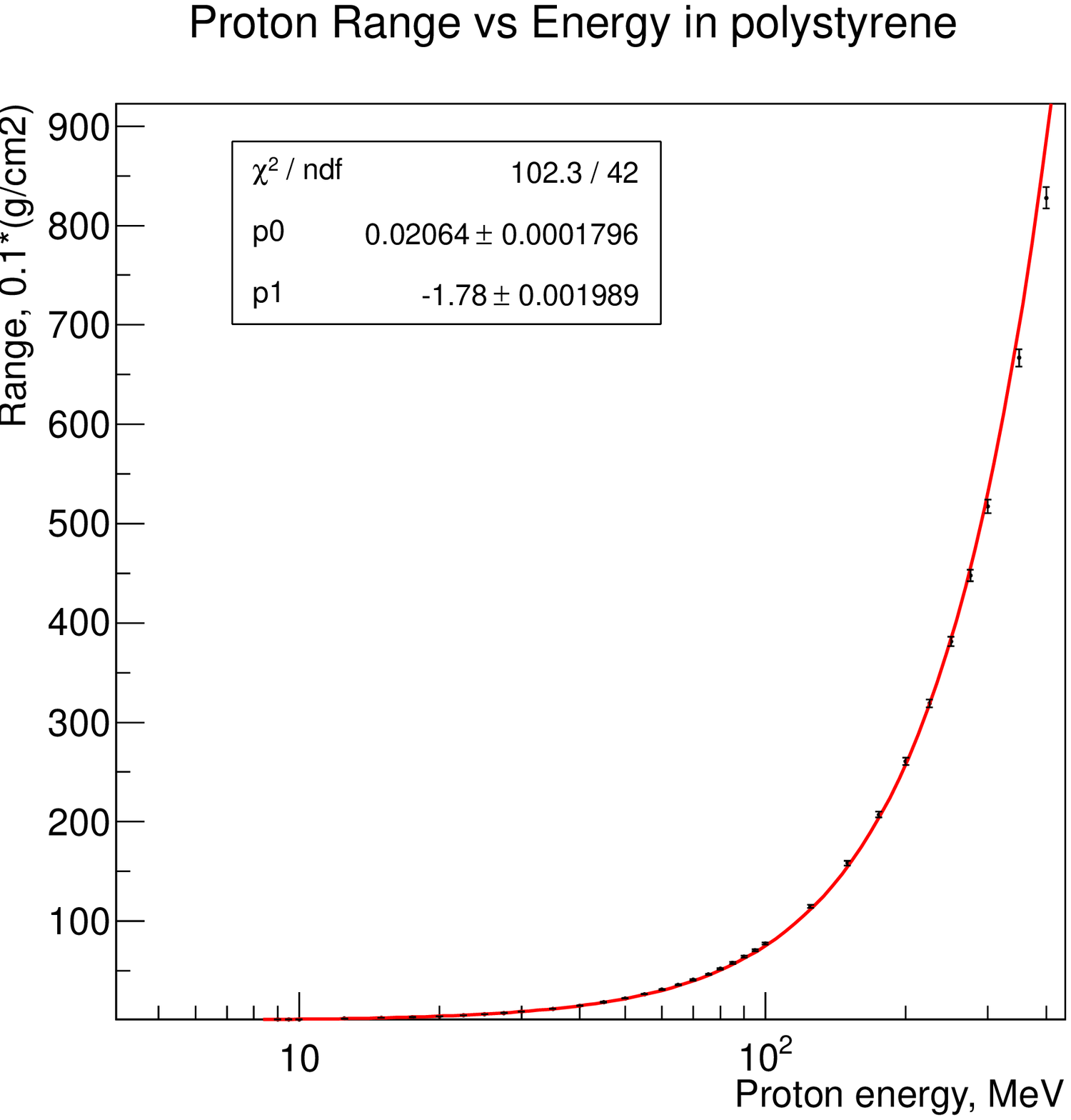}
\leftline{ \hspace{3.0cm}{\bf (a)} \hfill\hspace{2.5cm} {\bf (b)} \hfill}
\caption{\label{fig:mev_from_cm} a) The proton stopping range in water $R_w(g/cm^2)$ (black dots) versus proton energy $E (MeV)$,
as measured in~\cite{janni_tables}. The fit $R_w(g/cm^2) =  0.0022\times E_{MeV}^{(1.77)}$ conversion function (the red line) 
is used to find beam energies in $MeV$  that correspond to nominal CDH energies in $cm$.  b) The proton stopping range in polystyrene $R_{poly} (g/cm^2)$ (black dots) versus proton energy $E(MeV)$.}
\end{figure}

To obtain the energy $E_{beam}$ in $MeV$ we use proton energy-range tables (a.k.a. Janni's tables) \cite{janni_tables}.  
A fit of the stopping range  $R_w(g/cm^2)$ as a function of $E (MeV)$ is shown in Fig.~\ref{fig:mev_from_cm}(a).  
We use 
\begin{equation}
\label{bb1}  
R_w (g/cm^2) =  0.0022\times E_{MeV}^{(1.77)}
\end{equation}
to convert beam energies between $MeV$ and $g/cm^2$ units.
%new p

We calculate the proton stopping range in the range stack $R_{rs}(g/cm^2)$  using the 
measured proton stopping position as described in Section~\ref{energy-range}.
We compare the $R_{rs}(g/cm^2)$ with the range calculated from the total energy 
measured by the range stack using the energy-range dependence in polystyrene shown in Fig.~\ref{fig:mev_from_cm}(b). 

\section{Stack calibration procedure}
Energy deposition in each range stack scintillator tile is measured by two SiPMs connected
to the tile's single wavelength shifting (WLS) fiber. After passage of a proton,  for each of the two SiPMs the maximum digitized signal, $A^{max}_{SiPM}$, is collected
by the DAQ system.  Thus the measured energy deposition in each range stack tile,  $A^{ADC}_{tile}$, is obtained as a sum
of  $A^{max}_{SiPM}$  signals from SiPMs connected to this tile.  This measurement varies from tile to tile even for protons 
of similar energy due to differences in the SiPM's properties  and  the settings of corresponding 
readout channels. The following four step procedure is applied to calibrate the range stack detector.

\begin{figure}[ht]
\centering
 {
  \includegraphics[scale=0.27]{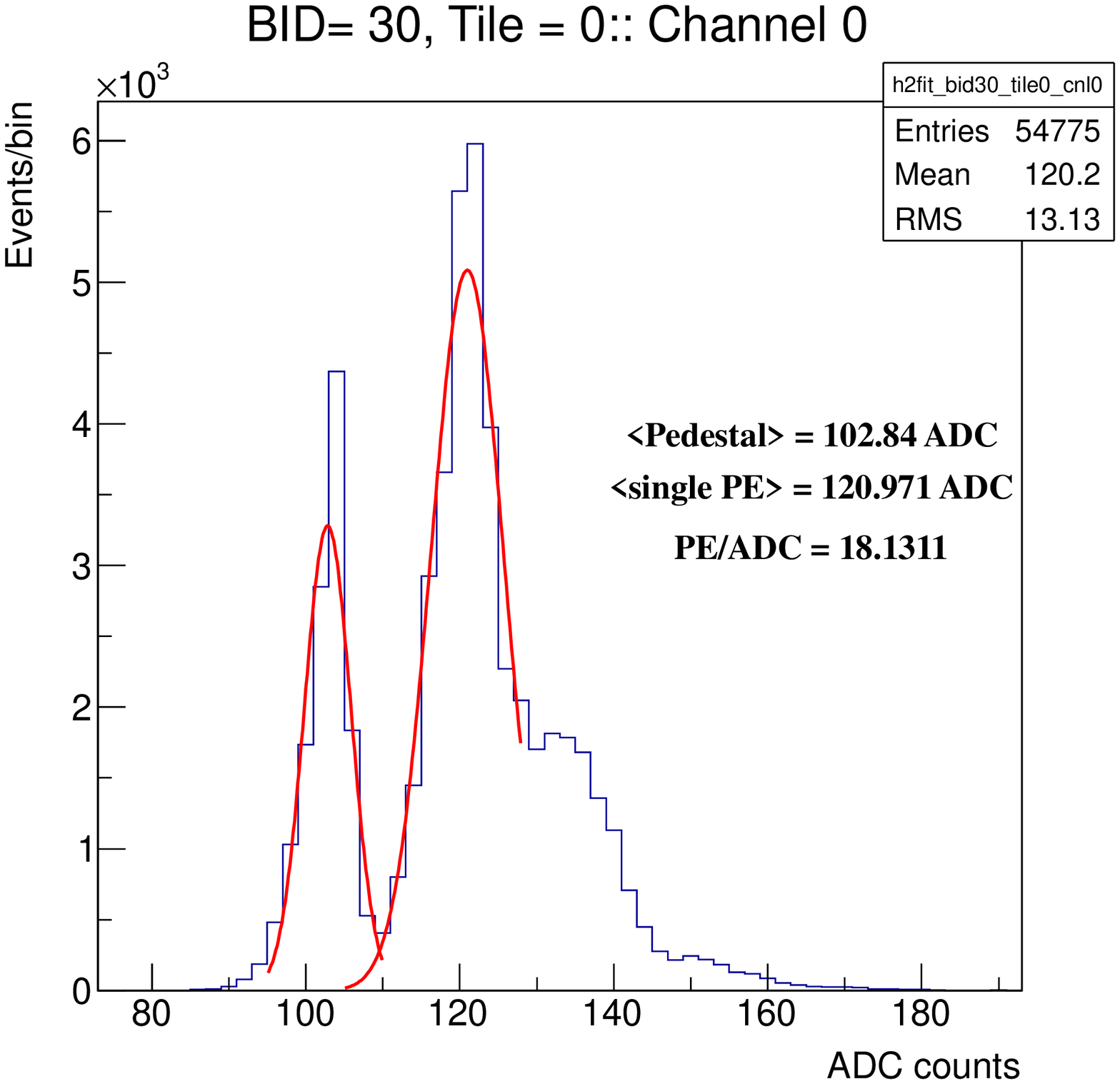}
  \includegraphics[scale=0.27]{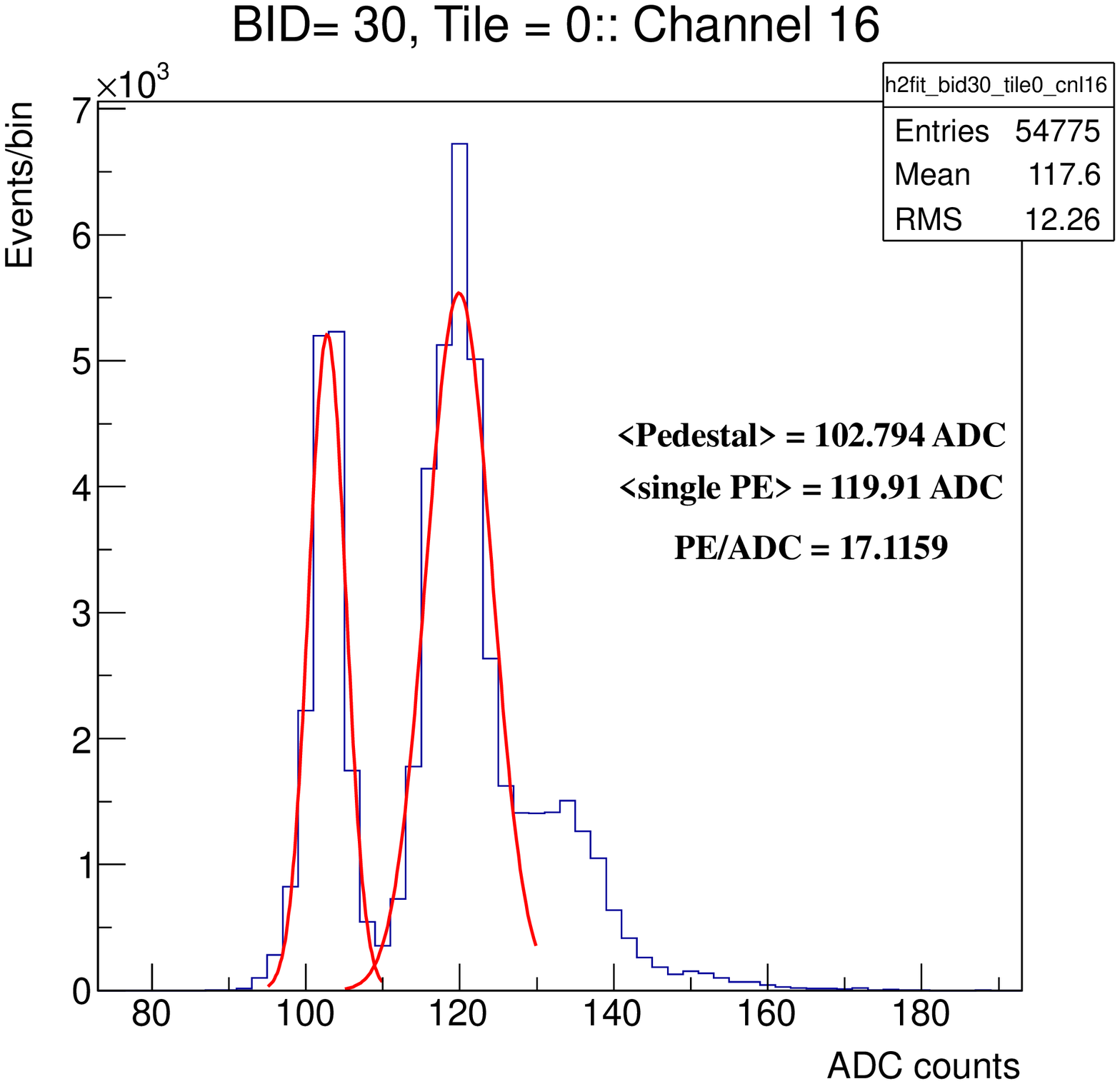}
  \includegraphics[scale=0.27]{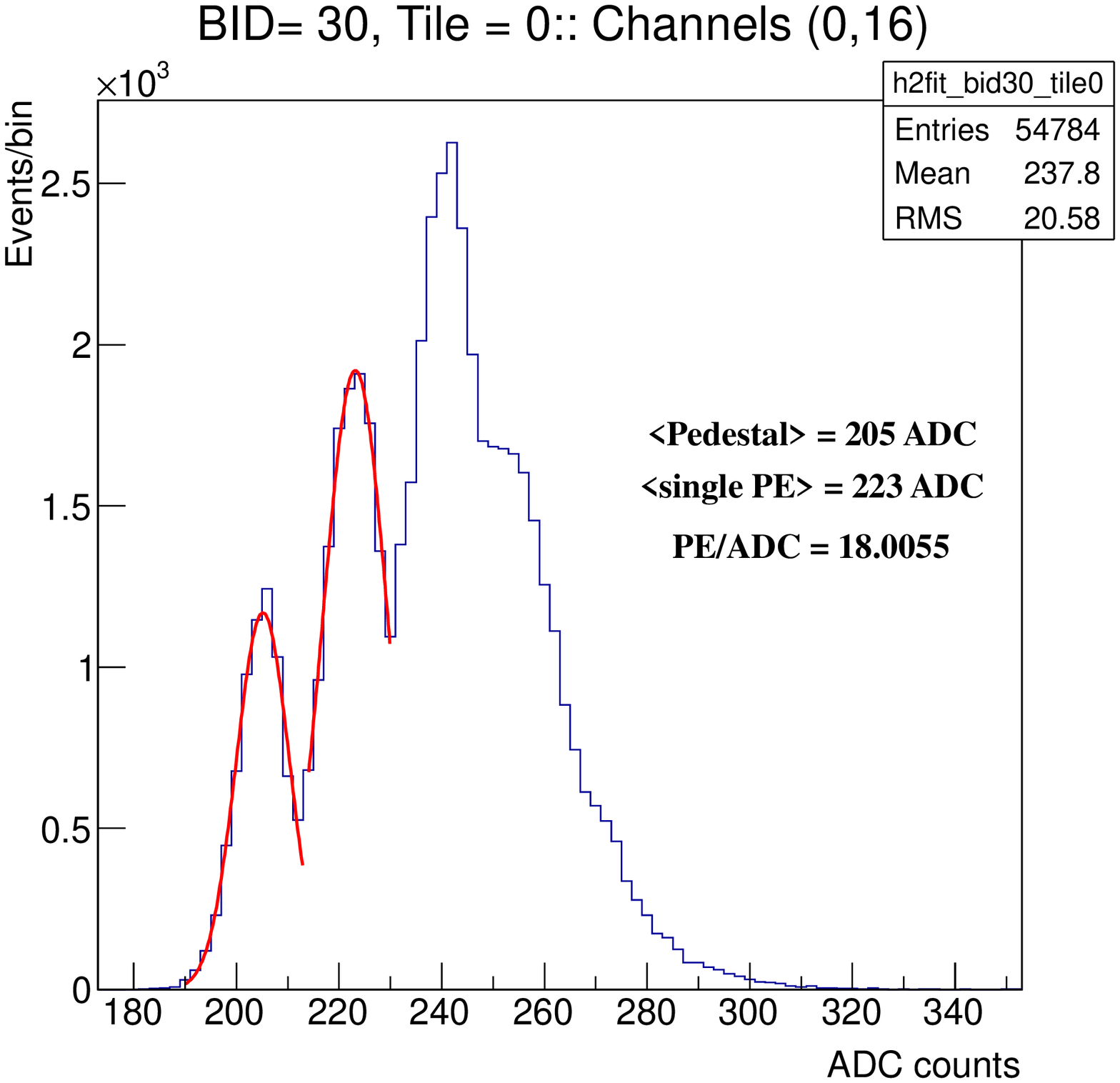}
}
\leftline{ \hspace{3.0cm}{\bf (a)} \hfill\hspace{2.5cm} {\bf (b)} \hfill\hspace{2.5cm} {\bf (c)} \hfill}
\caption{\label{fig:calibration-signals} Measured Tile0 signal amplitudes :
a) pedestal and the first  photo-electron (PE) peak in the SiPM1 of Tile 0 in events with no beam.
b) pedestal and the first  photo-electron (PE) peak in the SiPM2 of Tile 0 in events with no beam.
c) SiPM1+SiPM2 combined.
}
\end{figure}

\begin{figure}[]
\centering
 {
 \includegraphics[scale=0.19]{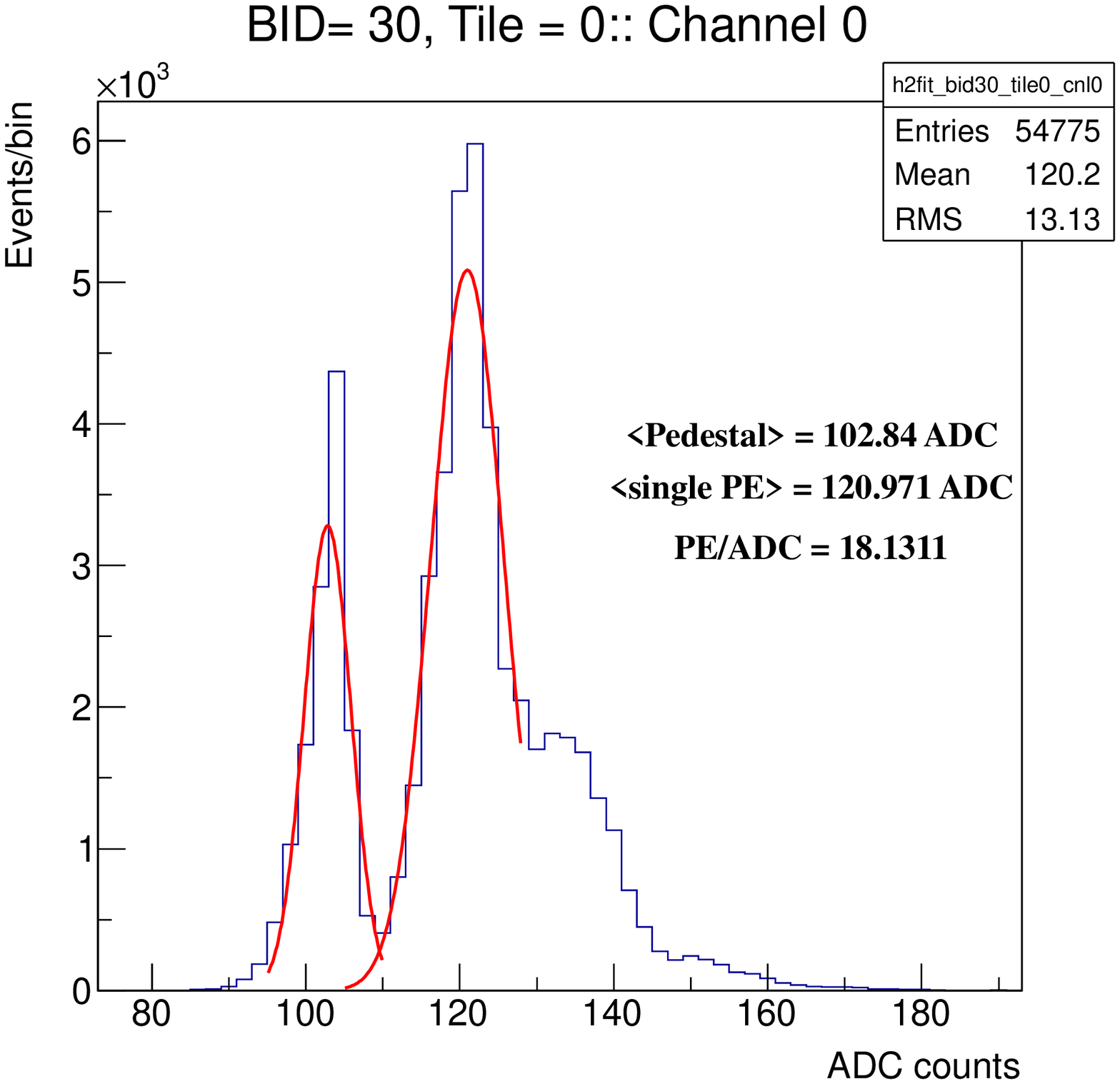}
 \includegraphics[scale=0.19]{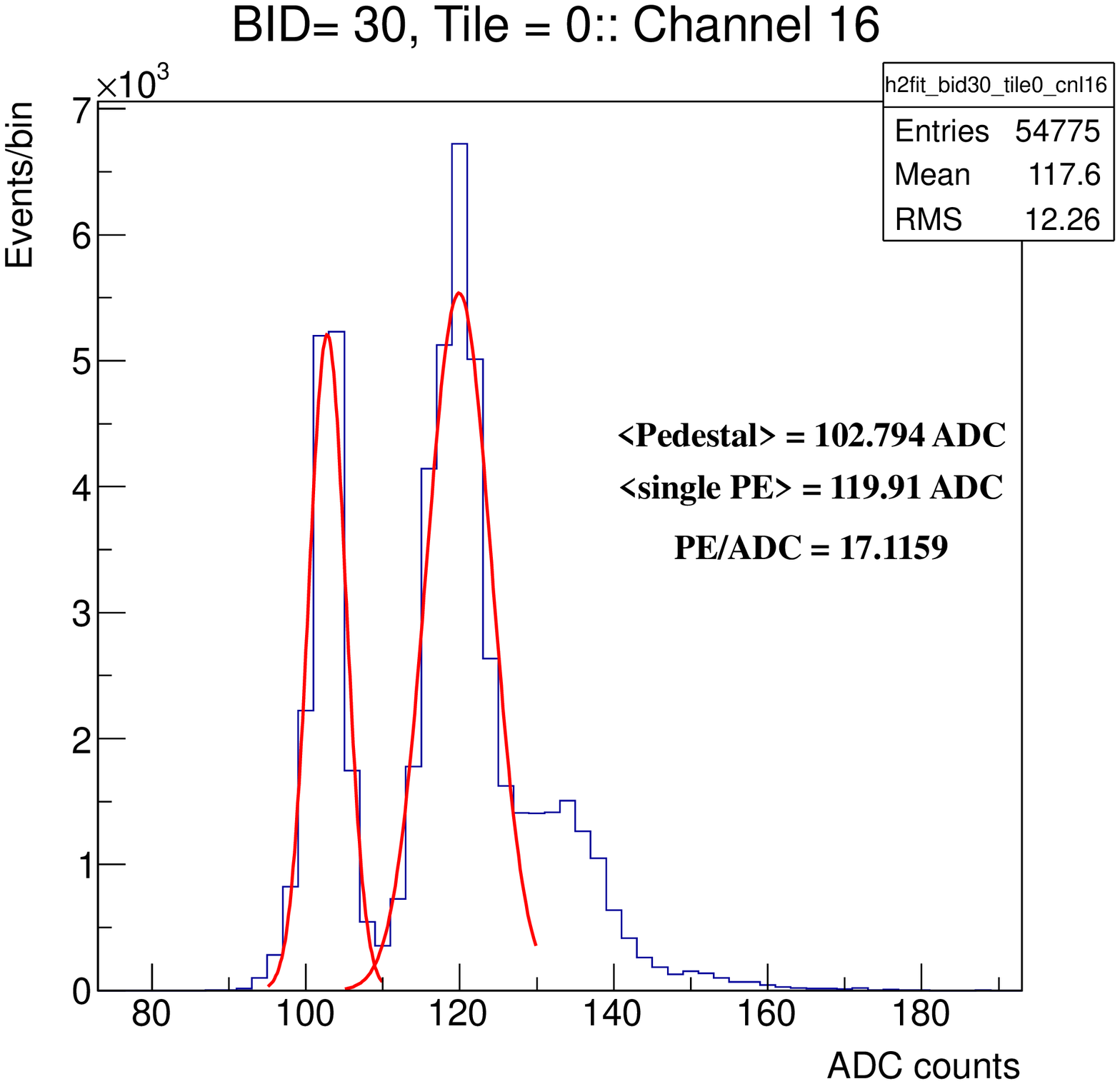}
 \includegraphics[scale=0.19]{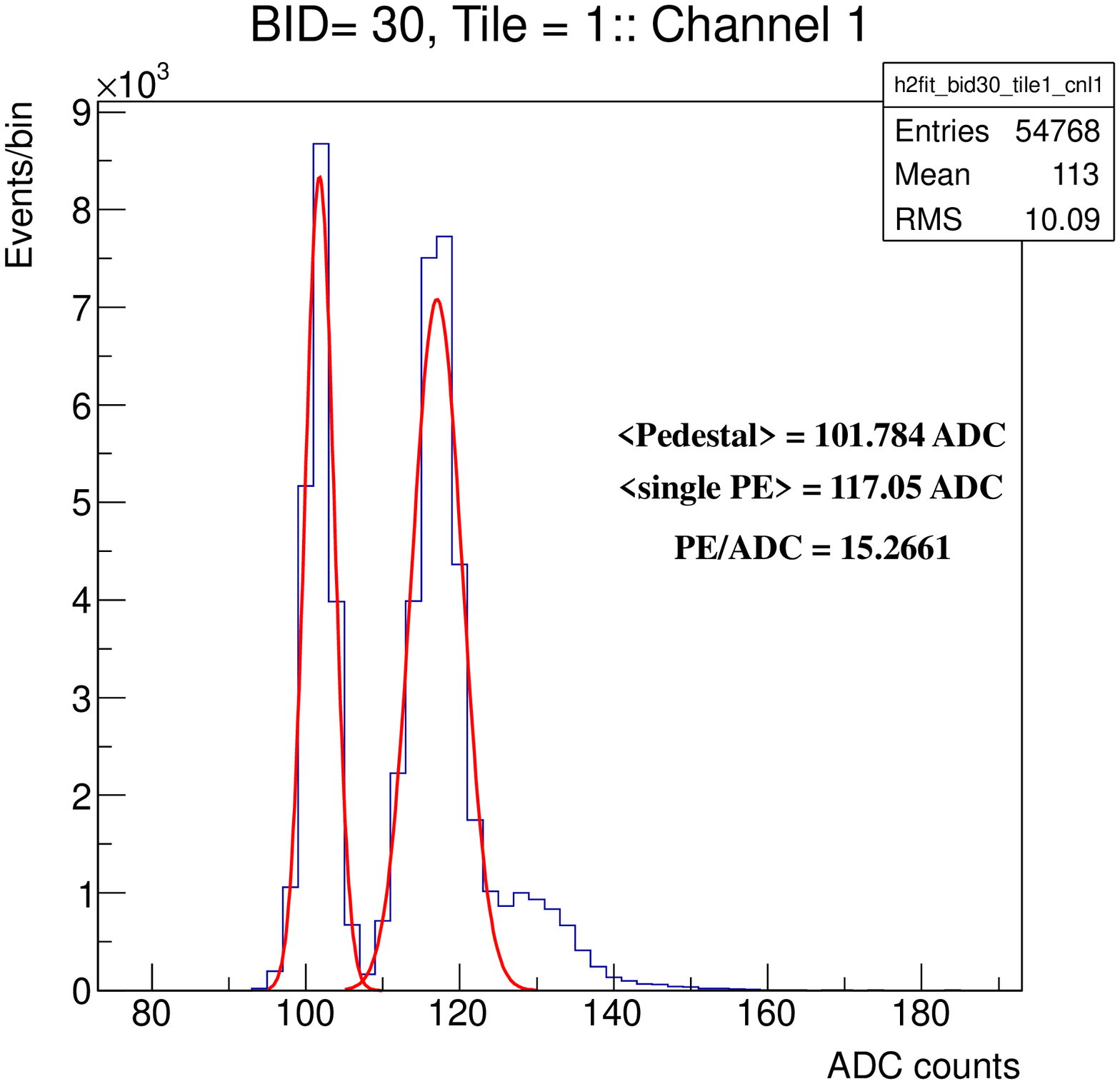}
 \includegraphics[scale=0.19]{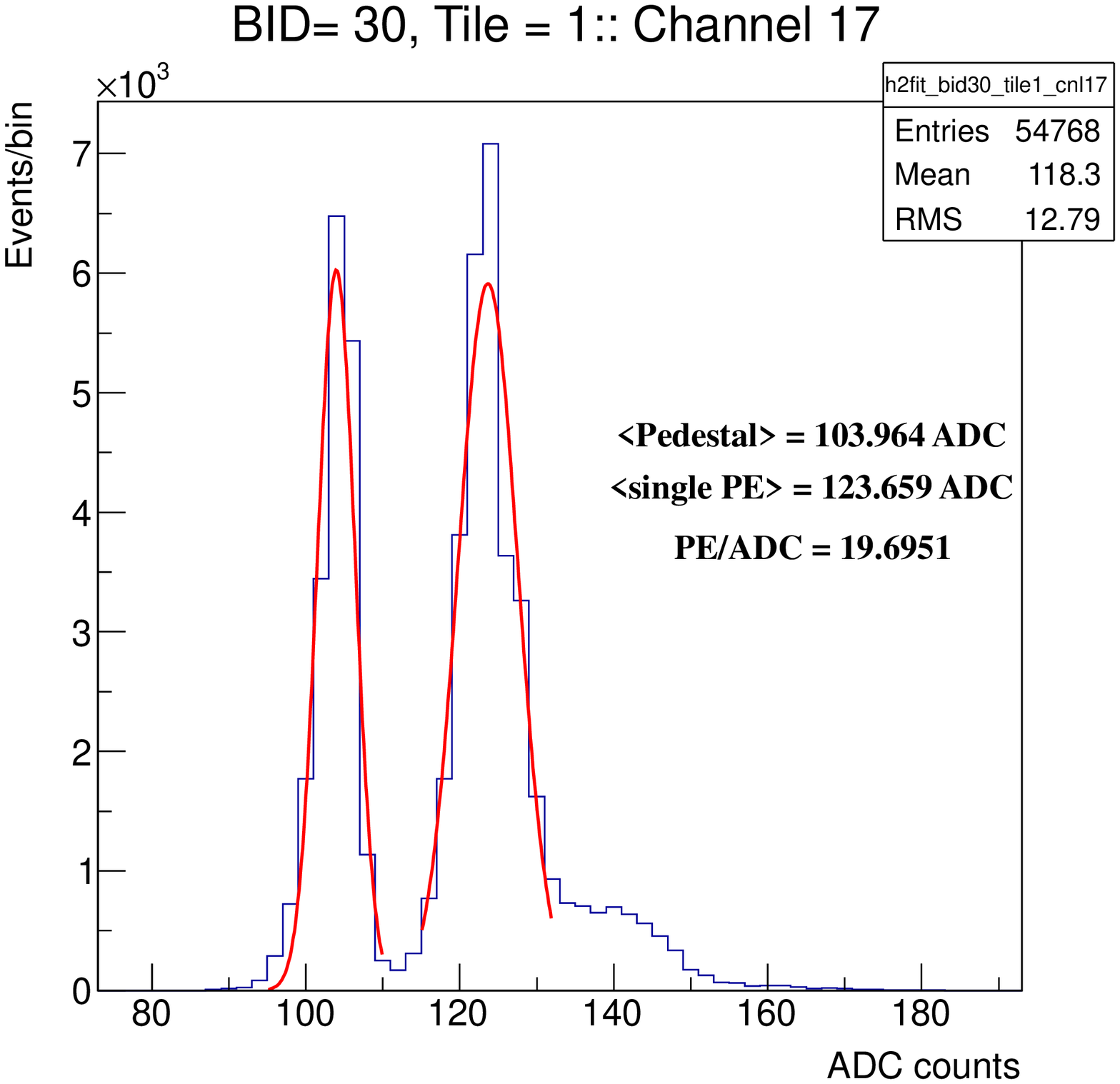}
\leftline{  \hfill }
 \leftline{  \hfill }
% \leftline{ \footnotesize \hspace{2.3cm}{\bf  (a)} \hfill\hspace{3.2cm} {\bf (b)} \hfill \hspace{3.2cm} {\bf (c)} \hfill }
 \includegraphics[scale=0.19]{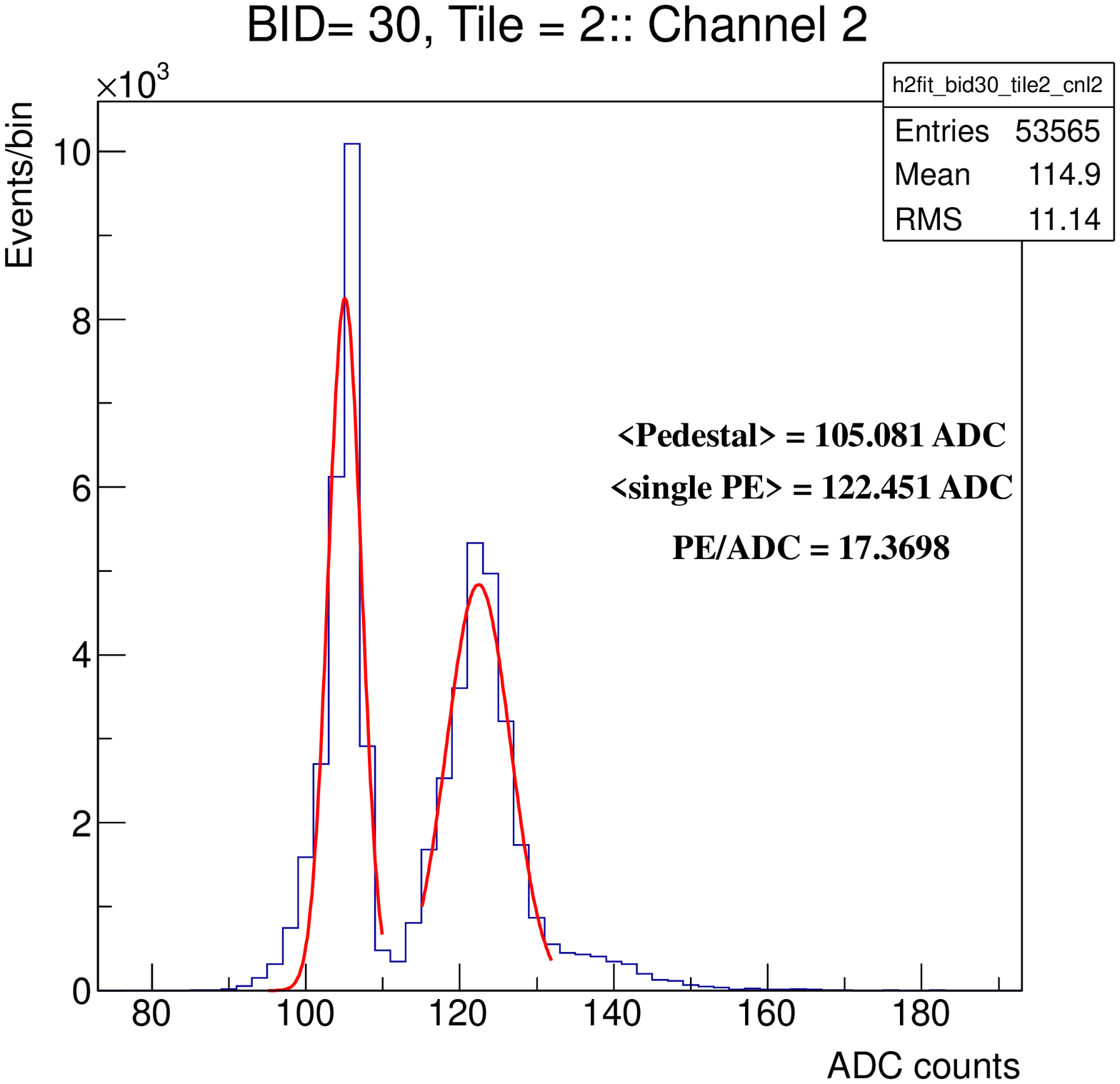}
 \includegraphics[scale=0.19]{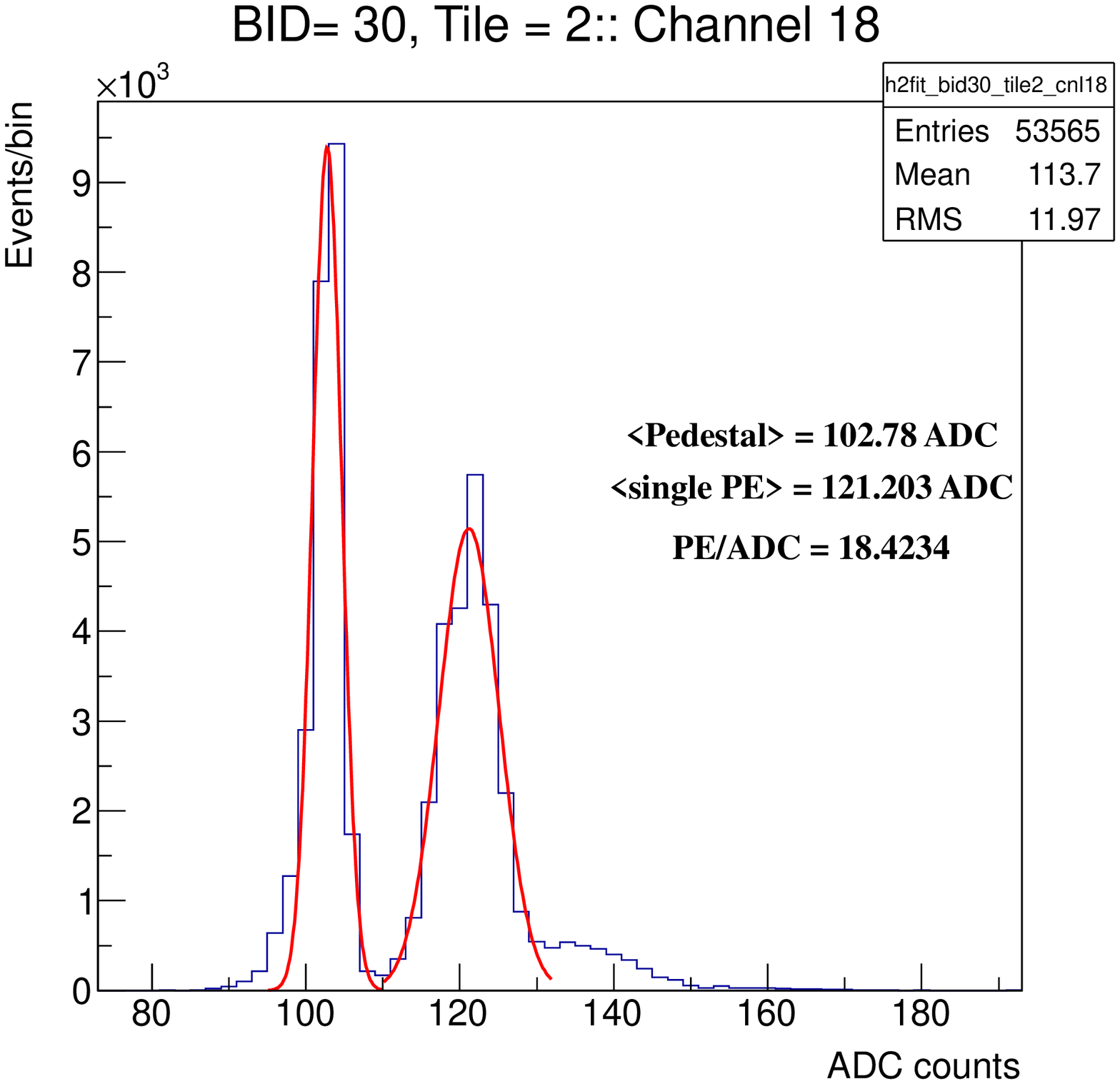}
 \includegraphics[scale=0.19]{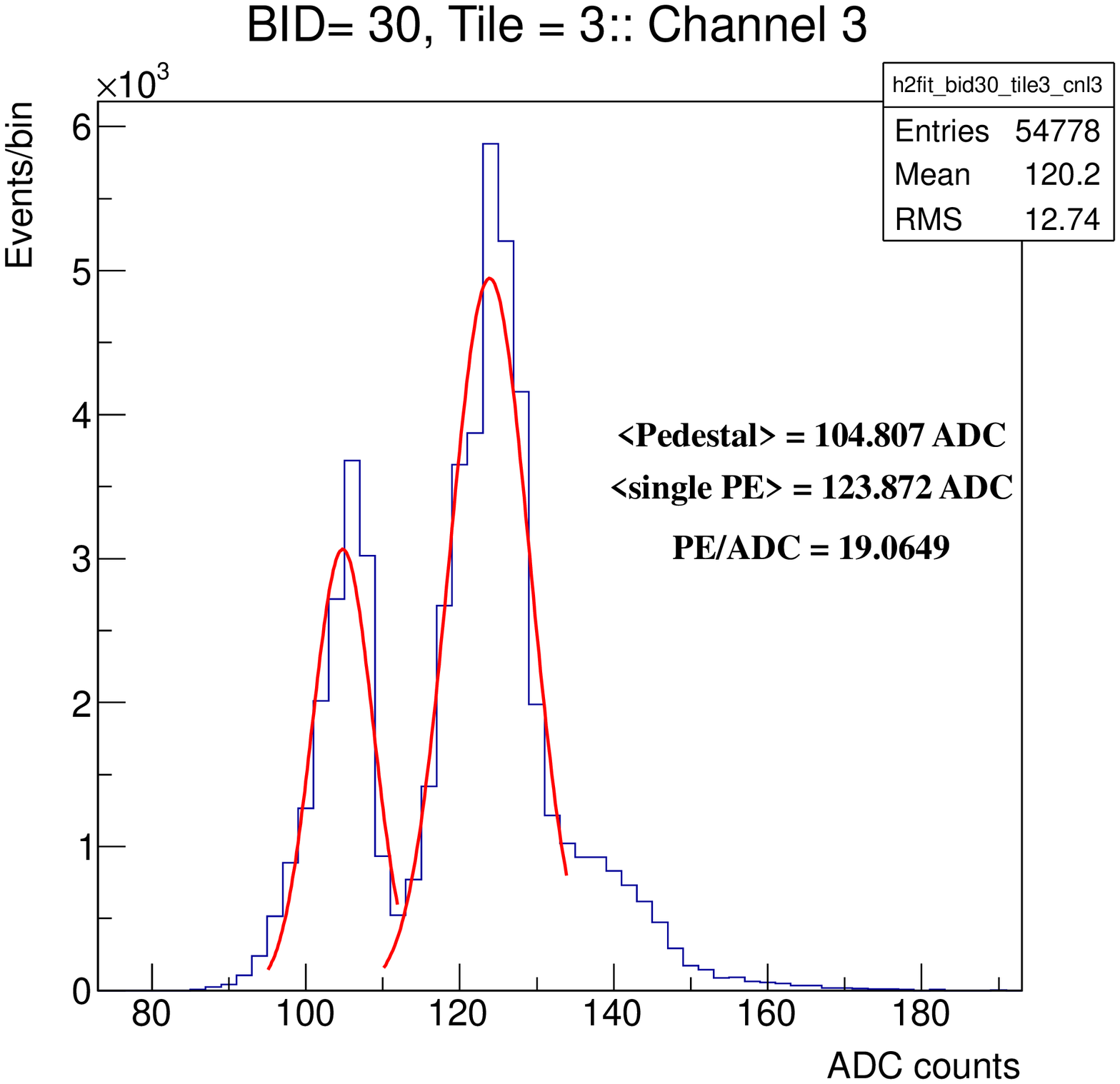}
 \includegraphics[scale=0.19]{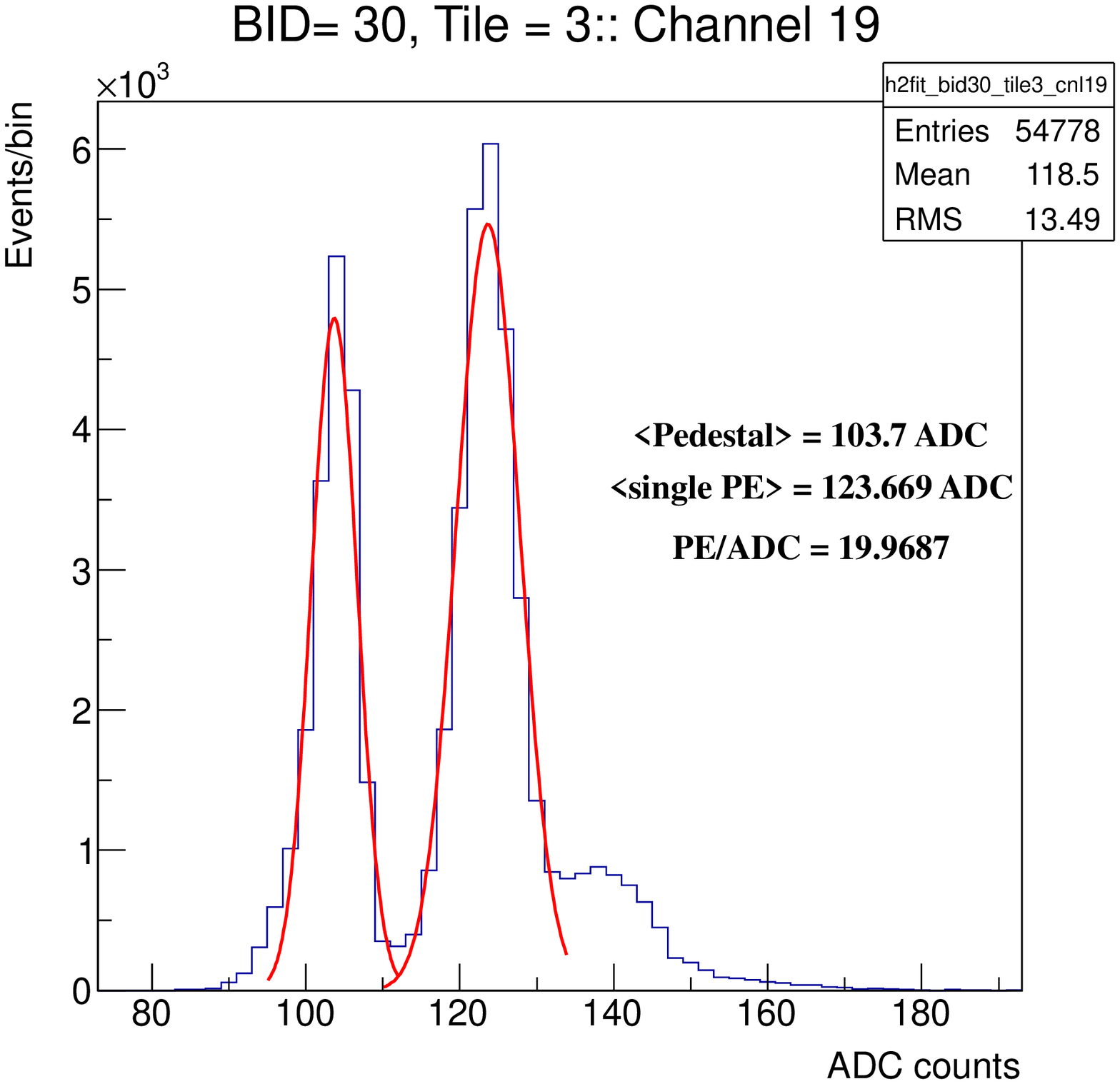}
% \leftline{ \footnotesize \hspace{2.3cm}{\bf  (a)} \hfill\hspace{3.2cm} {\bf (b)} \hfill \hspace{3.2cm} {\bf (c)} \hfill }
\leftline{  \hfill }
 \leftline{  \hfill }
\includegraphics[scale=0.19]{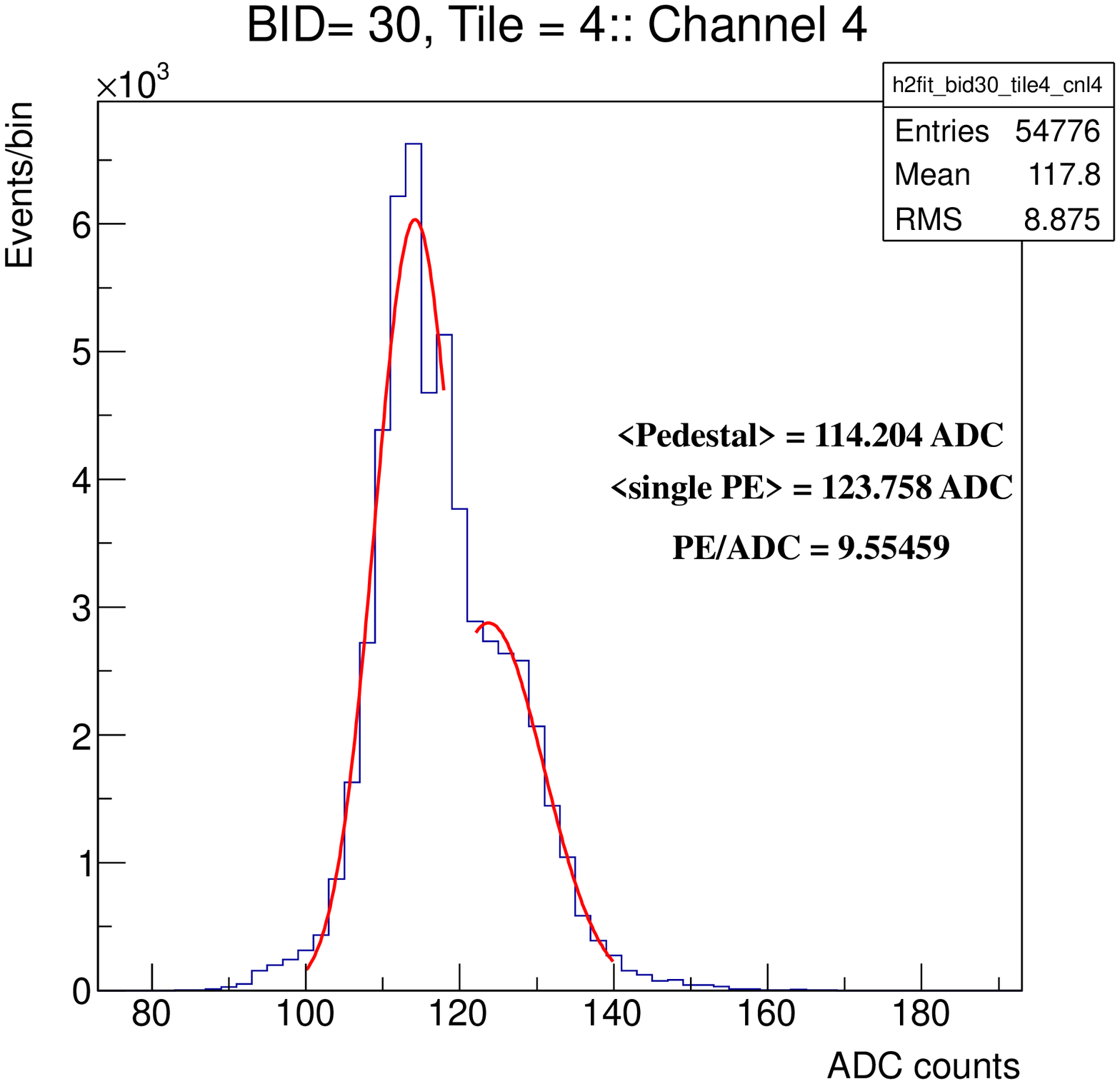}
 \includegraphics[scale=0.19]{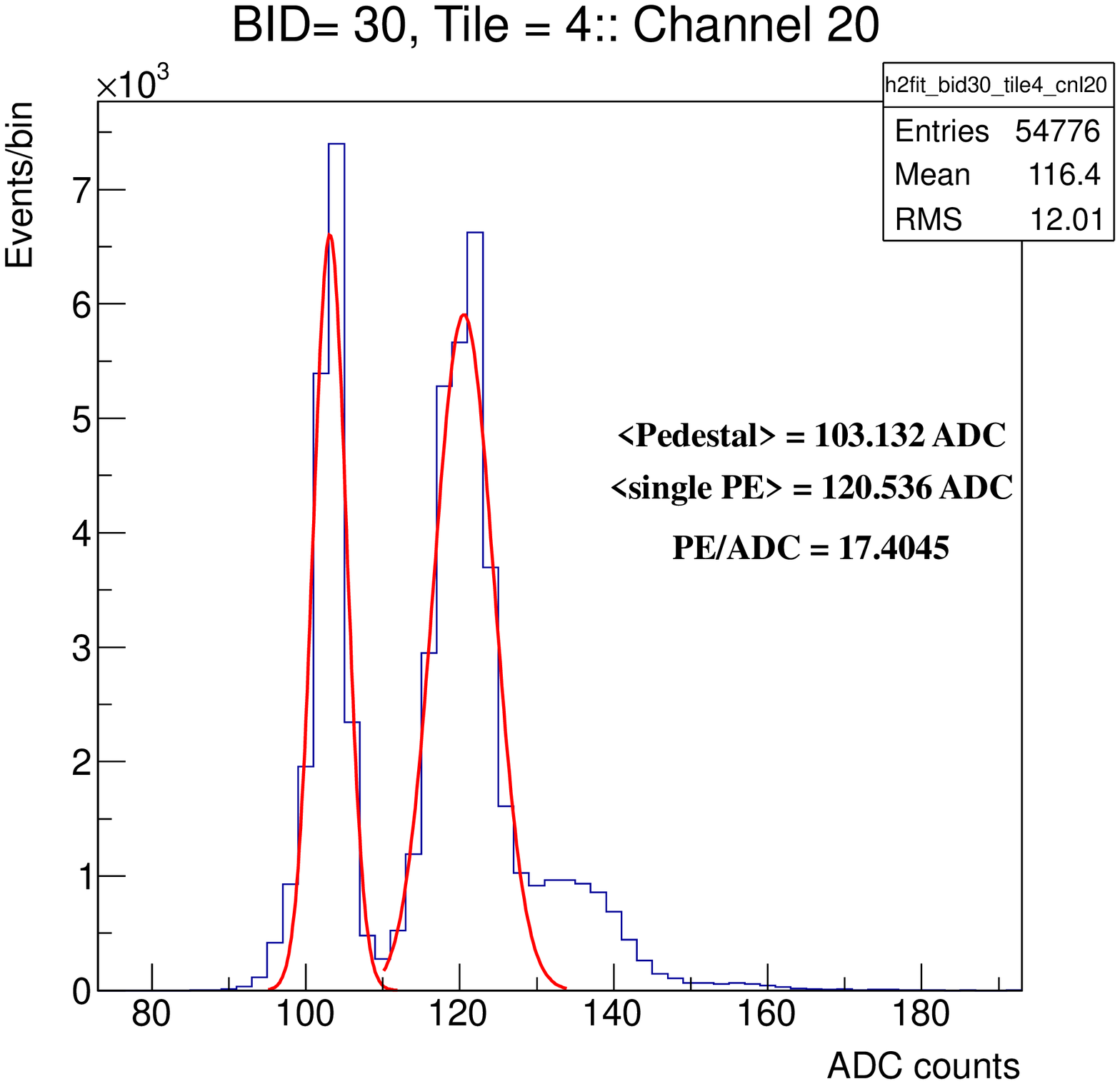}
 \includegraphics[scale=0.19]{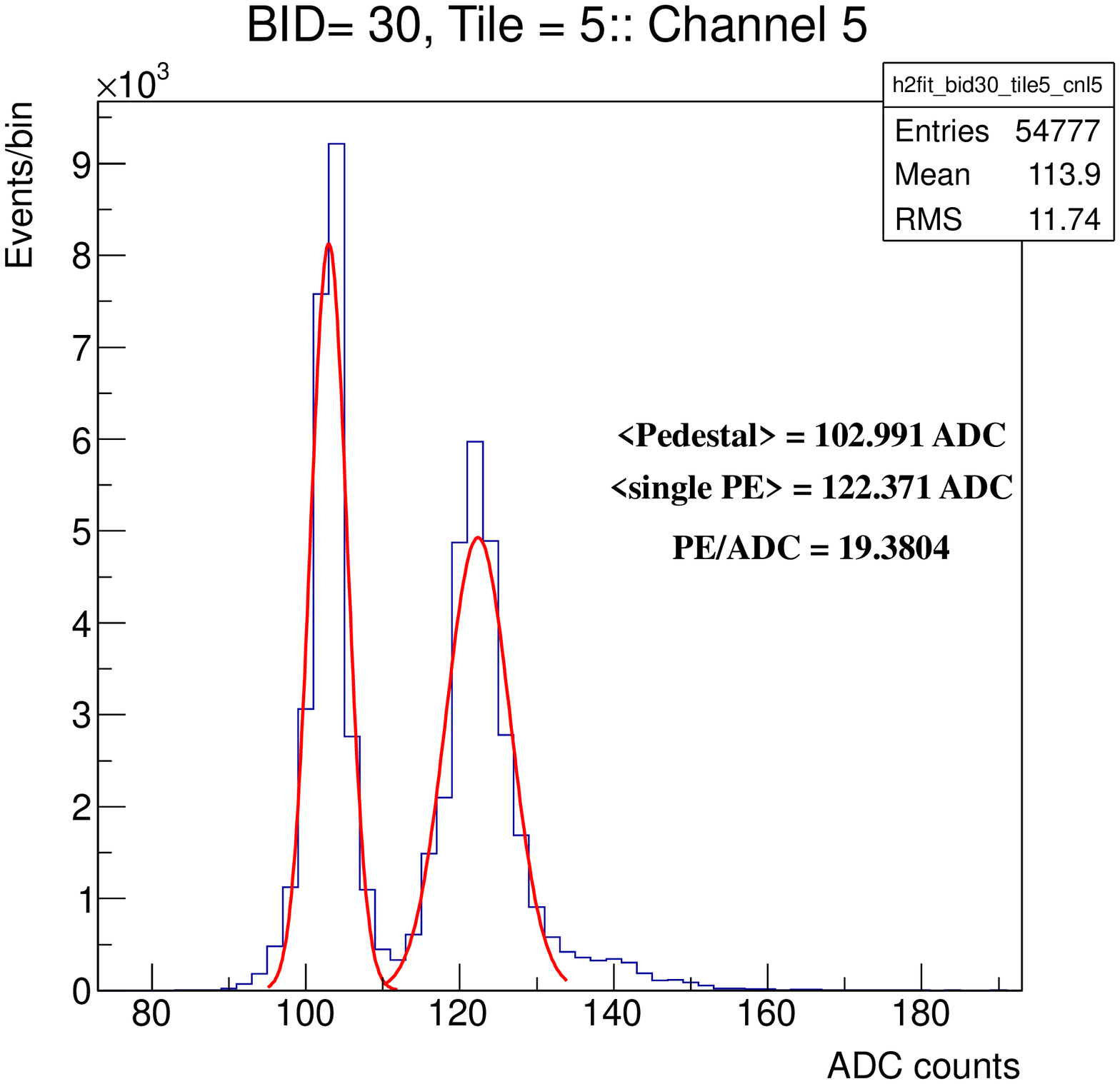}
 \includegraphics[scale=0.19]{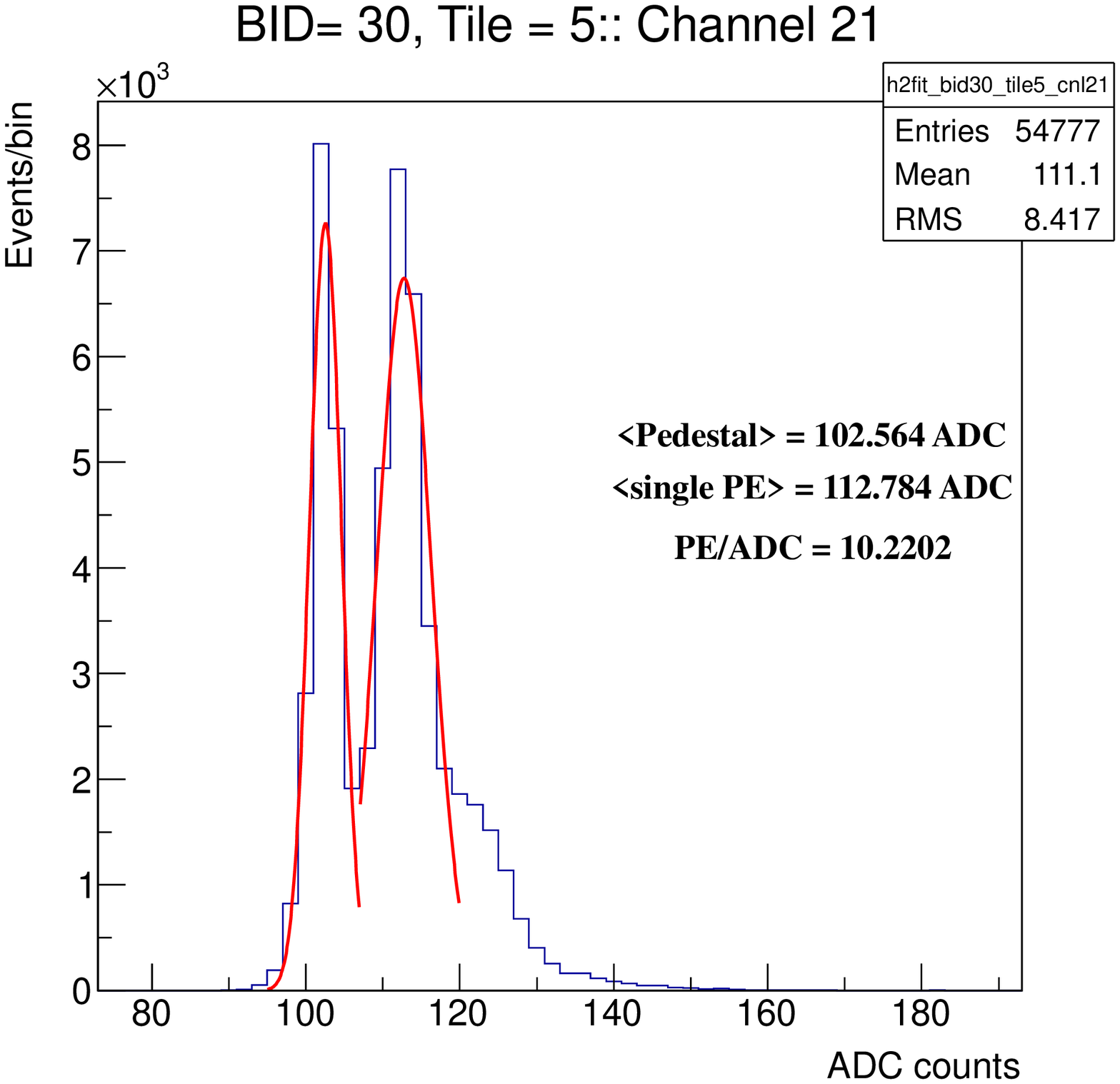}
 \leftline{  \hfill }
 \leftline{  \hfill }
\includegraphics[scale=0.19]{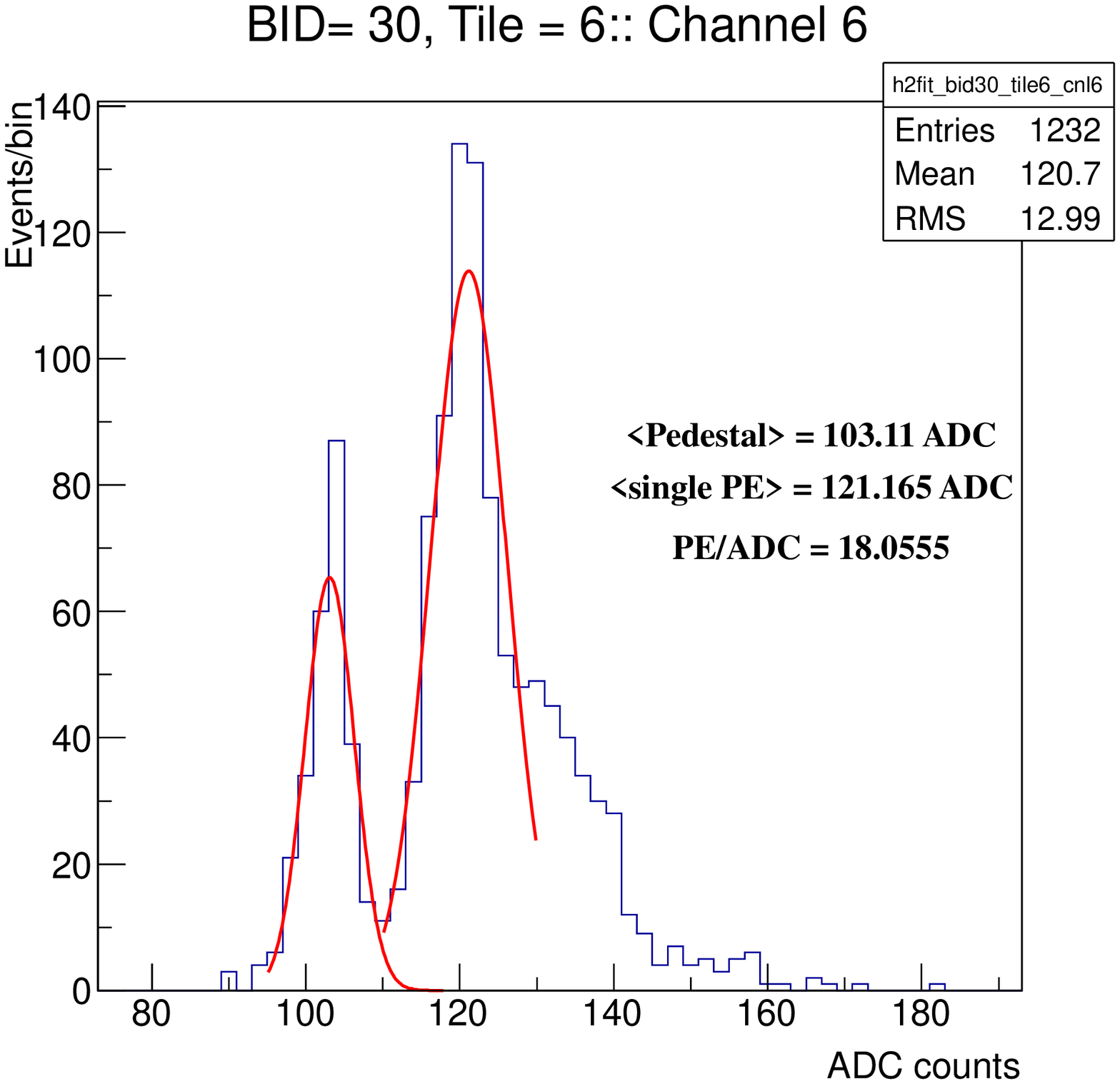}
 \includegraphics[scale=0.19]{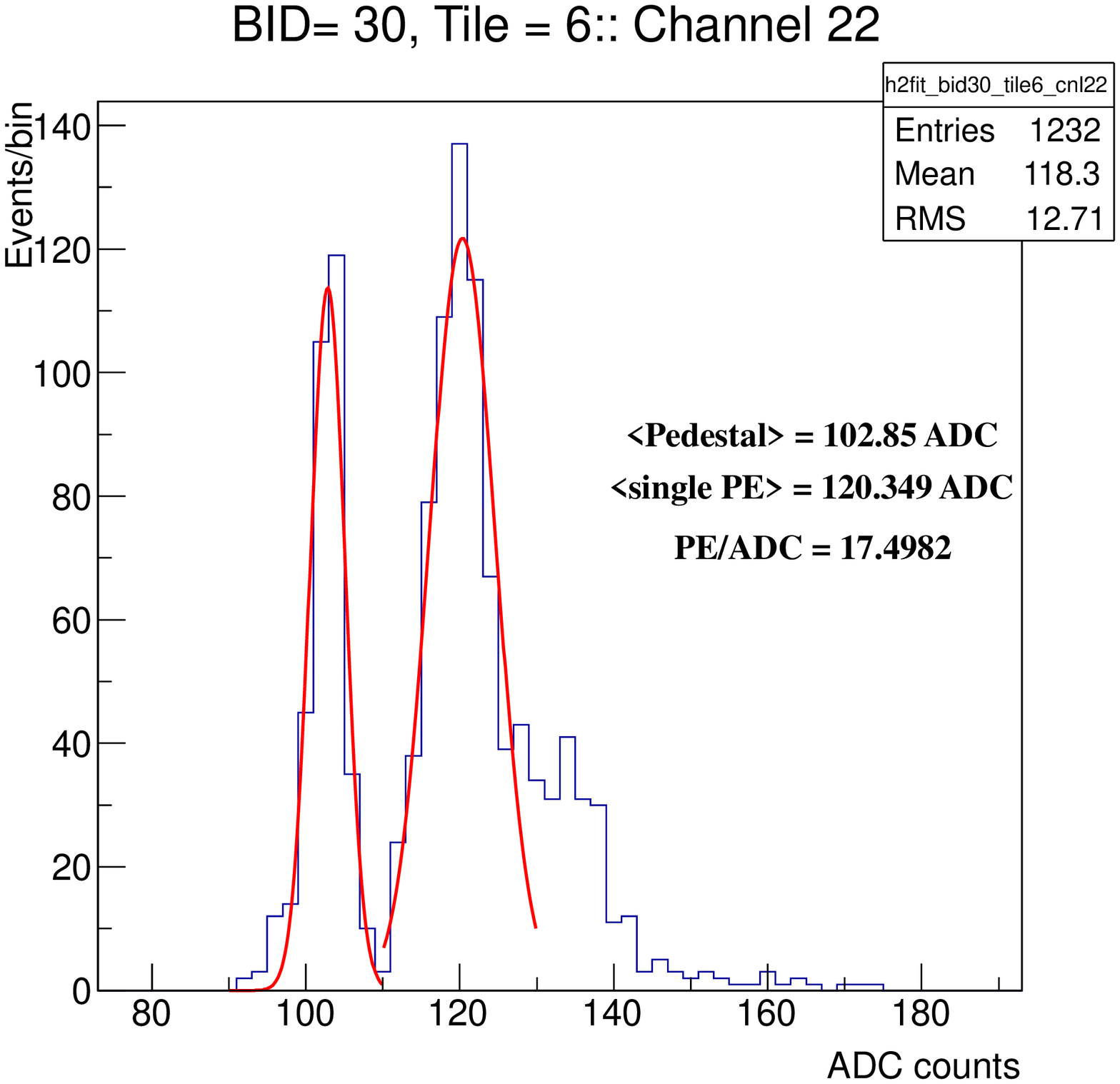}
 \includegraphics[scale=0.19]{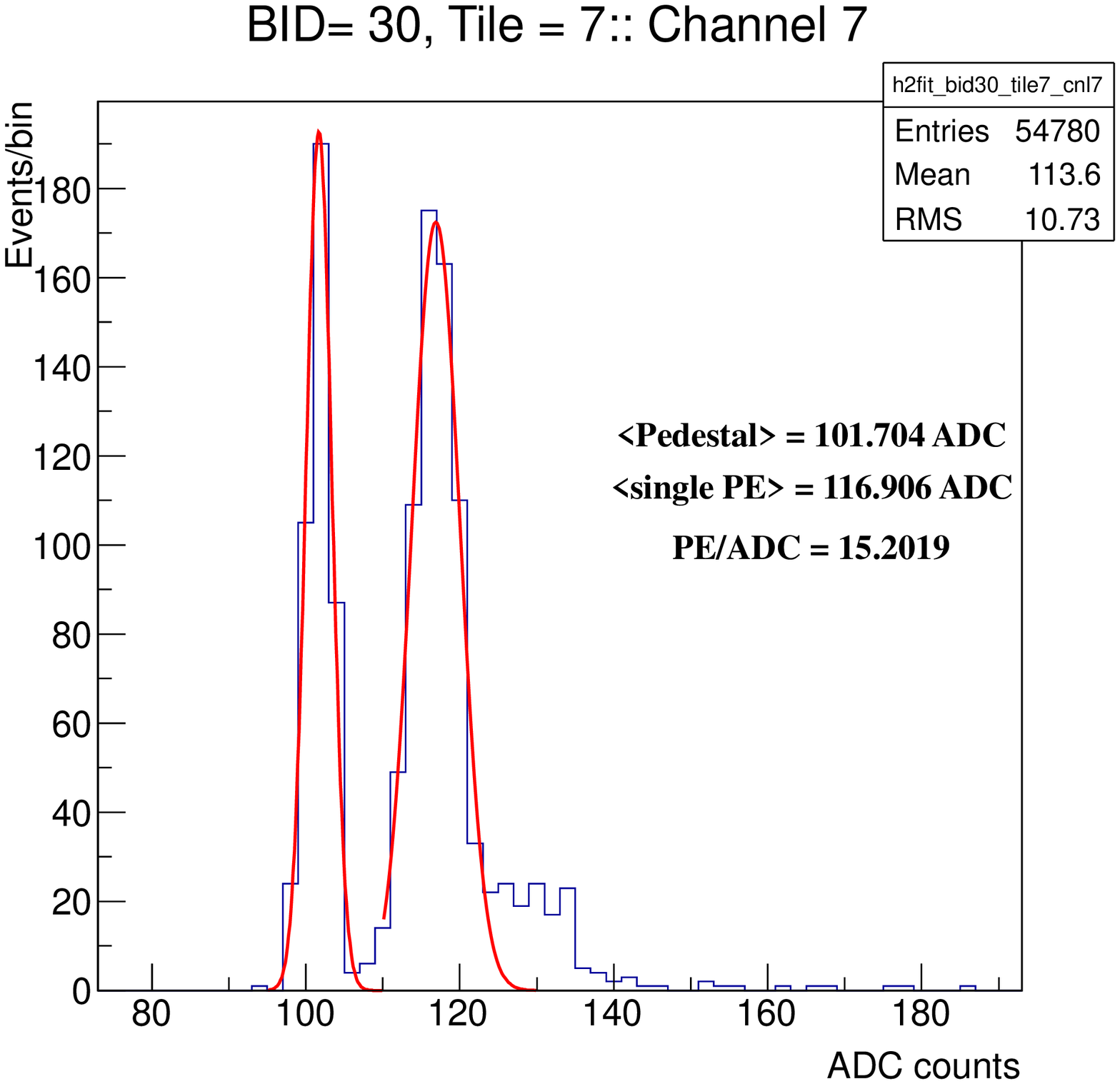}
 \includegraphics[scale=0.19]{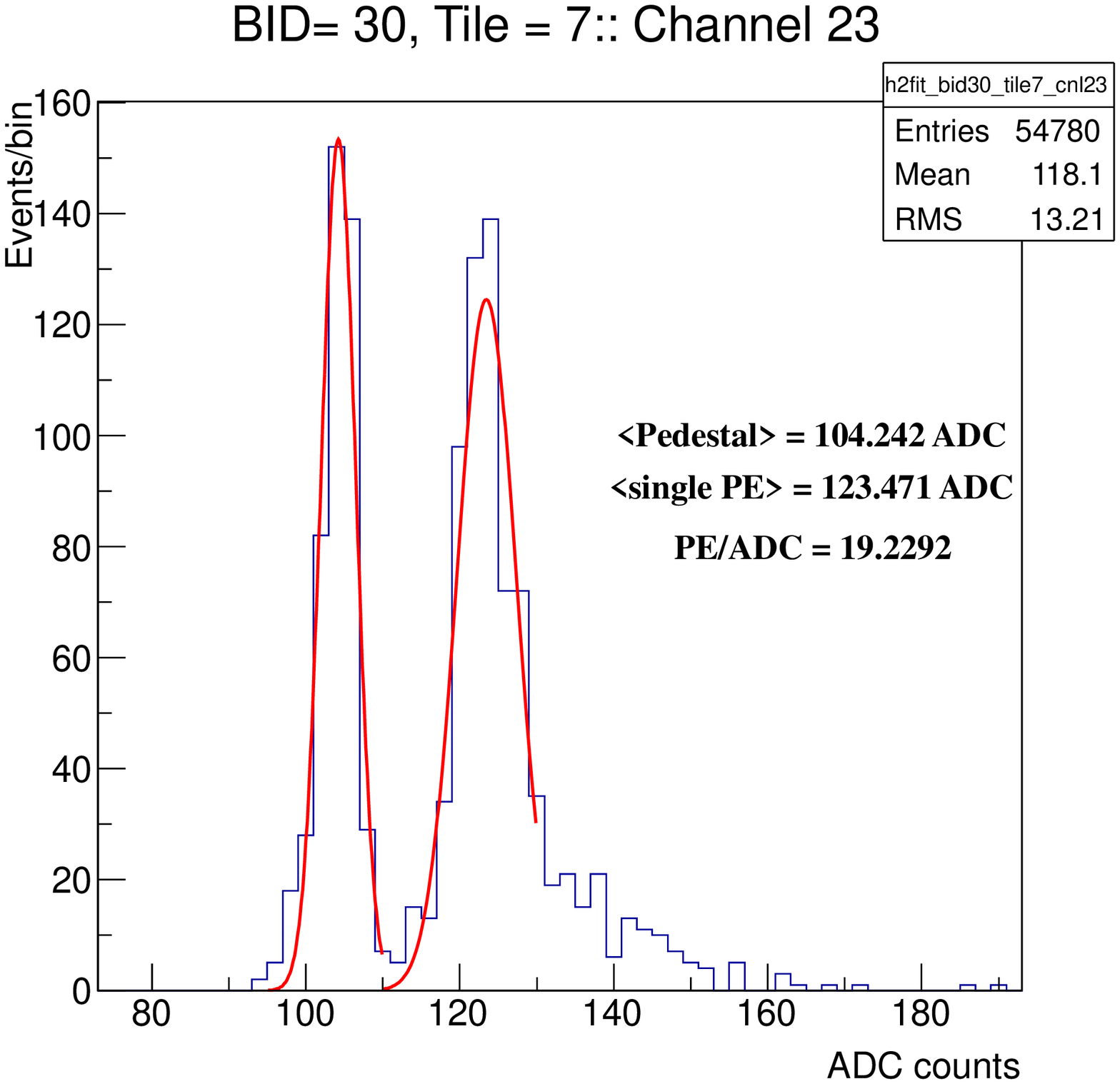}
 \leftline{  \hfill }

}
\caption{\label{fig:clb-bid30} No beam signals used for PE calibration for the first eight tiles of the range stack.}
\end{figure}

\begin{figure}[ht]
\centering
 {
  \includegraphics[scale=0.32]{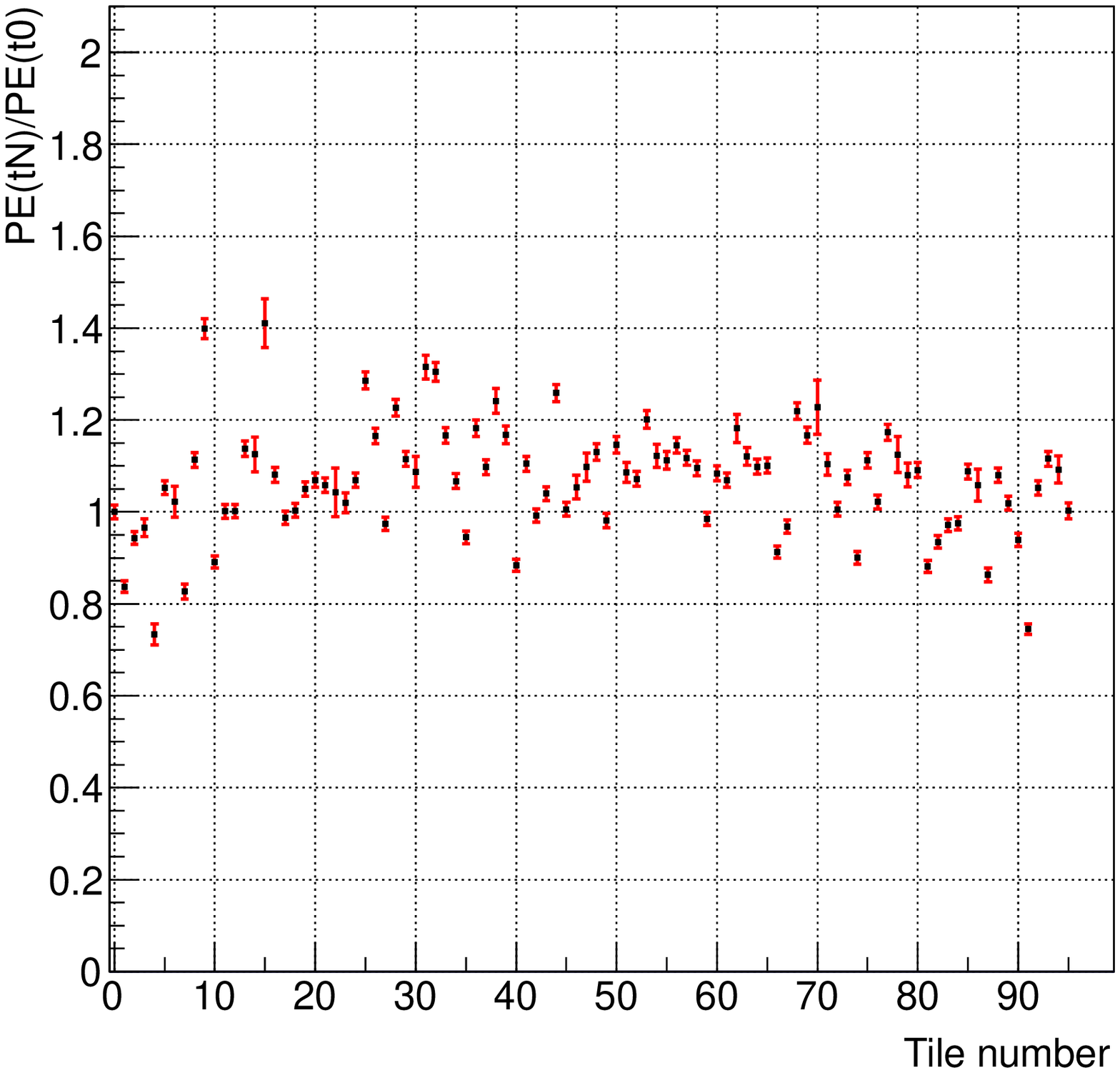}
  \includegraphics[scale=0.32]{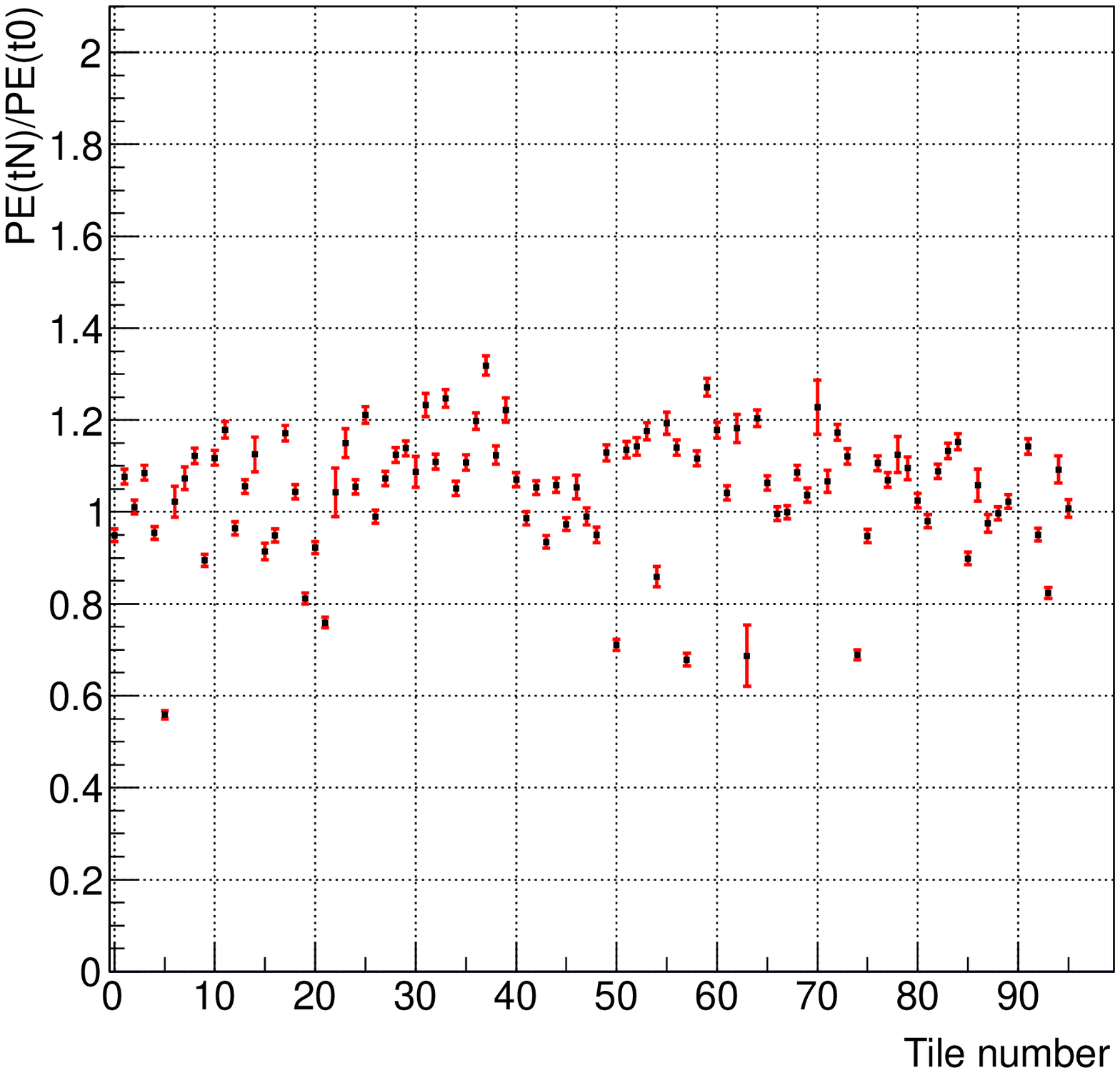}
}
\leftline{ \hspace{3.0cm}{\bf (a)} \hfill\hspace{2.5cm} {\bf (b)} hfill}
\caption{\label{fig:pe-ratio}
a) The ratio of PE conversion coefficients $K^{peSiPM0_{tn}}/ K^{peSiPM0_{t0}}$ for the first SiPMs.
b) The ratio of PE conversion coefficients $K^{peSiPM1_{tn}}/ K^{peSiPM0_{t0}}$ for the second SiPMs.
}
\end{figure}

1) We measure pedestal amplitudes $A^{pdSiPM1}_{tn}$, $A^{pdSiPM2}_{tn}$ and amplitudes $A^{1peSiPM1}_{tn}$, \\ $A^{1peSiPM2}_{tn}$
of the first  photo-electron (PE) peak for all range stack tiles, $tn$, by collecting events with no beam. Fig.~\ref{fig:calibration-signals}(a) and Fig.~\ref{fig:calibration-signals}(b) show these distributions  for SiPM1 and SiPM2 of Tile0.  
The combined  SiPM1+SiPM2 no-beam signal in Tile0 is shown in  Fig.~\ref{fig:calibration-signals}(c). 
Figure~\ref{fig:clb-bid30} shows calibration signals for all 16 SiPMs of the first range stack frame. 
From these data the ADC to PE conversion coefficients for each SiPM  are calculated as 
% \begin{equation}
% \label{bb3}    
\[
K^{peSiPM}_{tn} =   A^{1peSiPM}_{tn} - A^{pdSiPM}_{tn}
\]
% \end{equation}
% In Tile6 of each frame the individual SiPM signals (readout channels 6 and 22) 
% appear to be corrupted, however their combined signal still does look reasonable. Conversion  
% coefficients for SIPMs in Tile6 in each frame are calculated from the combined signal as \\
%                                          \centerline{$K^{peSiPM}_{tn} = 0.5 *( A^{1peSum}_{tn} - A^{pdSum}_{tn}$).}\\ 
Ratios of PE conversion coefficients  $K^{peSiPM0}_{tn}/ K^{peSiPM0}_{t0}$ of the first and second SiPM in each tile to the 
conversion coefficient in the first SiPM in Tile0 are  shown  in  Fig.~\ref{fig:pe-ratio}(a) and  Fig.~\ref{fig:pe-ratio}(b). Most sensors
have a response within 10\% of one another.
%\clearpage
%

\begin{figure}[ht]
\centering
 {
  \includegraphics[scale=0.27]{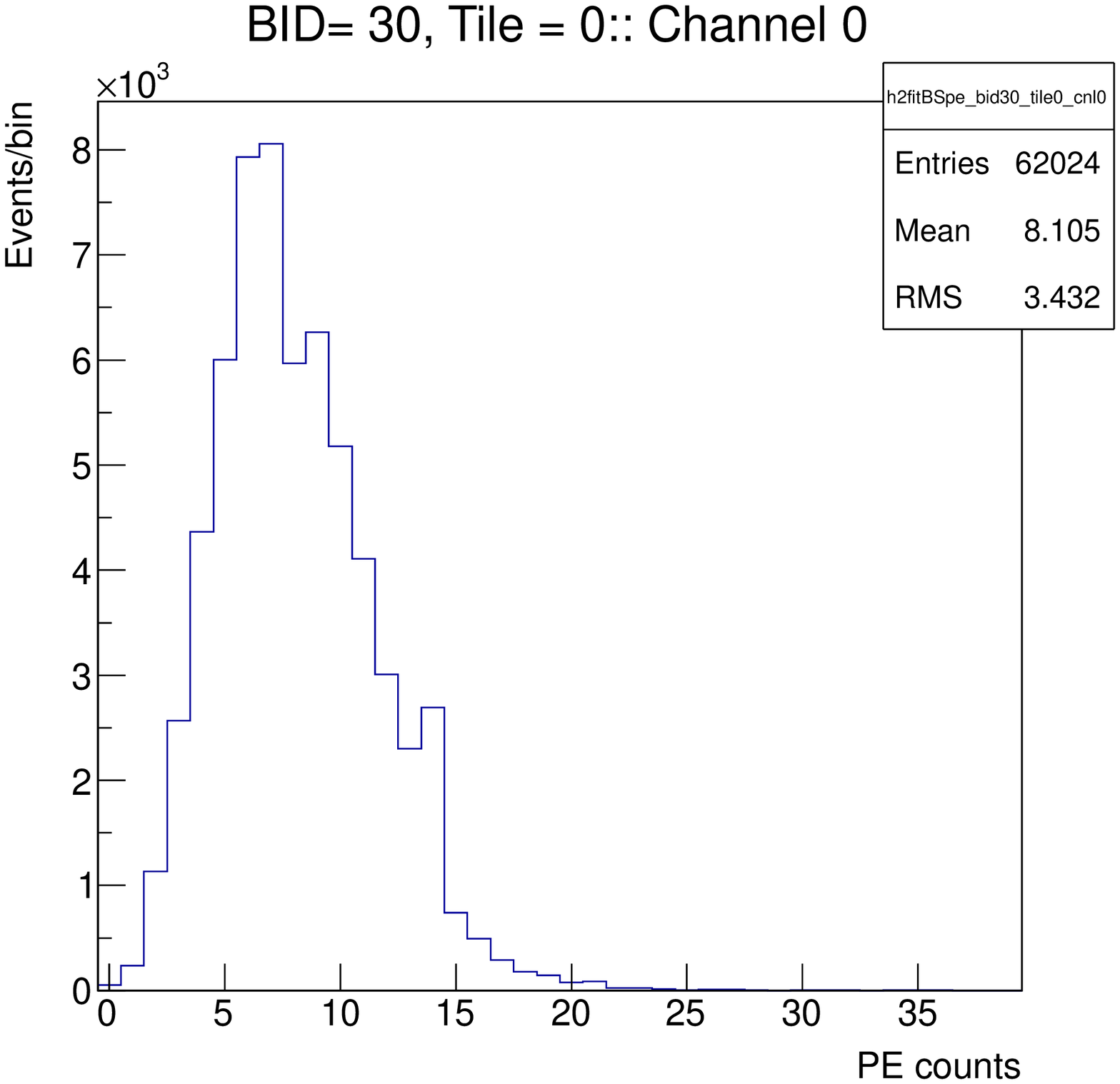}
  \includegraphics[scale=0.27]{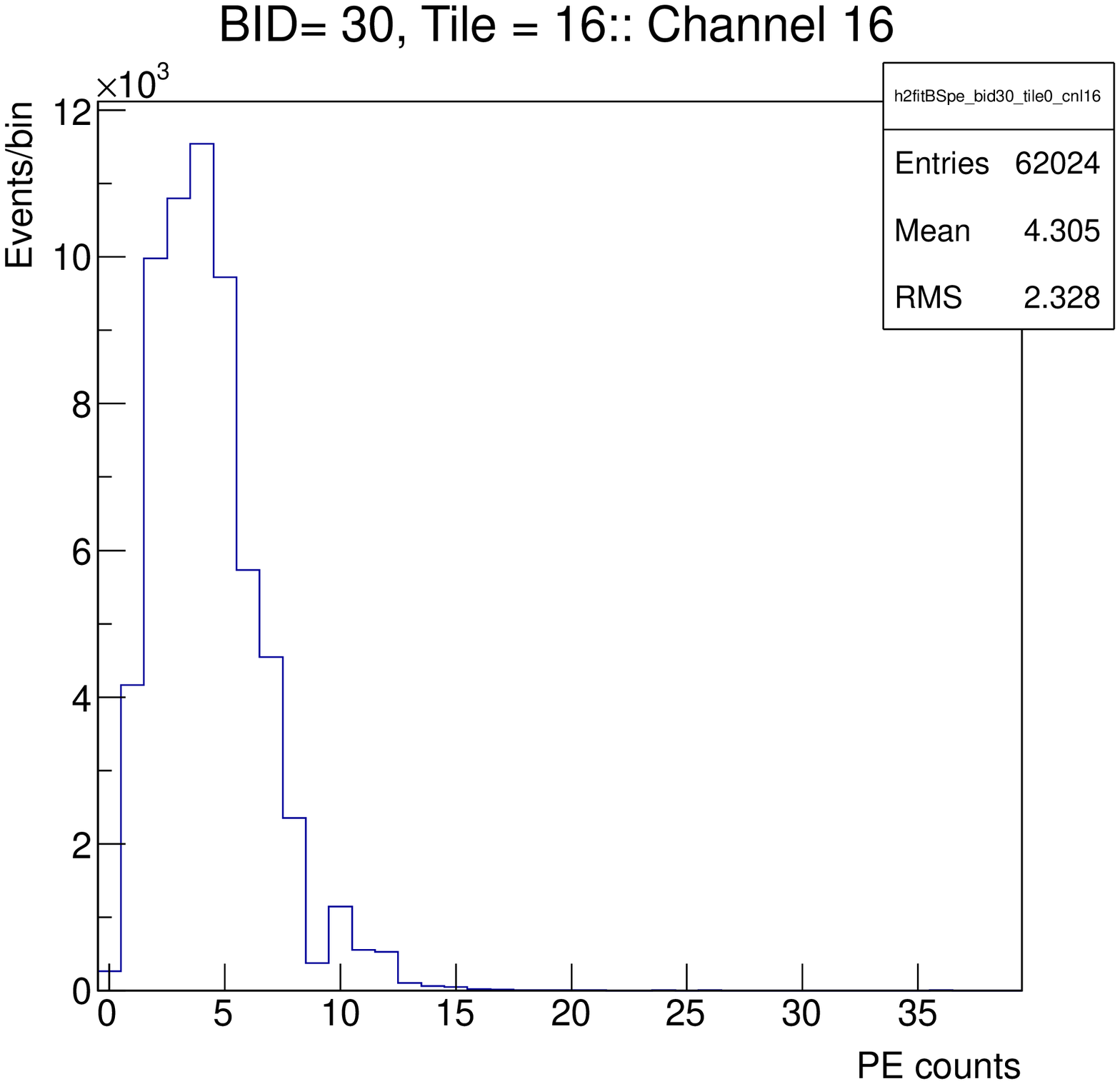}
  \includegraphics[scale=0.27]{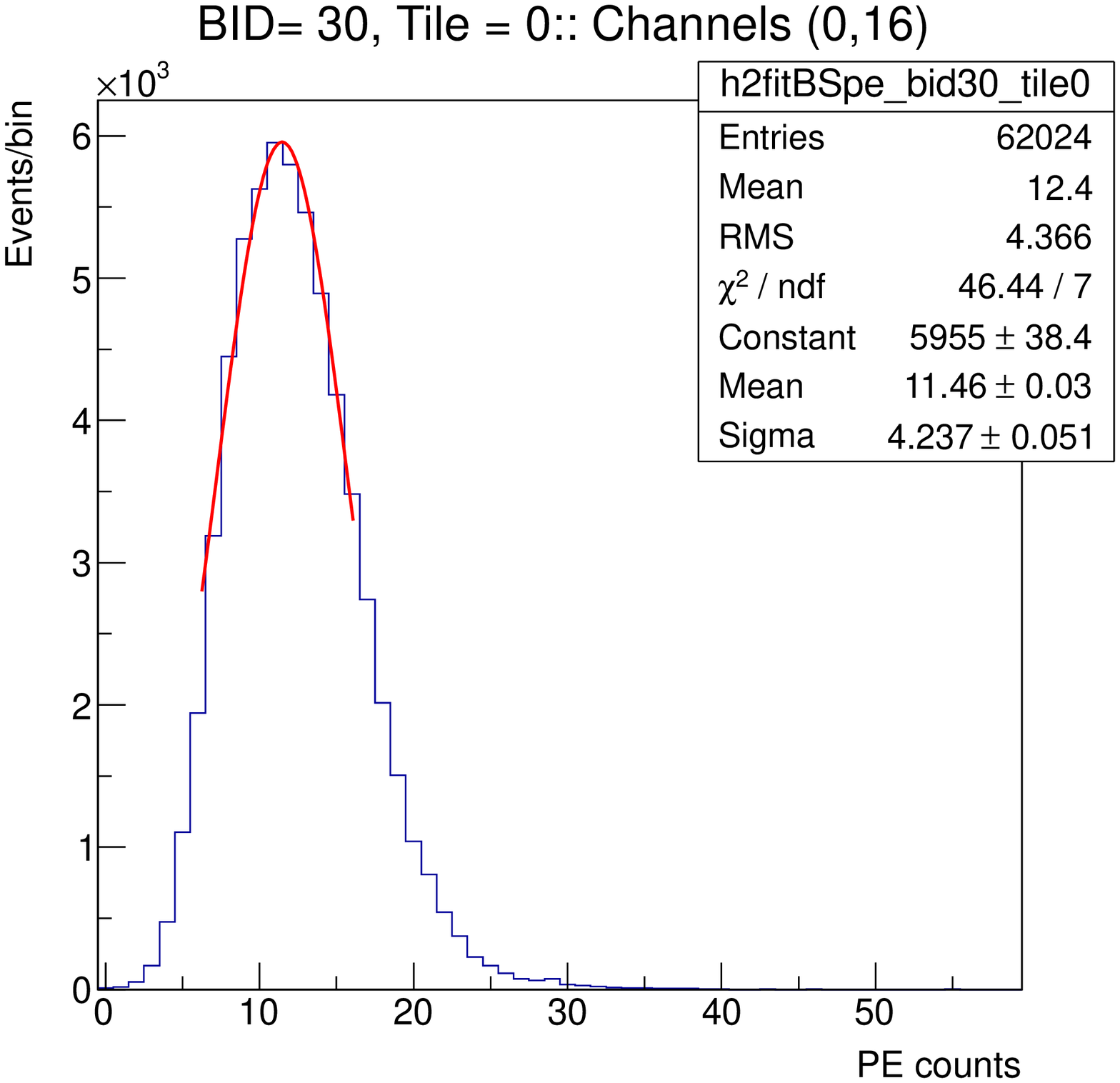}
}
\leftline{ \hspace{3.0cm}{\bf (a)} \hfill\hspace{2.5cm} {\bf (b)} \hfill\hspace{2.5cm} {\bf (c)} \hfill}
\caption{\label{fig:signal-example} Measured signal amplitudes (PE units) in Tile0 at a beam energy of  26~cm (200 MeV) after subtracting pedestals : 
a) in  SiPM1;
b) in  SiPM2;
c) sum of SiPM1 and SiPM2.  The means of Gaussian fits of combined signals away from the 
Bragg peak at a beam energy of  32~cm (225 MeV) 
were used  to extract the normalization coefficients for the range stack tiles.
}
\end{figure}

2) The proton energy deposition in each tile  $E^{pe}_{tn}$ in PE units is obtained via
\[
    \centerline{$E^{pe}_{tn} = (A^{sSiPM1}_{tn}- A^{pdSiPM1}_{tn})/ K^{peSiPM1}_{tn} + (A^{sSiPM2}_{tn}- A^{pdSiPM2}_{tn})/ K^{peSiPM2}_{tn}$.}
\]
 Fig.~\ref{fig:signal-example}(a)  and Fig.~\ref{fig:signal-example}(b)  show the PE signals of SiPM1 and SiPM2 in Tile0, Fig.~\ref{fig:signal-example}(c) shows the combined  SiPM1+SiPM2  signal, $A^{pe}_{t0}$, in Tile0.\\
%New paragraph

3)  We measure signals $E^{clbExp}_{tn}$ of all range stack tiles in the region far away from the Bragg peak.
We conducted two calibration runs at an energy of 32~cm (225 MeV).  For the second run, the assembled scanner 
was turned  180~degrees to expose the back tiles to the beam first.  The ``front'' run is used 
to calibrate the first front 48 tiles of the stack, while  the ``back'' run is used to calibrate the 48 back tiles.  
We assume that the ``true'' $E^{clbTrue}_{tn}$ 
amplitudes of the tile signals follow energy profiles calculated from proton energy-range tables 
for polystyrene (the material used for the range stack tiles). 
%new Prg

\begin{figure}[ht]
\centering
 {
  \includegraphics[height=4.8 cm]{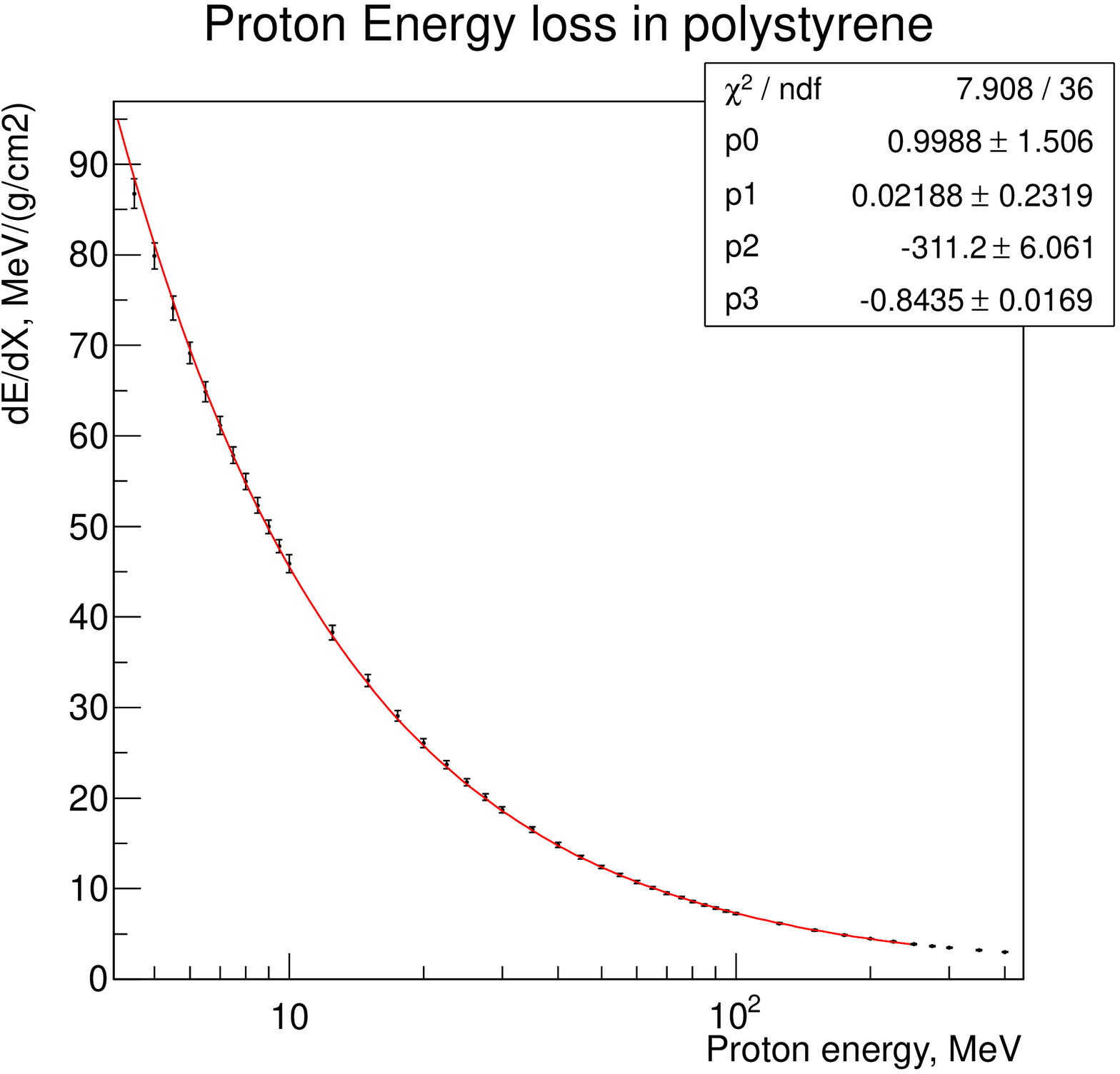}
  \includegraphics[height=5 cm]{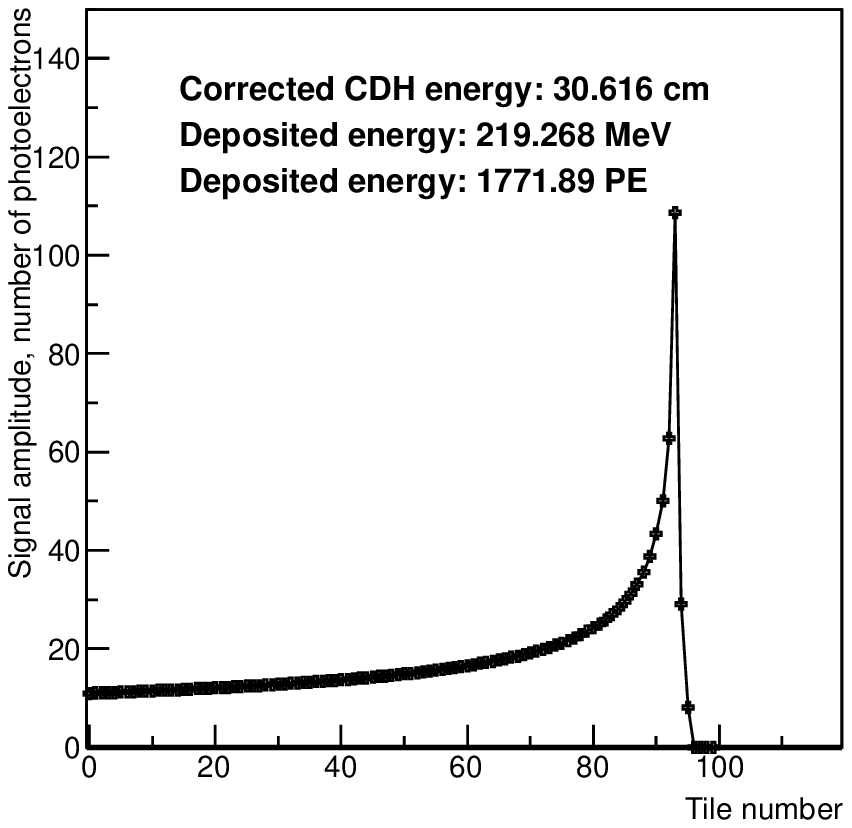}
  \includegraphics[height=5 cm]{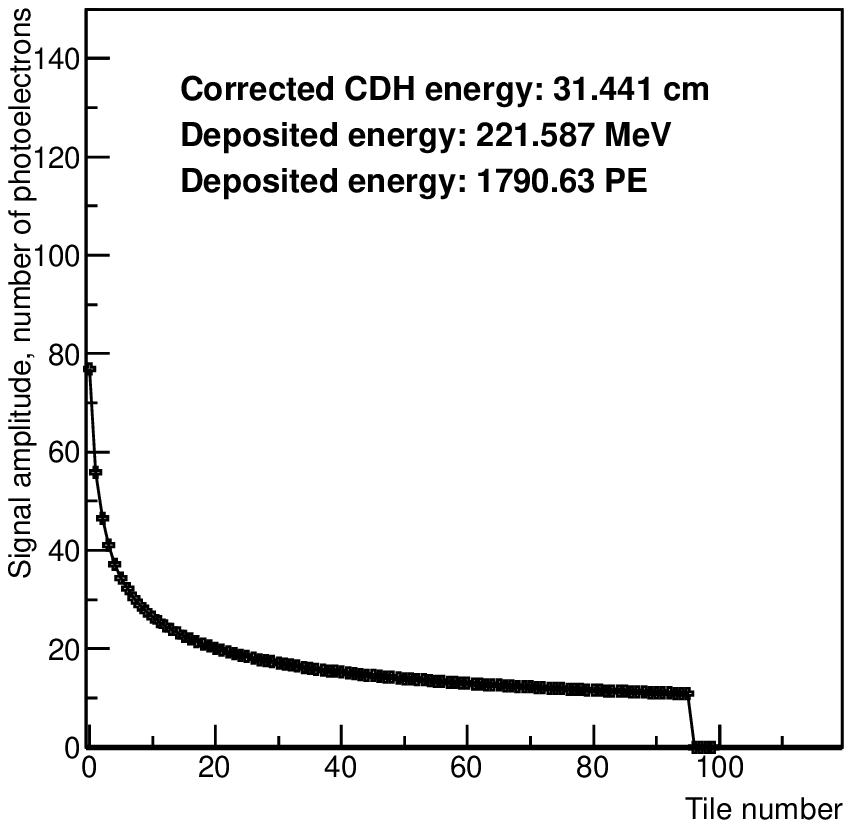}
}
\leftline{ \hspace{3.0cm}{\bf (a)} \hfill\hspace{2.5cm} {\bf (b)} \hfill\hspace{2.5cm} {\bf (c)} \hfill}
\caption{\label{fig:janni_profile} a) The proton dE/dX dependency in polystyrene as tabulated in Janni's proton energy-range tables.
b) The ``true'' front run signal profile used for calibration of tiles (0-47) of the range stack.
c) The ``true'' back run signal profile used for calibration of tiles (48-95) of the range stack.
}
 \end{figure}

Figure~\ref{fig:janni_profile}(a) shows  
the tabulated proton dE/dx dependence. Fig.~\ref{fig:janni_profile}(b) and Fig.~\ref{fig:janni_profile}(c) show 
energy profiles calculated for protons entering the range stack with energies of 30.6~cm  in the ``front'' run 
and 31.4~cm in the ``back'' run (corrections to the nominal CDH accelerator energy were applied to account 
for material in the tracker which is only present in the ``front'' run configuration and material in the CDH beam transport line, as discussed in Section~\ref{ecorr}). 
All ``true'' $E^{clbTrue}_{tn}, tn=0,95$   amplitudes are normalized to the 
signal $E^{clbExp}_{t0}$  of the Tile0 in the ``front'' run.  That is, we take the observed energy in Tile0 as to be correct.
%new Prg

\begin{figure}[ht]
\centering
{
 \includegraphics[scale=0.50]{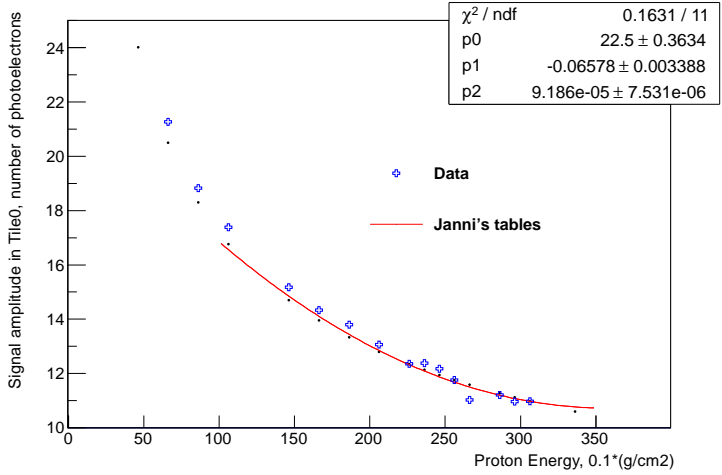}
}
\caption{\label{fig:tile0_dataVSjnt}Measured signal amplitudes (blue crosses) and expected amplitudes 
calculated from Janni's tables (red line) in Tile0 of the range stack for different proton energies. 
The expected signals normalized to the mean (over 10000 protons) Tile0 data signal in the 32~cm run. }
\end{figure}

\begin{figure}[ht]
\centering
 {
 \includegraphics[scale=0.40]{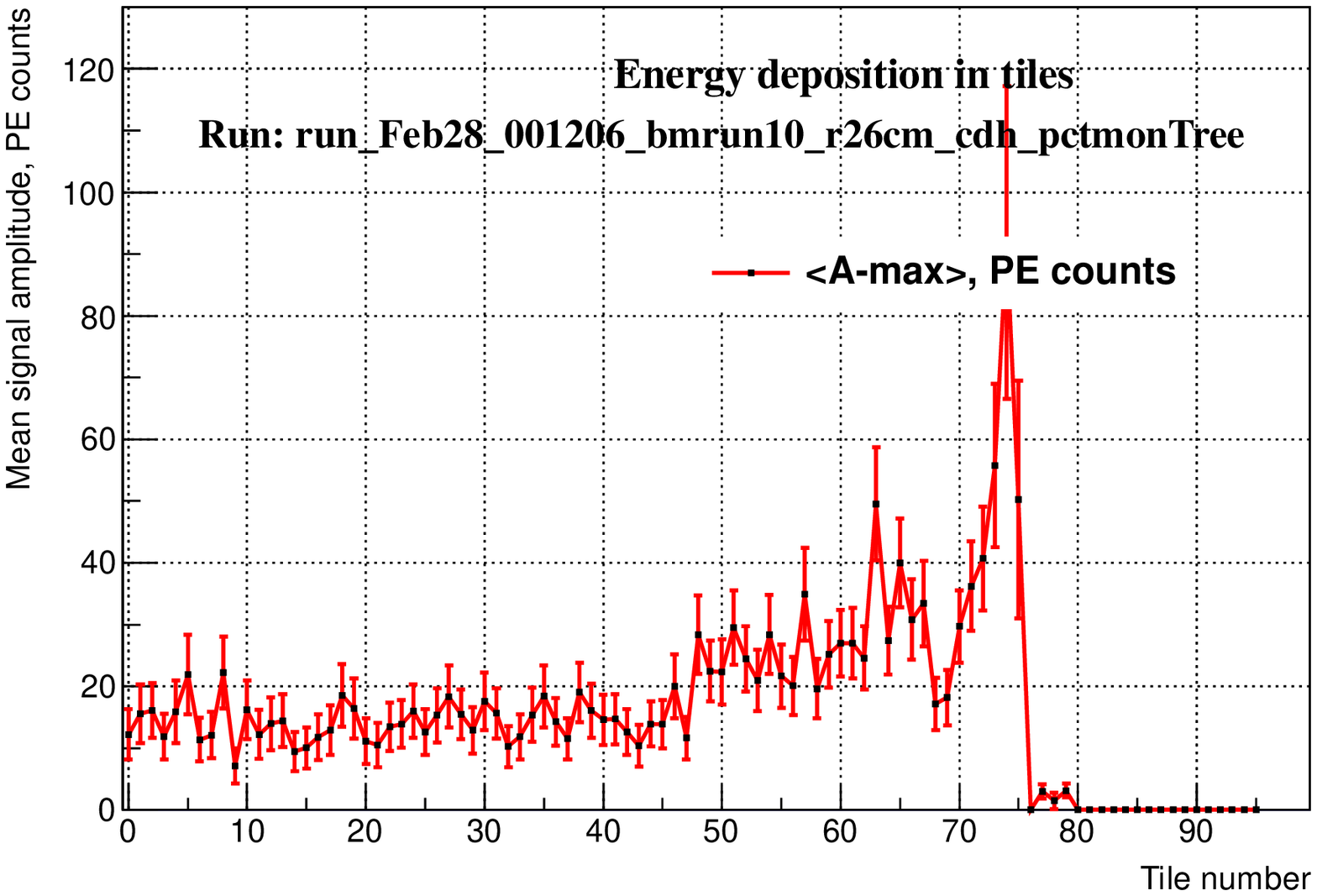}
 \includegraphics[scale=0.40]{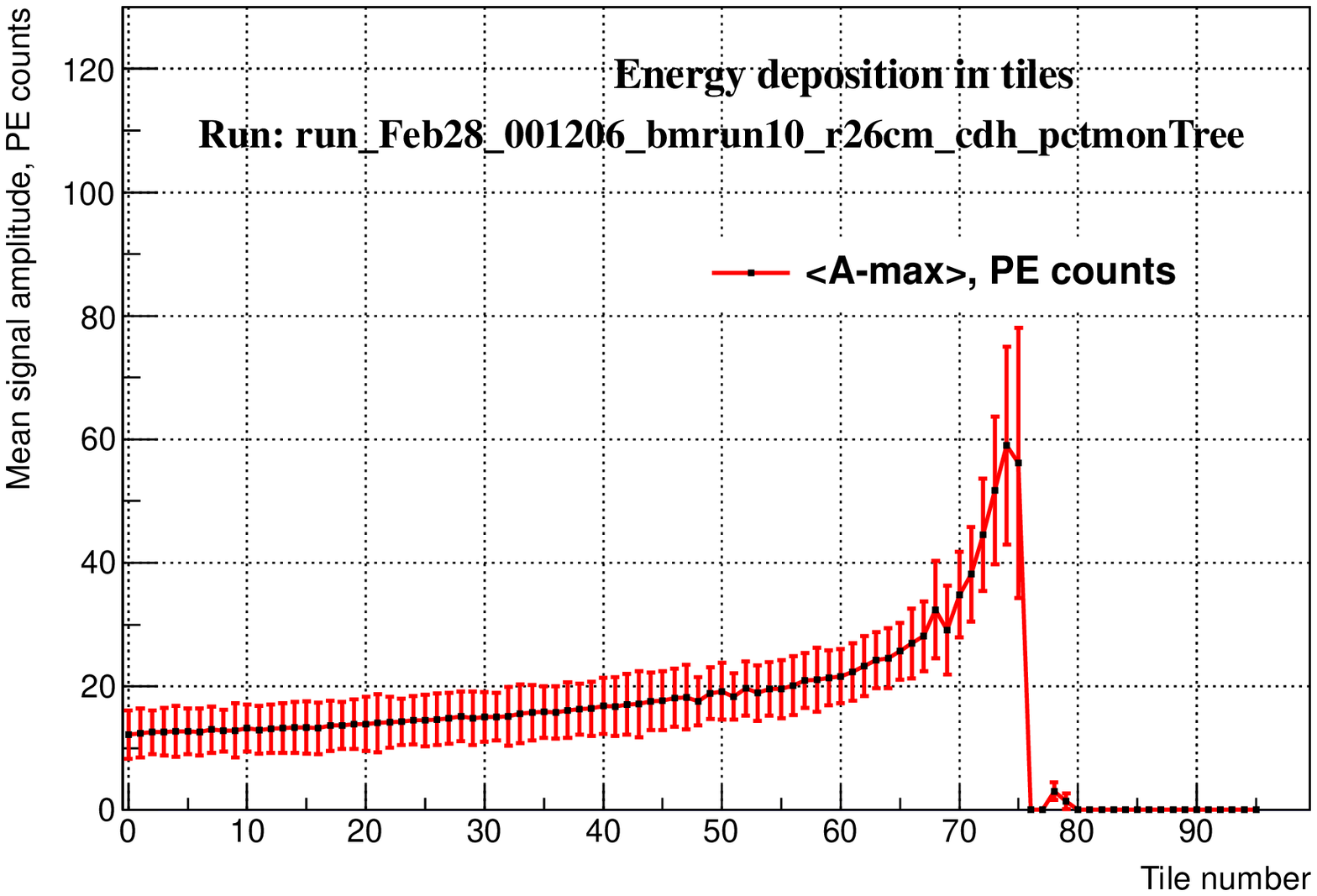}
 }
\leftline{ \hspace{4.5cm}{\bf (a)} \hfill\hspace{3.5cm} {\bf (b)} \hfill}
\caption{\label{fig:stack-profiles-clb} The mean number of photoelectrons in the range stack tiles as a function of tile number
produced by protons with energy 26~cm (200~MeV)  (a) raw; (b) calibrated. 
The errors bars reperesent $\pm 1$ sigmas of Gaussian fits about the average, for an example see  Fig.~\ref{fig:signal-example}(c) .
}
\end{figure}

\begin{figure}[ht]
\centering
 {
 \includegraphics[scale=0.40]{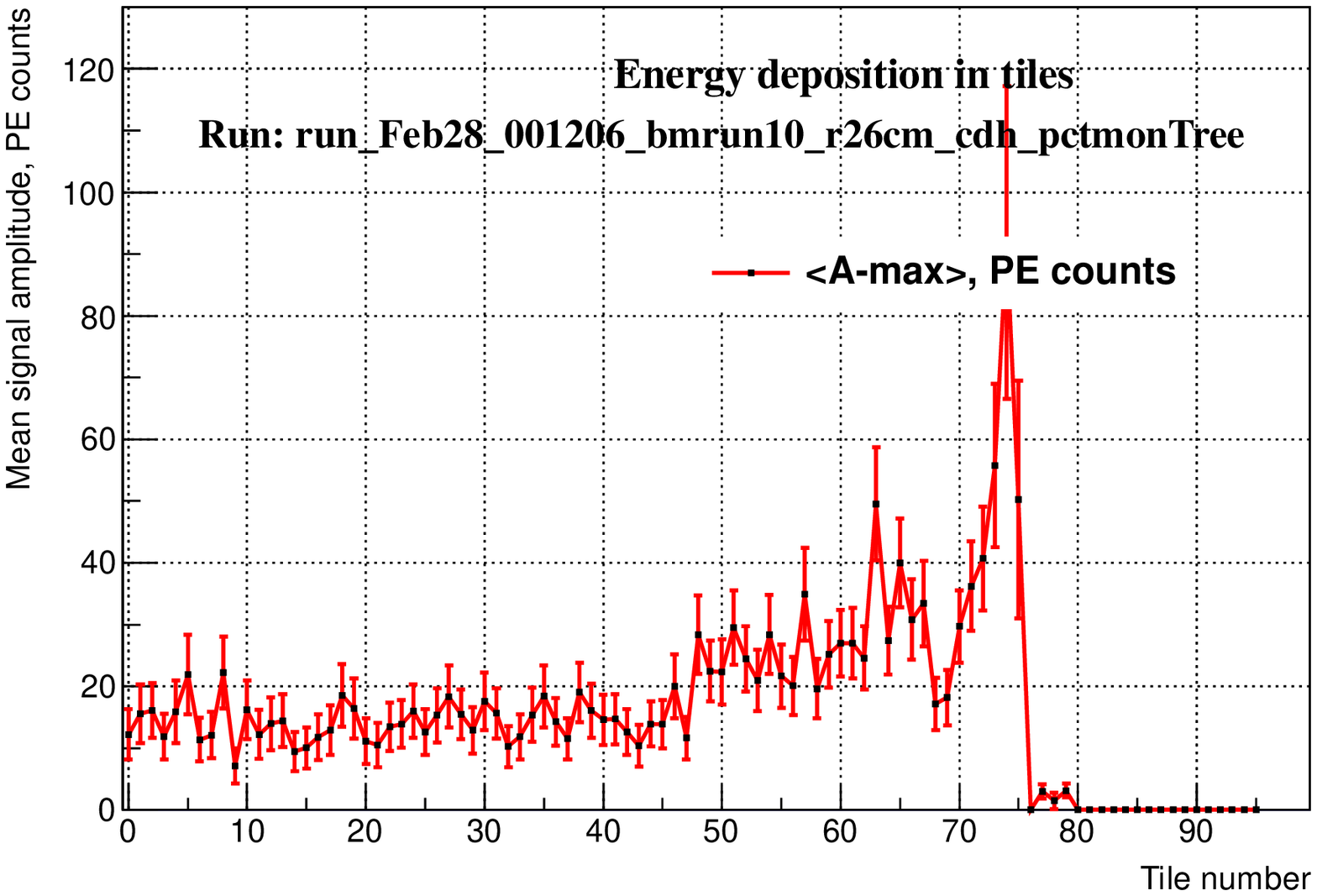}
 \includegraphics[scale=0.40]{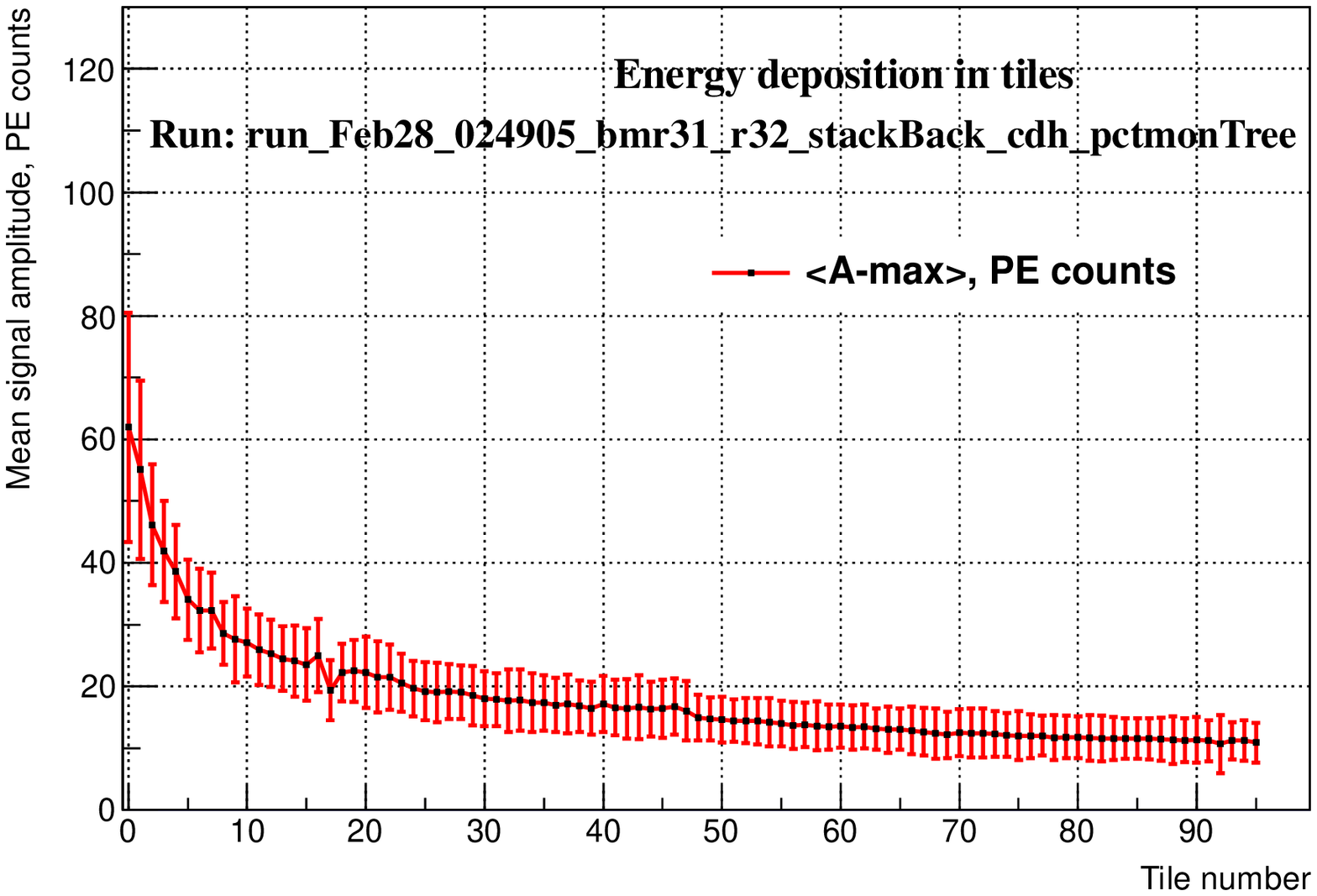}
 }
 \leftline{ \hspace{4.5cm}{\bf (a)} \hfill\hspace{3.5cm} {\bf (b)} \hfill}
\caption{\label{fig:stack-profiles-clb_r32} The mean number of photoelectrons in the range stack tiles as a function of tile number
produced by protons with energy (a) 32~cm (225~MeV), ``front'' run ; (b) 32~cm (225 MeV), ``back'' run, no tracker. 
The errors bars reperesent $\pm 1$ sigmas of Gaussian fits about the average.
%.
}
\end{figure}

\begin{figure}[ht]
\centering
 {
 \includegraphics[scale=0.40]{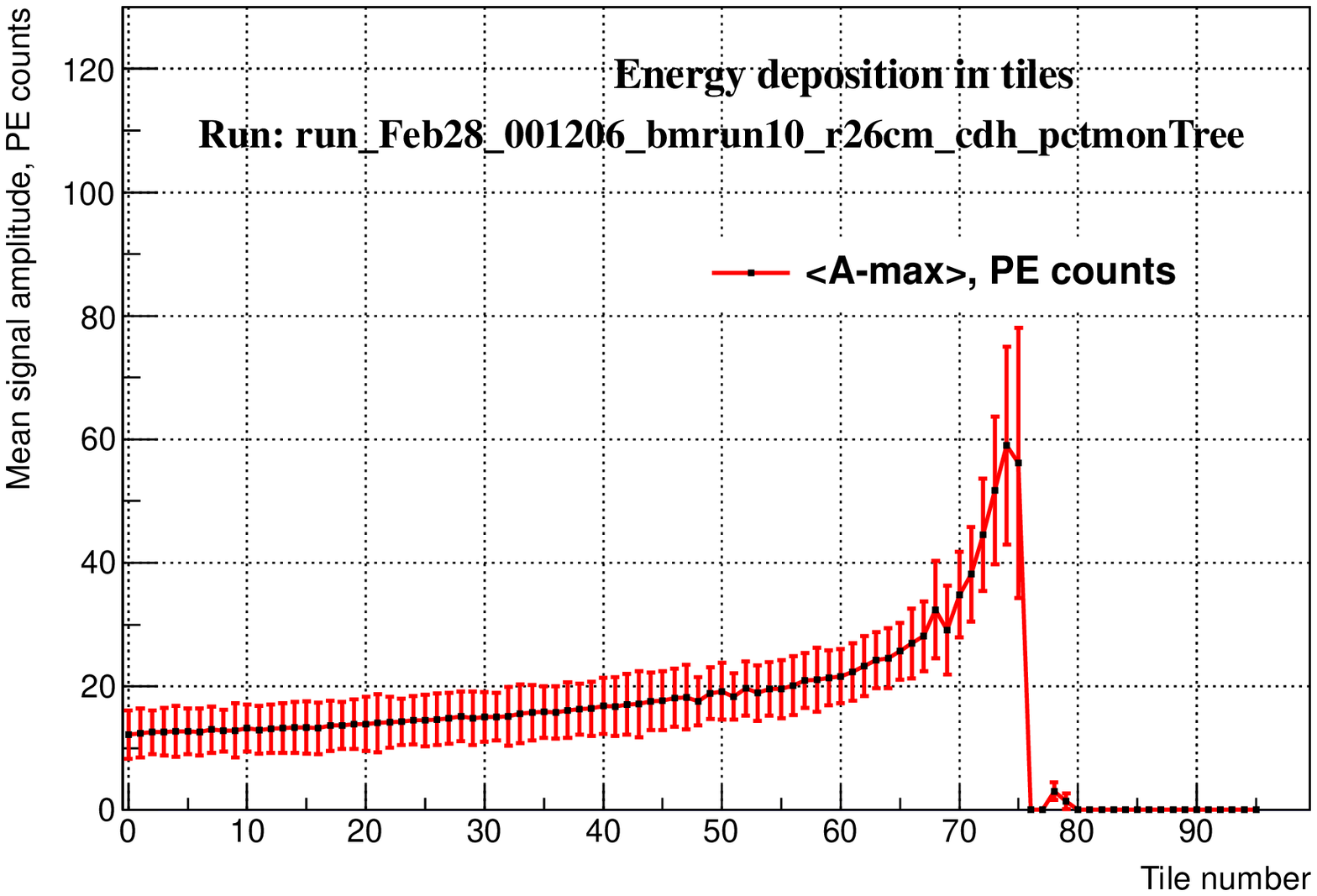}
 \includegraphics[scale=0.40]{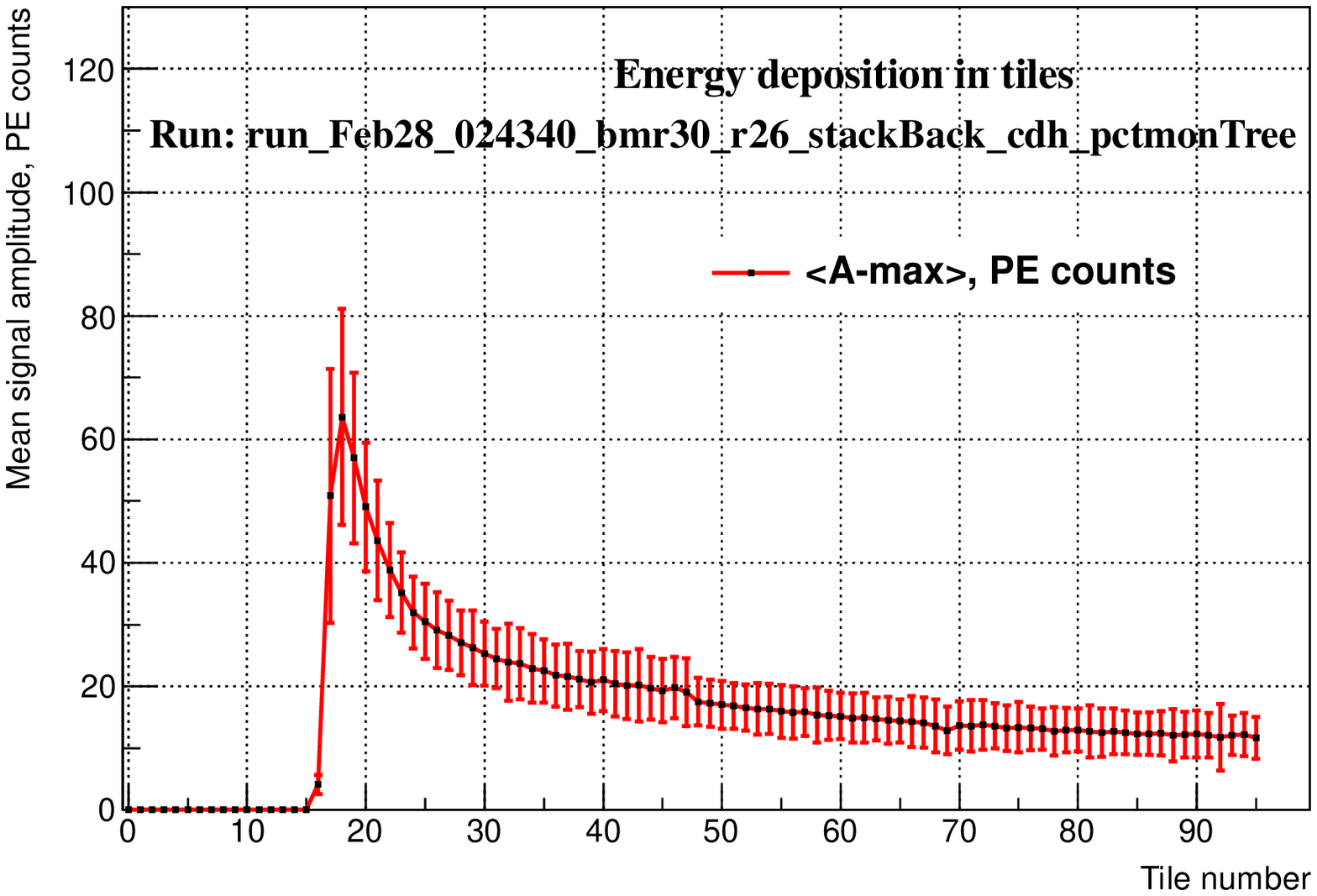}
 }
 \leftline{ \hspace{4.5cm}{\bf (a)} \hfill\hspace{3.5cm} {\bf (b)} \hfill}
\caption{\label{fig:stack-profiles-clb_r26} The mean number of photoelectrons in the range stack tiles as a function of tile number
produced by protons with energy (a) 26~cm (200 MeV, ``front run'' ) ; (b) 26~cm (200 MeV, ``back run'' ), no tracker. 
The errors bars reperesent $\pm 1$ sigmas of Gaussian fits about the average.
The three tile difference in the stopping position ($nt_{stop}$ = 74 in the front run, while $nt_{stop}$=77 in
the back run run without the  tracker) agrees with the tracker stopping power, ~$3.2\times1.01\times3 = 0.97$~cm.}
\end{figure}

\begin{figure}[ht]
\centering
 {
 \includegraphics[scale=0.27]{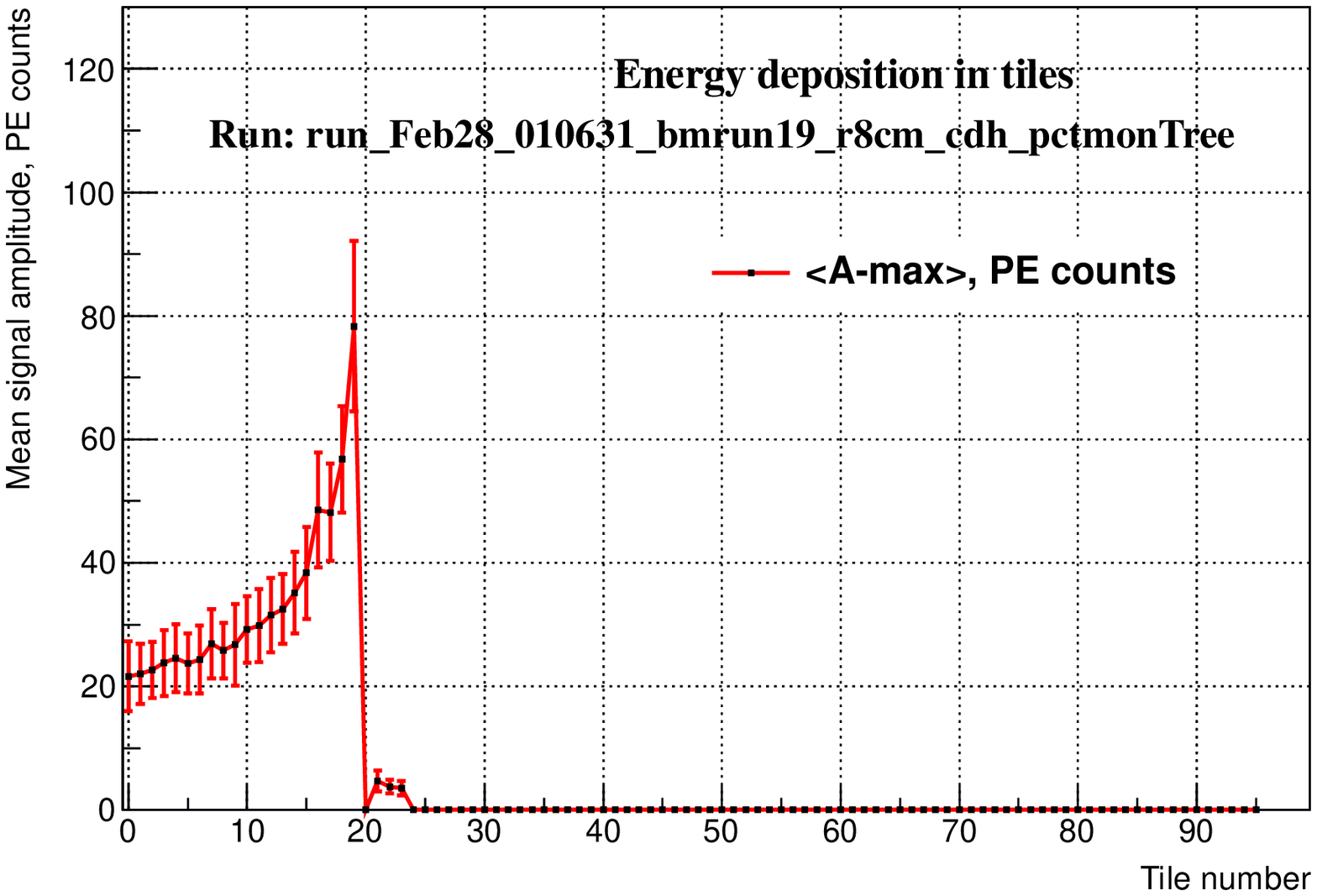}
 \includegraphics[scale=0.27]{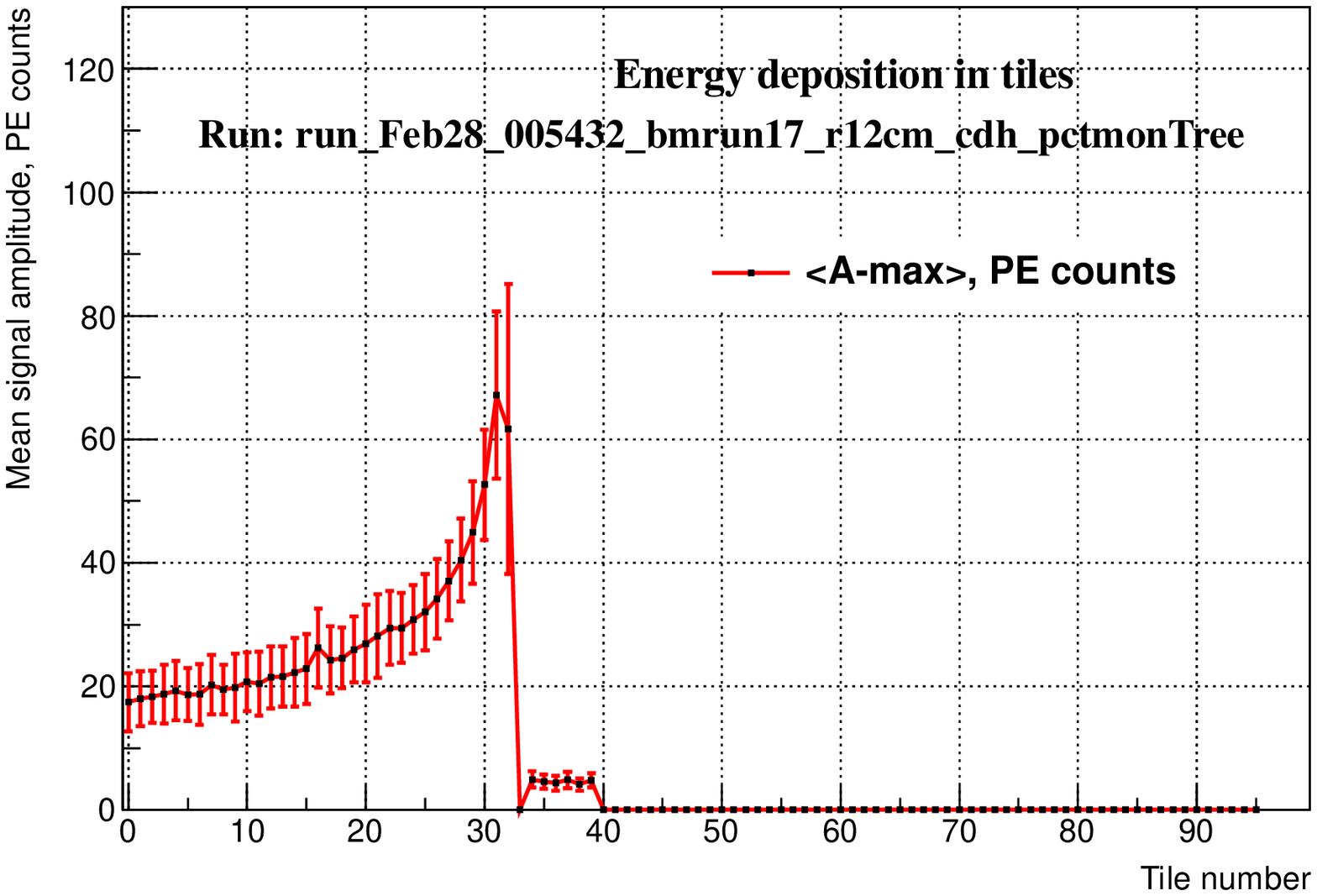}
 \includegraphics[scale=0.27]{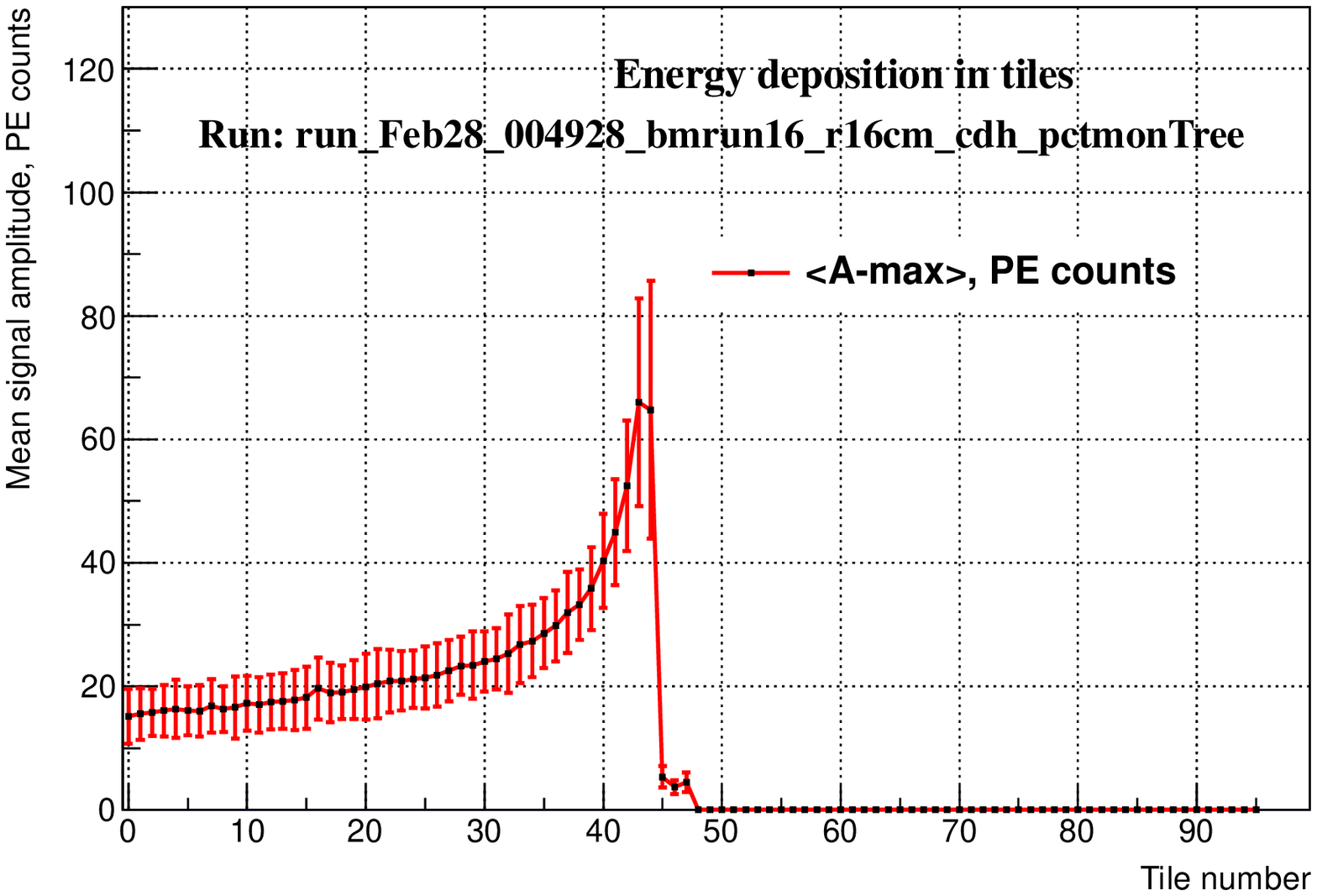}
 \leftline{ \footnotesize \hspace{2.3cm}{\bf  (a)} \hfill\hspace{3.2cm} {\bf (b)} \hfill \hspace{3.2cm} {\bf (c)} \hfill }
 \includegraphics[scale=0.27]{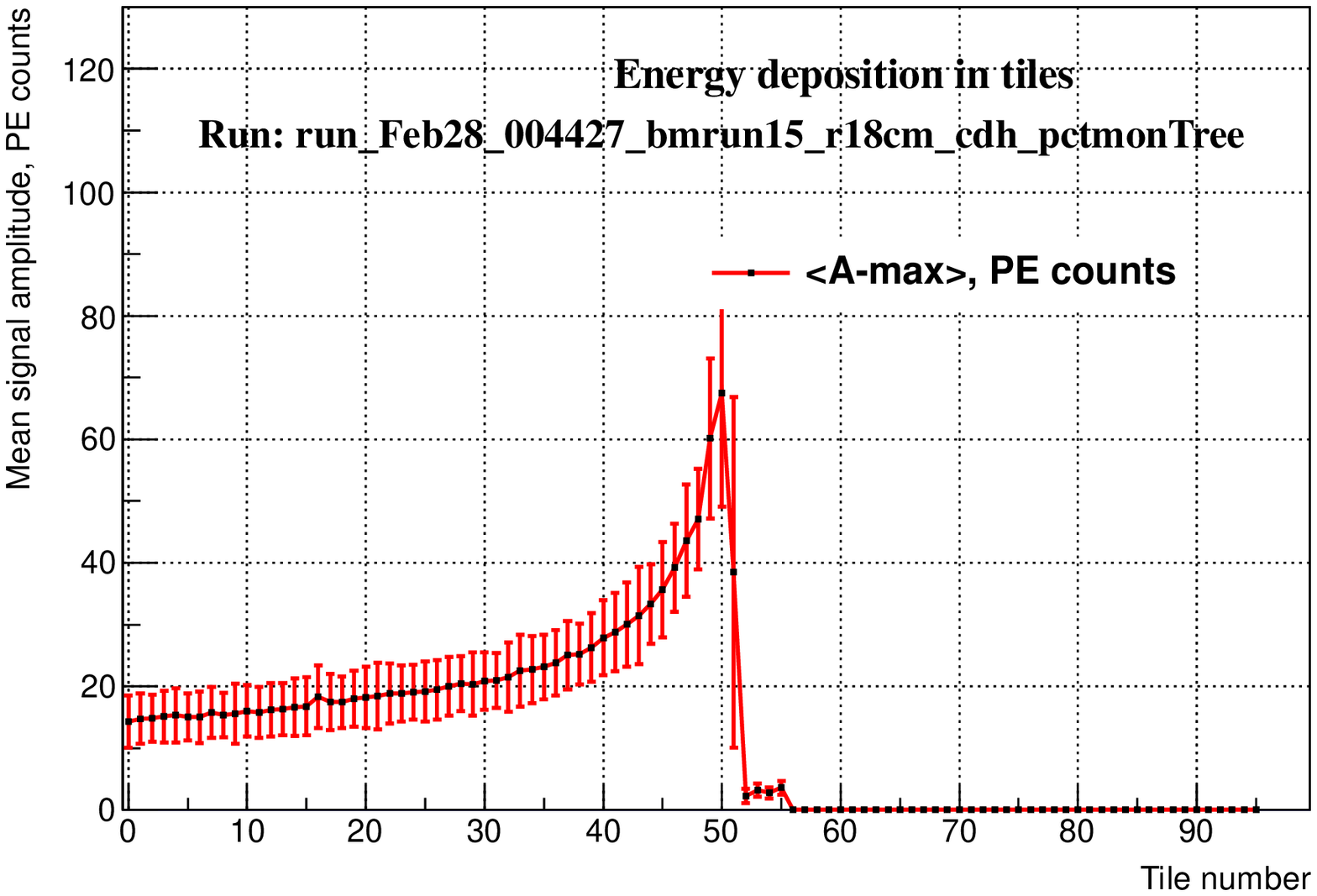}
 \includegraphics[scale=0.27]{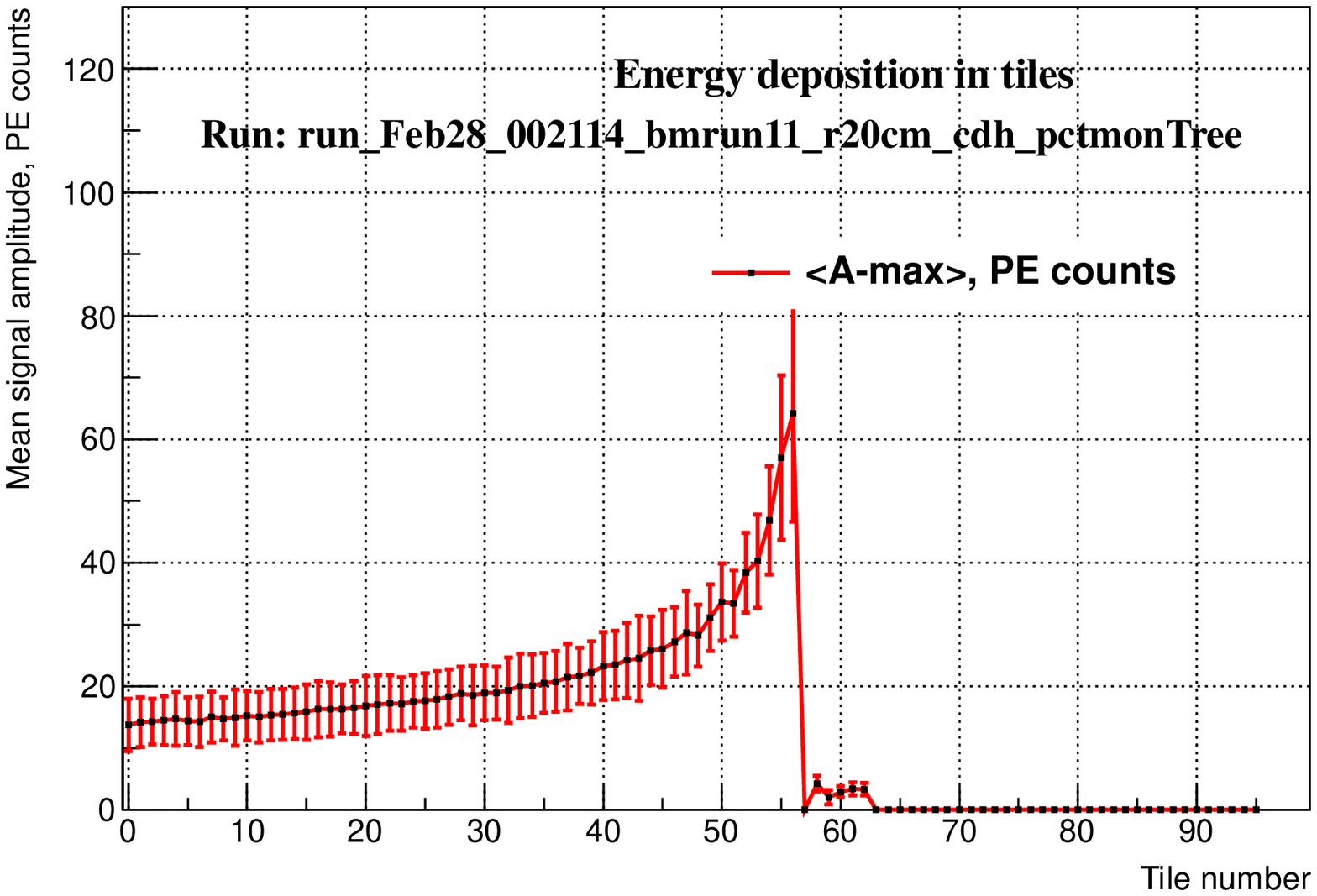}
 \includegraphics[scale=0.27]{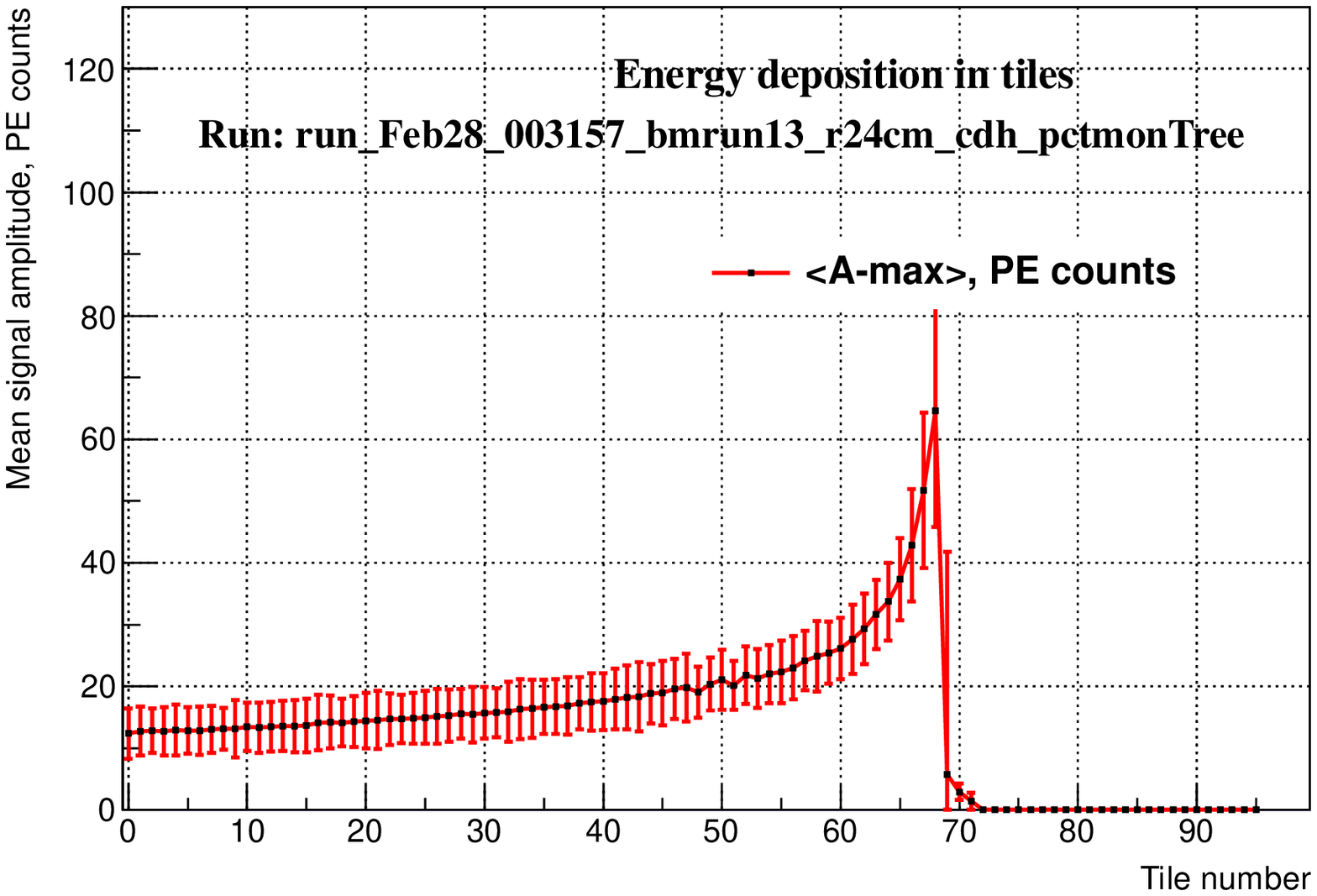}
 \leftline{  \footnotesize \hspace{2.3cm}{\bf (d)} \hfill\hspace{3.2cm} {\bf (e)} \hfill \hspace{3.2cm} {\bf (f)} \hfill}
 \includegraphics[scale=0.27]{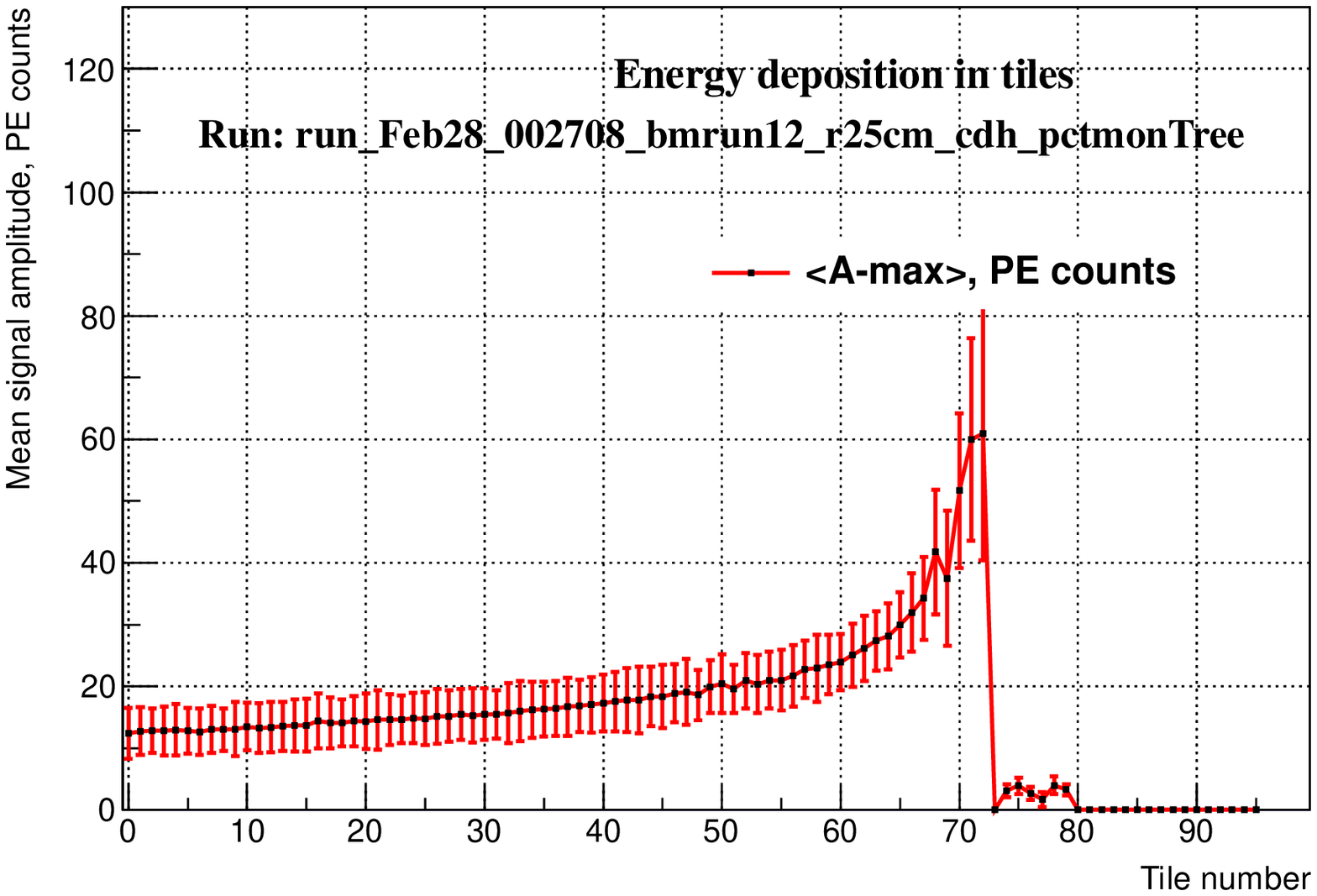}
 \includegraphics[scale=0.27]{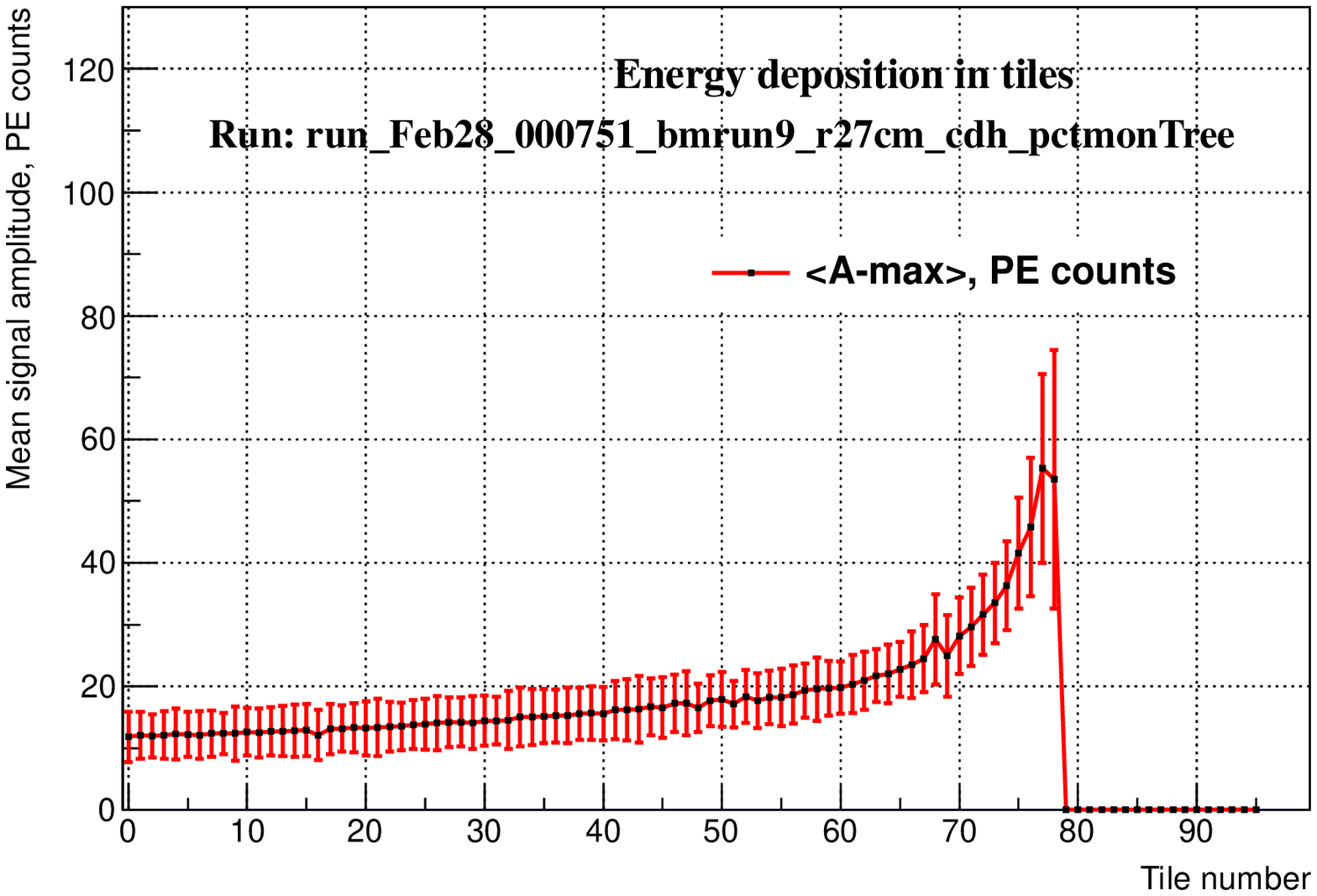}
 \includegraphics[scale=0.27]{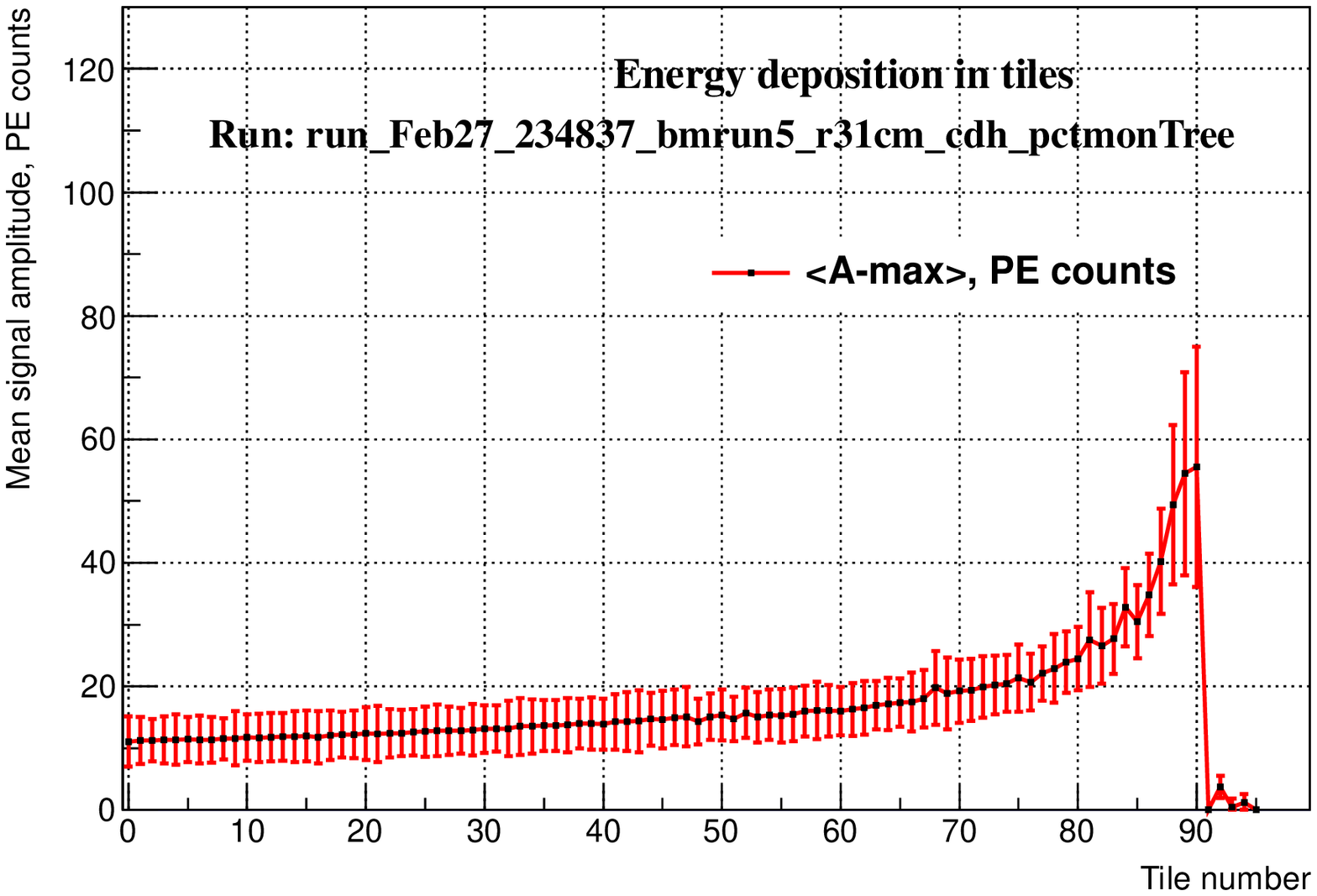}
 \leftline{ \footnotesize  \hspace{2.3cm}{\bf (g)} \hfill\hspace{3.2cm} {\bf (h)} \hfill \hspace{3.2cm} {\bf (i)} \hfill}
}
\caption{\label{fig:stack-profiles-clb-8-31} The mean number of photoelectrons in the range stack tiles as a function of tile number
produced by protons with energy (a) 8~cm (103~MeV); (b) 12~cm (117~MeV);  (c) 16~cm (129~MeV); (d) 18~cm (162~MeV) ; (e) 20~cm (172~MeV);   (f) 24~cm (191~MeV); (g) 25~cm (196~MeV) ; (h) 27~cm (204~MeV);   (i) 31~cm (221~MeV).}
\end{figure}

The comparison of signals observed in Tile0 in runs of different energies and expected signals obtained by integration of 
the tabulated proton dE/dx dependence are shown in Fig.~\ref{fig:tile0_dataVSjnt}. The expected signals are normalized to the mean  Tile0 data signal in the 32~cm run. The measured and calculated amplitudes are in good agreement, however the data signals are about 5\% higher at low proton energies. \\
4)  We extract normalization coefficients $K^{clb}_{tn} \equiv  {E^{clbTrue}_{tn}\over E^{clbExp}_{tn}}$ and use 
them in all data runs to correct the observed  signals in the range stack tiles.
Figures~\ref{fig:stack-profiles-clb}(a) and  (b) show 
the corrected energy deposition profiles (the mean number of photoelectrons from about 10000 protons per tile as function of tile number) for 200~MeV protons.
Corrected energy profiles for different beam energies are shown in Fig.~\ref{fig:stack-profiles-clb_r32} through Fig.~\ref{fig:stack-profiles-clb-8-31}. Slight variations are attributed to statistical effects.
% In  all profiles several tiles ( 48, 49, 50, 52, 69) show excess of signal out of the smooth line. The reason for this a sudden change of response in SiPm's of these tiles in the ``back'' calibration run compare to the ``front'' calibration run.
%cfroot dedx_jnt.C
 %cfroot dedx_jnt.C
%=------------------------------------------------------------------------------------------------

%

% %
\section{\label{energy-range} Range and energy  measurements}
The NIU image reconstruction software uses the WEPL of a scanned object, $wepl_{obj}$.  For each proton  the $wepl_{obj}$ can be 
obtained from WEPL of the range stack, $wepl_{rs}$,\\
                                             \centerline{$wepl_{obj} = E_{beam} (cm) - wepl_{rs}$,}\\
% To find $wepl_{rs}$  one need calibrate the stopping range $R_{rs}$ or the total energy $E_{rs}$ measured by the range stack detector using
% a set of phantoms with known WEPL~\cite{wepl_clb}.  Here we compare the accuracy of  $R_{rs}$ and $E_{rs}$ to
% choose what measurement is preferrable for the WEPL calibration.
To find $wepl_{rs}$  one need calibrate the total energy  $E_{rs}$ or the stopping range $R_{rs}$ measured by the range stack detector using
a set of phantoms with known WEPL~\cite{wepl_clb}.  Here we compare the accuracy of  $R_{rs}$ and $E_{rs}$ to
choose what measurement is preferrable for the WEPL calibration.
%new paragraph

\begin{figure}[ht]
\centering
{
\includegraphics[scale=0.27]{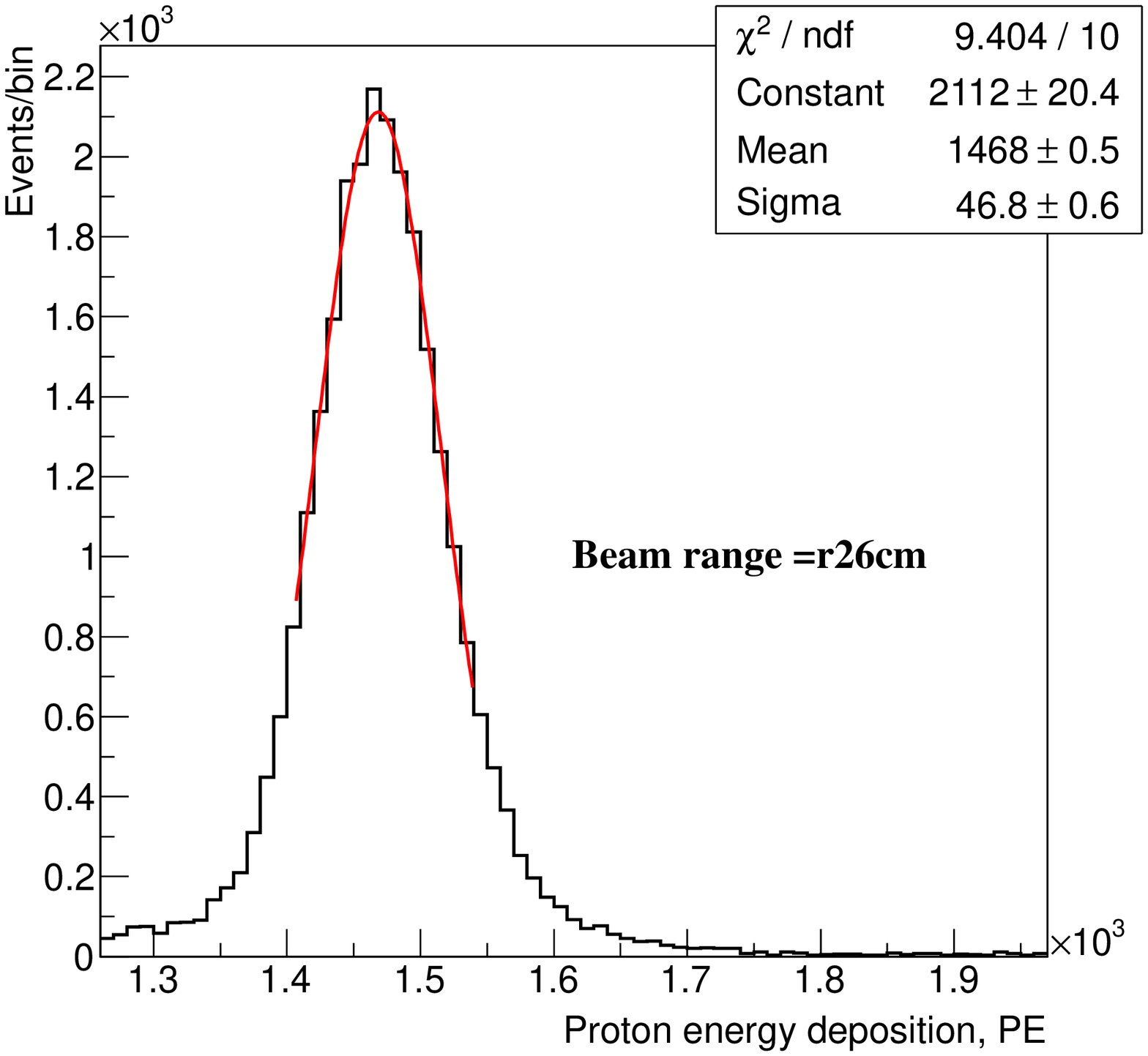}
\includegraphics[scale=0.27]{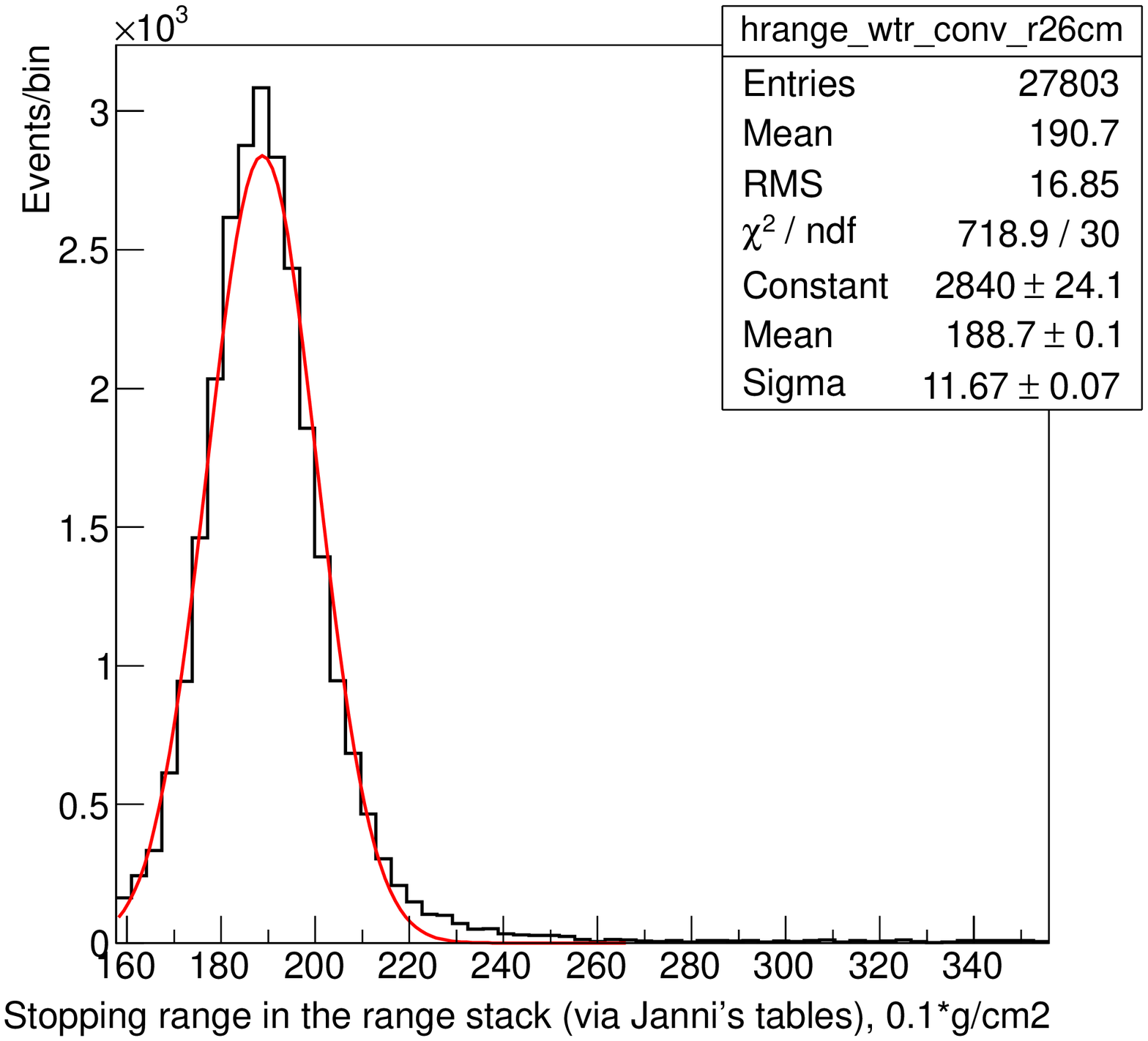}
\includegraphics[scale=0.27]{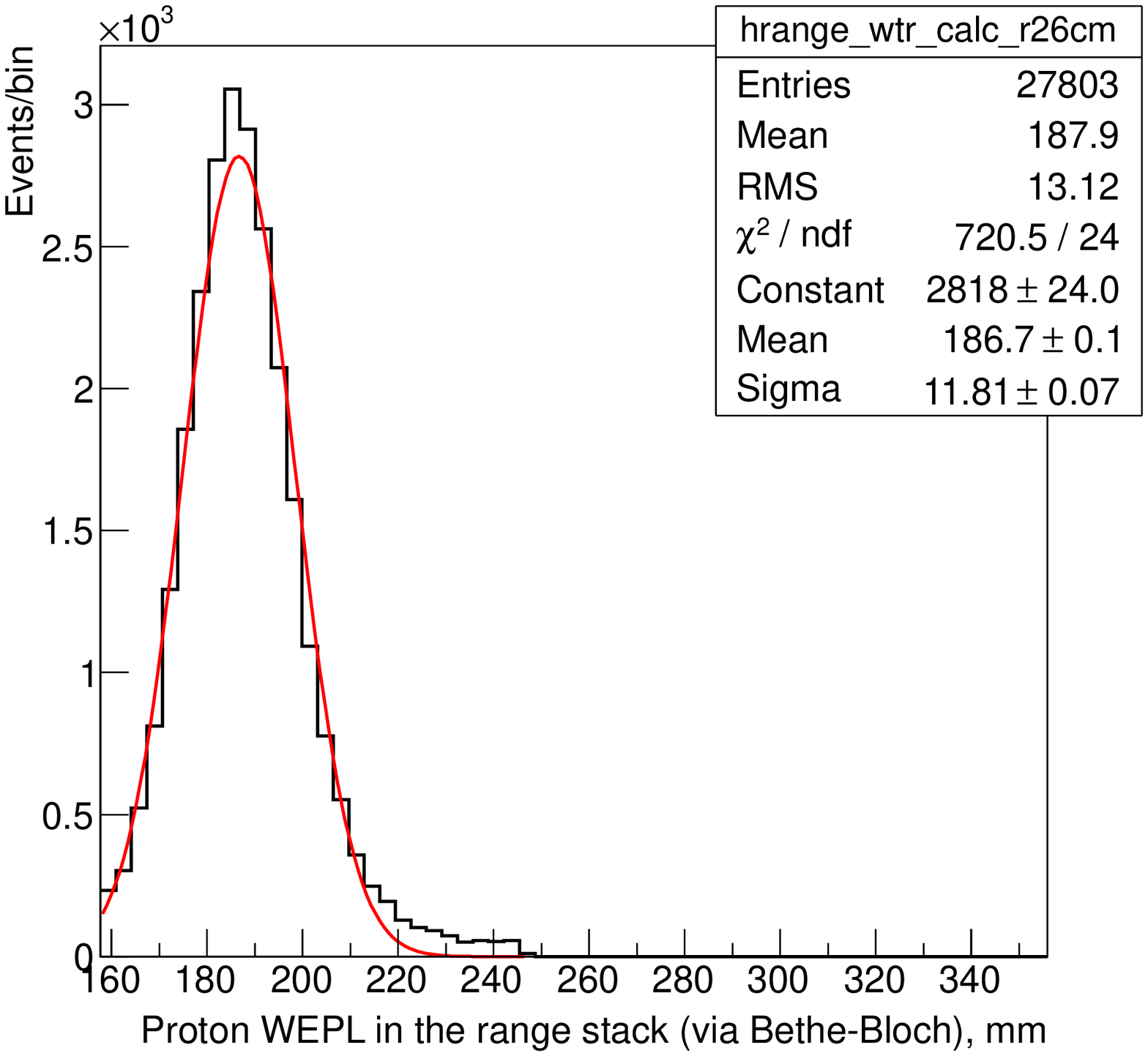}
\leftline{ \hspace{2.3cm}{\bf (a)} \hfill\hspace{3.2cm} {\bf (b)} \hfill \hspace{3.2cm} {\bf (c)} \hfill}
}
\caption{\label{fig:rs-edep} (a) the total energy, $E_{rs}$, in $PE$ measured with the range stack detector in a $E_{beam} = 26~cm$ run; 
(b) the proton stopping range in the range stack  $R_{rs}^{conv}(g/cm^2)$ obtained from  $E_{rs}$ via  $R_{rs}^{conv} =  0.0022\times E_{rs}^{(1.77)}$;
(c) the proton  WEPL in the range stack  $wepl^{calc}_{rs}$, $mm$ calculated from $E_{rs}$ via energy loss equation.
}
\end{figure}

To find the total energy $E_{rs}$  deposited in the range stack we first search for the frame with a stopping tile (the ``stopping'' frame),
then sum signals (in PE units) from all tiles and all frames including the stopping frame.  Only events with no missing frames before the 
stopping frame were selected. Figure~\ref{fig:rs-edep}(a) shows the total $E_{rs}$ in PE measured in a run with $E_{beam} = 26$~cm. 
%new paragraph

To find  $R_{rs}$  we use the $Z$-position of the tile with the maximum signal 
(the ``stopping'' tile, labeled as $nt_{stop}$).  We calculate  $R_{rs}$ as
\begin{equation}
\begin{split}
       R_{rs} = (nt_{stop}+1) \times (tileW \times tileD  +  alW \times alD + mlrW \times mlrD ) + \\
               nframe \times (alW \times alD + mlrW \times mlrD ), nt_{stop}= [0,95]; nframe=[0,11] \nonumber
\end{split}
\end{equation}
where $(tileW,alW,mlrW) = (3.2,  0.00022, 0.00625)~mm$ are the widths of scintillator and wrapper (aluminized mylar) layers, and
 $(tileD,alD,mlrD) = (1.011,  2.700,  1.397)~g/cm^{3}$  are the densities of these materials.  The second term accounts for the extra layer 
 of the wrapper in the end of each range stack frame.  
 %New paragraph
 
\begin{figure}[ht]
\centering
{
 \includegraphics[scale=0.40]{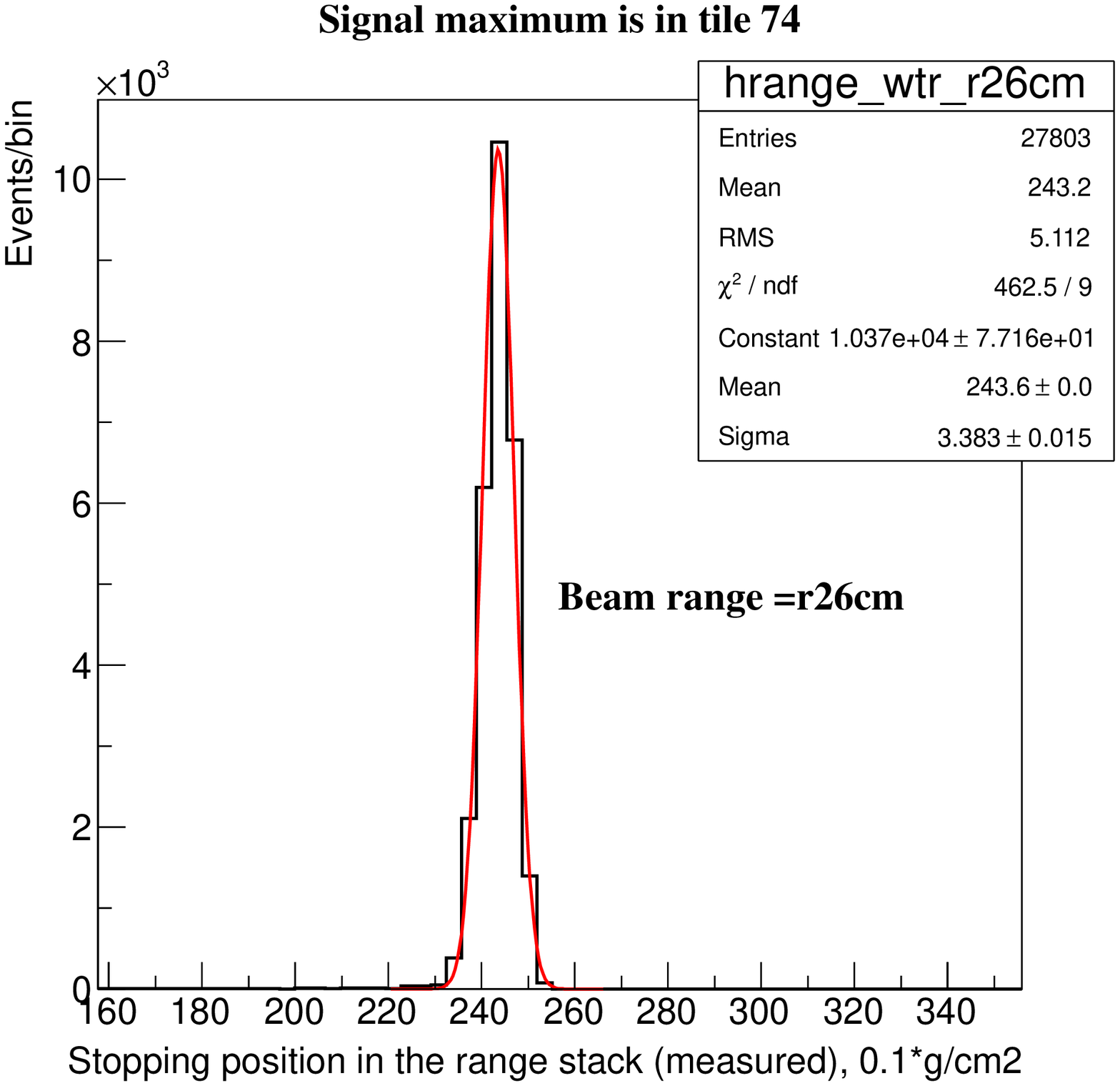}
%\leftline{ \hspace{4.5cm}{\bf (a)} \hfill\hspace{3.5cm} {\bf (b)} \hfill}
}
\caption{\label{fig:rs-wepl}The measured proton stopping range, $R_{rs}$, in the range stack, $E_{beam} = 26~cm$ run. }
\end{figure}

 Note, the stopping ranges in polysterene and mylar  expressed in $g/cm^2$
 are approximately equal to the proton stopping range in water  $R_{w}$:\\  
 \begin{equation}
\label{bb2}         
                             R_w (cm)  = \int^{R_m}_{0} RSP_m dL
\end{equation}
where $RSP_m$ is the proton stopping power of the medium relative to water and $L$ is the physical proton path length along the calorimeter and $R_m$
is the physical depth at which the proton stops in the range stack. 
Then, neglecting small variations ( $< 0.5 \%$) 
in mean ionization potential between water, polystyrene and mylar, as used in the Bethe Bloch equation,  the water equivalent range of the proton becomes
\[
 R_w (cm)  \simeq \int^{R_m}_{0} \rho_{m}/ \rho_w  dL ,  and, for \rho_{w}=1.0~g/cm^3,  R_w (g/cm^2)  \simeq \int^{R_m}_{0} \rho_{m} dL
\]
Thus we expect $R_{rs}$ to have linear dependency on the beam energy,  $E_{beam}$, expressed in $cm$.
Figure~\ref{fig:rs-wepl} shows  $R_{rs}$ in a run with $E_{beam} = 26$~cm.  
%new paragraph

We also can find the $R_{rs}$ from the total energy $E_{rs}$  using Janni's range-energy  tables 
This method requires expression of $E_{rs}$ in MeV, and 
we use the conversion coefficient calculated as the ratio of the mean amplitude of the data signal (in number of photoelectrons)
to the mean amplitude of the estimated MC signal (in MeV) in Tile0, in  26~cm runs. 
%Figure~\ref{fig:rs-edep}(a) shows the total $E_{rs}$ in PE measured in a run with $E^{beam}_{Acc} = 26$~cm. 
Figure~\ref{fig:rs-edep}(b) shows the proton stopping range  in the range stack $R_{rs}^{conv}(g/cm^2)$ calculated  from 
$E_{rs}$ via the  $R^{conv}_{rs} =  0.0021\times E_{rs}^{(1.78)}$ conversion function obtained from Janni's tables.
Finally, we can find $wepl_{rs}$ directly,  from the WEPL of a scanned object calculated from $E_{rs}$ using
the Bethe-Bloch equation.  Howewer, this will also require calibration,  as the measured $E_{rs}$ only includes 
the visible part of deposited energy.
Figure~\ref{fig:rs-edep}(c) shows the WEPL of the range stack  $wepl^{calc}_{rs}(cm)$ calculated  from 
$E_{rs}$ via \\
\begin{equation}
\label{bb}         
                             wepl^{calc}_{rs} = E_{beam} (cm) - \int^{E_{rs}}_{E_{beam}} \frac{1}{ S(E_{p}) }dE 
\end{equation}
where $S(E_p) = - dE/dx $ is a water stopping power for proton with energy $E_p$. 
%New paragraph

\begin{figure}[ht]
\centering
 {
 \includegraphics[scale=0.40]{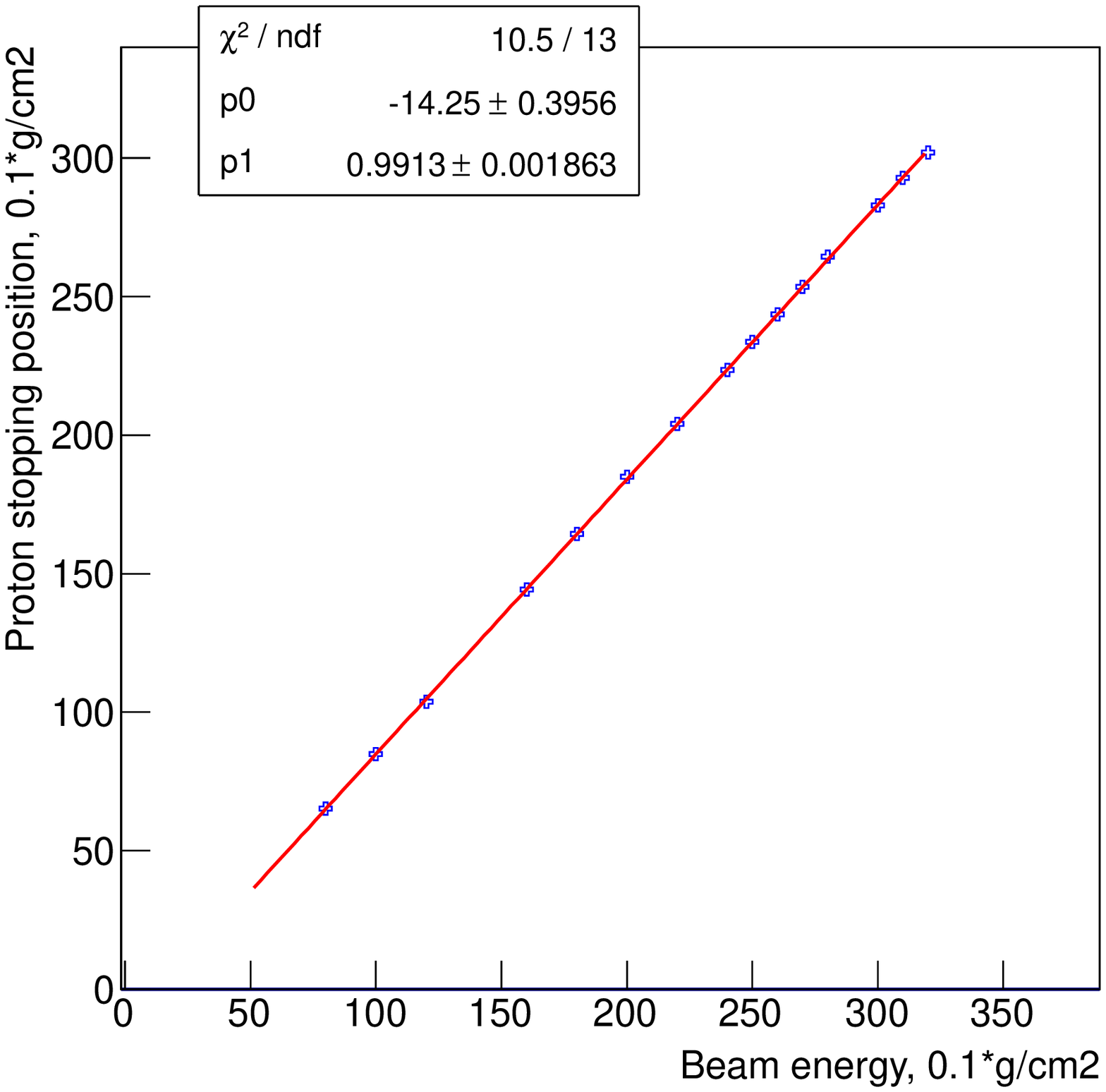}
 \includegraphics[scale=0.40]{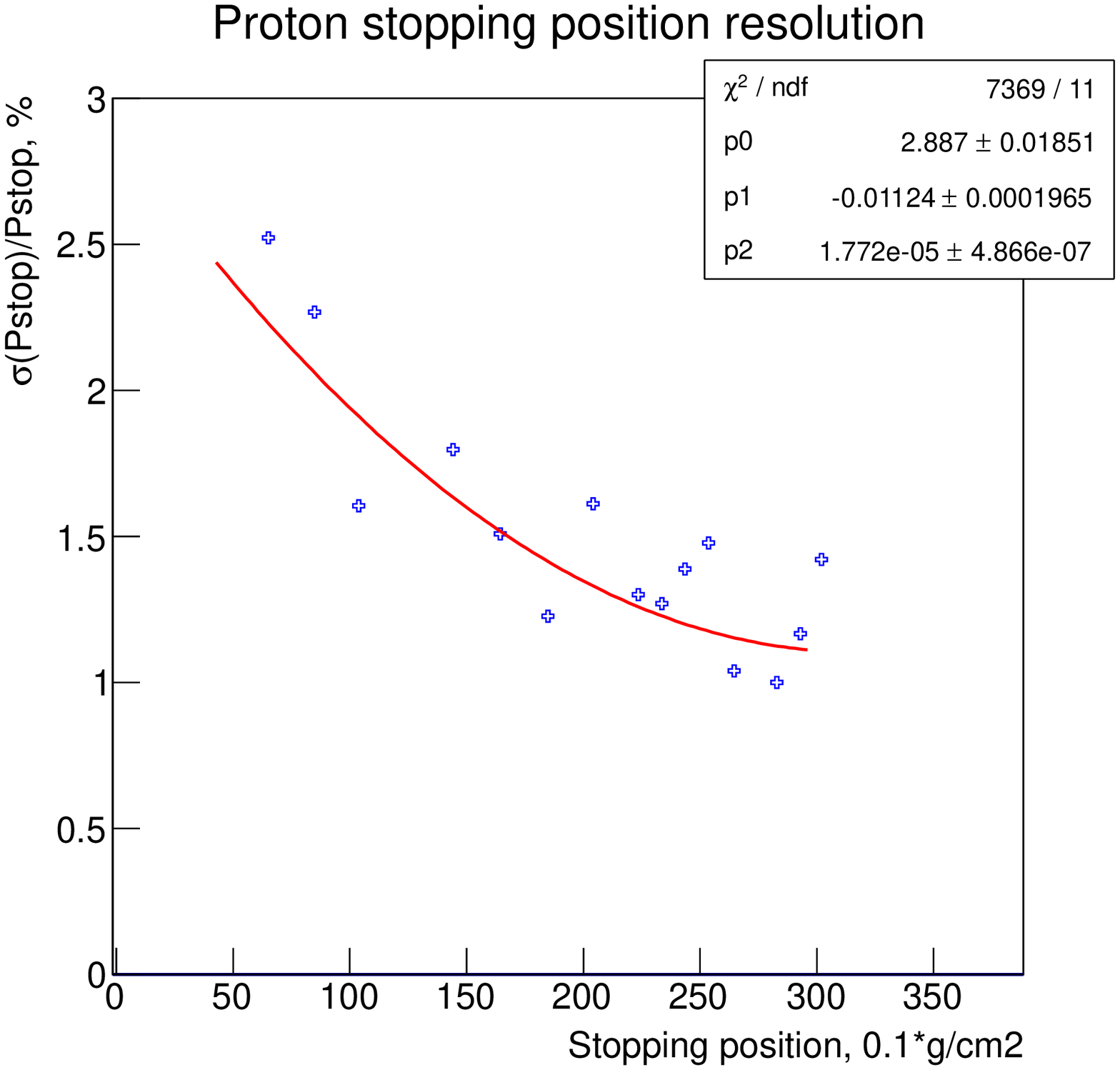}
 }
\leftline{ \hspace{4.5cm}{\bf (a)} \hfill\hspace{3.5cm} {\bf (b)} \hfill}
\caption{\label{fig:wepl-lin-and-res}  (a) The linearity of the directly measured proton stopping position measurement $R_{rs}$ ; (b)  the $R_{rs}$ resolution. The linearity fit allows an estimate of the width of  the extra material in front of the range stack ($\approx$14.0~mm).}
\end{figure}

\begin{figure}[ht]
\centering
 {
 \includegraphics[scale=0.40]{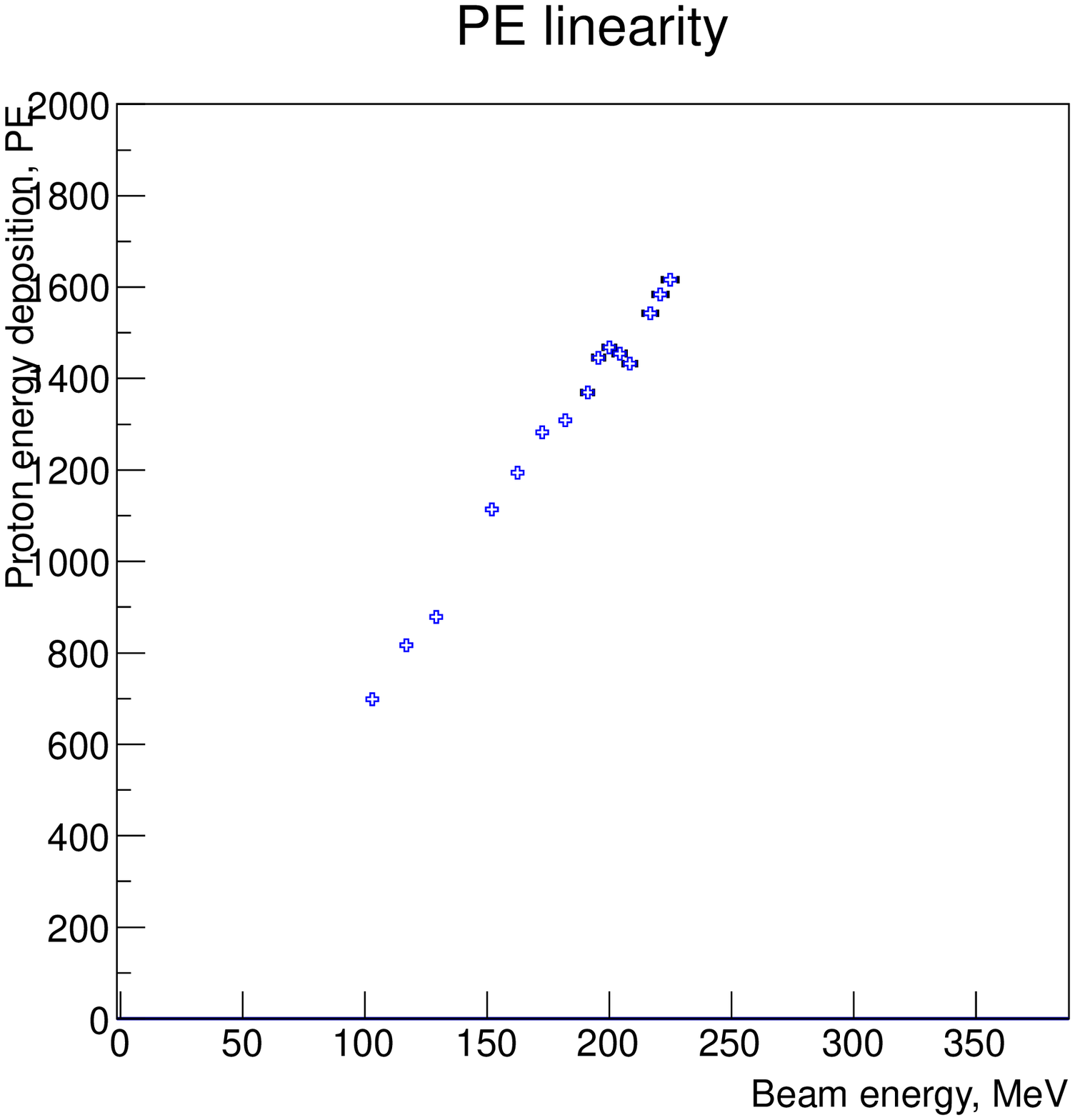}
 \includegraphics[scale=0.40]{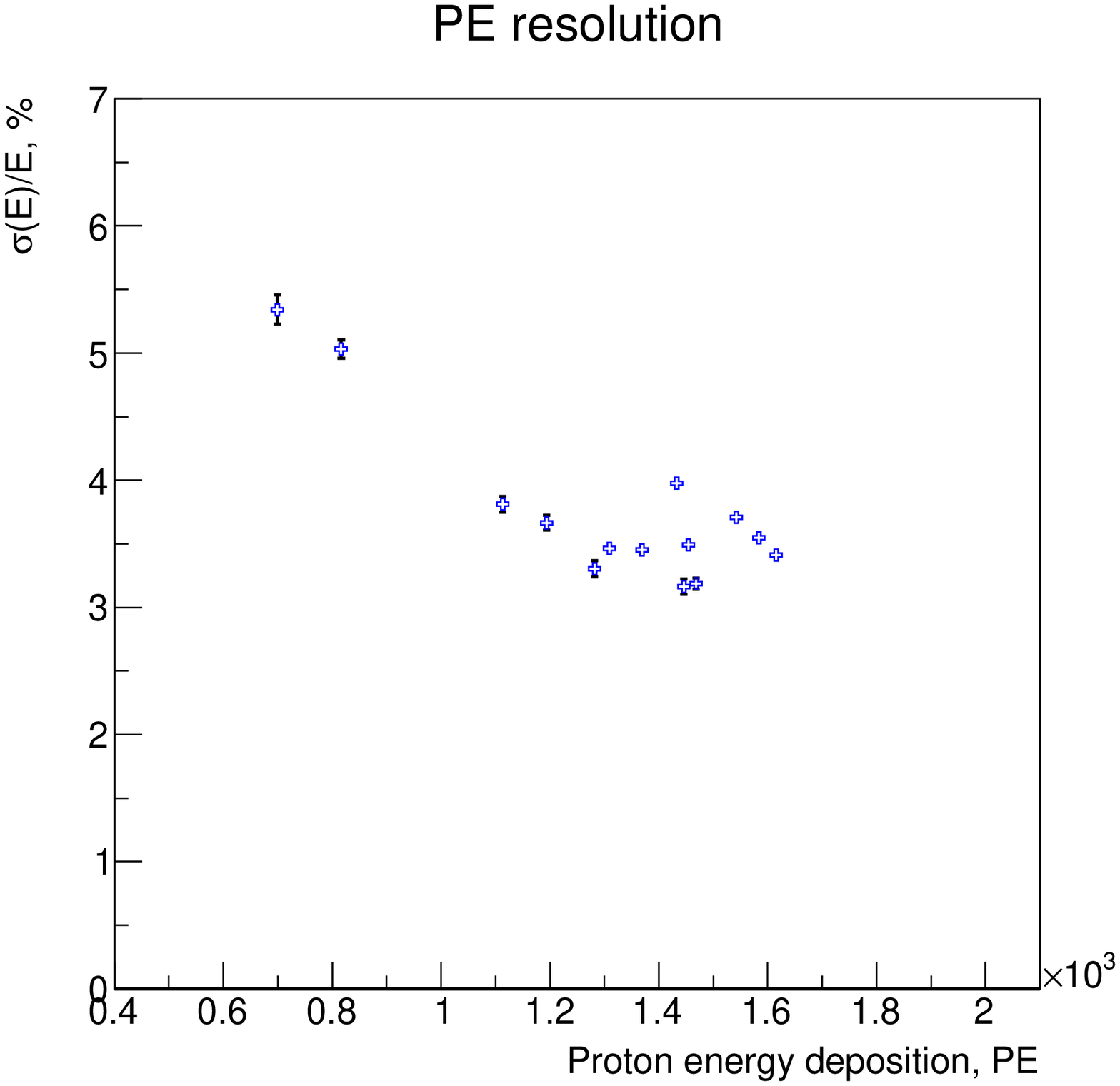}
 }
\leftline{ \hspace{4.5cm}{\bf (a)} \hfill\hspace{3.5cm} {\bf (b)} \hfill}
\caption{\label{fig:etot-lin-and-res} (a) the linearity of the energy measurement $E_{rs}$; (b) the  $E_{rs}$ resolution. }
\end{figure}

We fit peaks of the $R_{rs}$ and  $E_{rs}$ distributions with a Gaussian and use the mean and $\sigma$ parameters of the fits to study 
the linearity ($R_{rs}$ and $E_{rs}$  as functions of the beam energy) and resolution  ($\sigma(E_{rs})/E_{rs}$  and  $\sigma(R_{rs})/R_{rs}$ as functions of  $E_{rs}$ and $R_{rs}$) of the range stack detector.  The linearity and resolution plots for  the proton stopping position $R_{rs}$ are shown in Fig.~\ref{fig:wepl-lin-and-res} and  the linearity and resolution plots for the energy measurement are shown in  Fig.~\ref{fig:etot-lin-and-res}.  
The good linearity with a non zero intercept of the $R_{rs}$ shows there is material in front of the range stack at all energies.
%The intercept indicates that at all energies there is material in front of the range stack.
The energy measurement has lower accuracy (energy resolution ranges from 5.5\% to 3.5\% , compared to 2.2-1.2\% for  $R_{rs}$) and also
shows an unexpected suppression at beam energies of  27~cm and 28~cm. Additionally, Figures~\ref{fig:rs-edep}(a) and (b)  show that if we 
try to extract the stopping range or WEPL in the range stack from the direct energy measurement, the $R^{conv}_{rs}$  and distributions 
with $\sigma(R^{conv}_{rs})= 11.7~mm$  and $\sigma(wepl^{calc}_{rs})= 11.8~mm$ are  significantly wider than $R_{rs}$ distribution with $\sigma({R}_{rs})= 3.3~mm$. 
 %And even more, the mean values of the $wepl^{conv}_{rs}$ and $wepl^{calc}_{rs}$  distributions are smaller, as the range stack collects only a part of  the proton energy.  
Thus, the  direct $R_{rs}$ measurement is preferred for the WEPL calibration.
%
% \begin{figure}[ht]
% \centering
%  {
%  \includegraphics[scale=0.60]{figures/Stack_signal_profiles.eps}
%  }
% \caption{\label{fig:Mean Profiles} Profiles of energy deposition in tiles for  different proton energies.}
% \end{figure}


\subsection{Comparison with GEANT~4 simulations}
A GEANT~4 simulation of the pCT detector was used to obtain the energy deposition $E^{g4}_{tn}$ in the range stack tiles  
for different beam energies. We converted the range $R_p$ to energy $E_{p}$  using the inverse of the Janni fit:\\
 \centerline{ $E_{p} =  (R_{p}/0.0022)^{(1/1.77)}$.} \\
To compare energy profiles and total energy deposition in the range stack, the G4 signals in the tiles were expressed in the number of photoelectrons by normalizing to the data signal in Tile0 in the 26~cm beam run.  
\subsection{Smearing of simulated tile signals}
\begin{figure}[ht]
\centering
 {
 \includegraphics[scale=0.27]{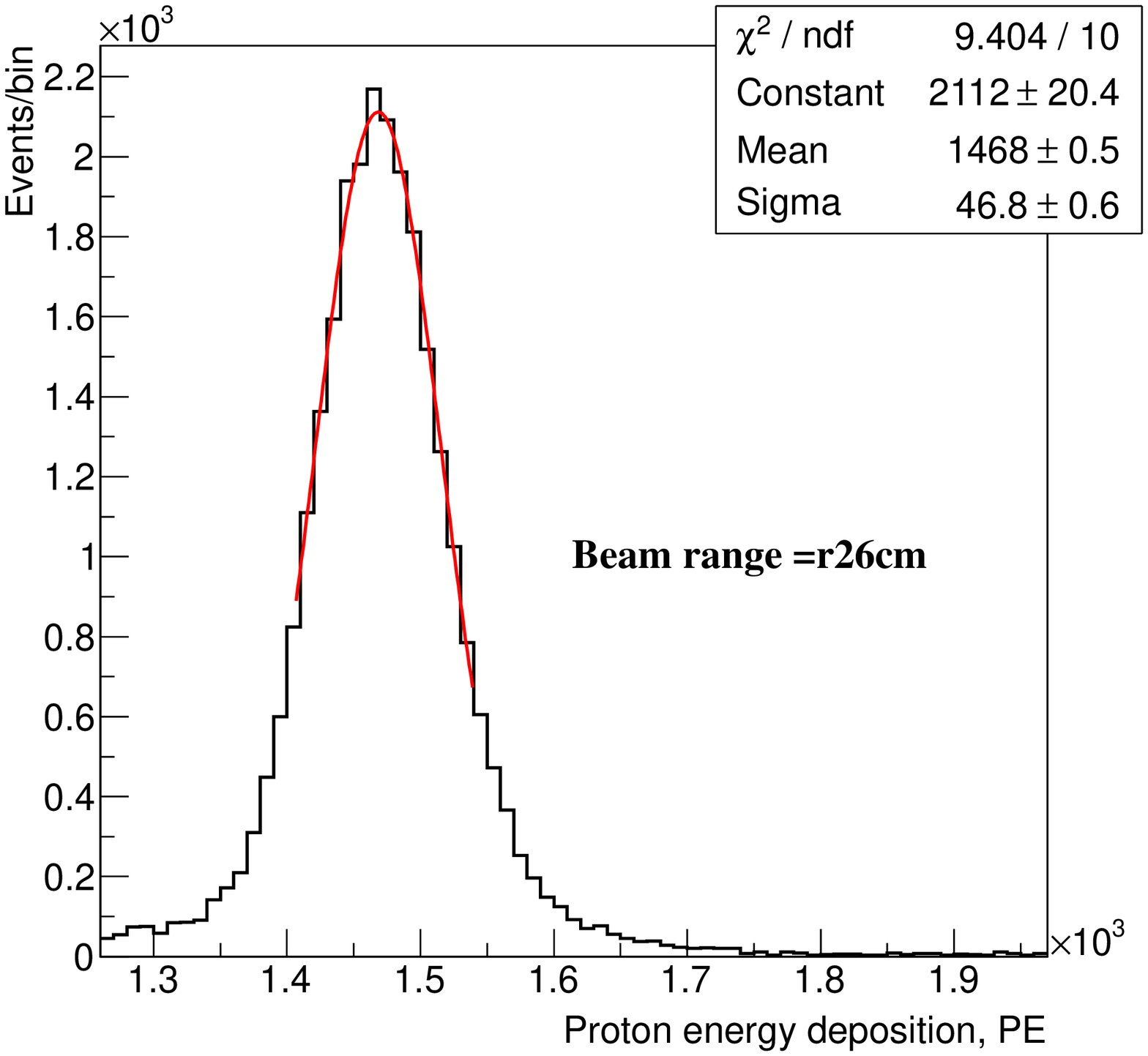}
 \includegraphics[scale=0.27]{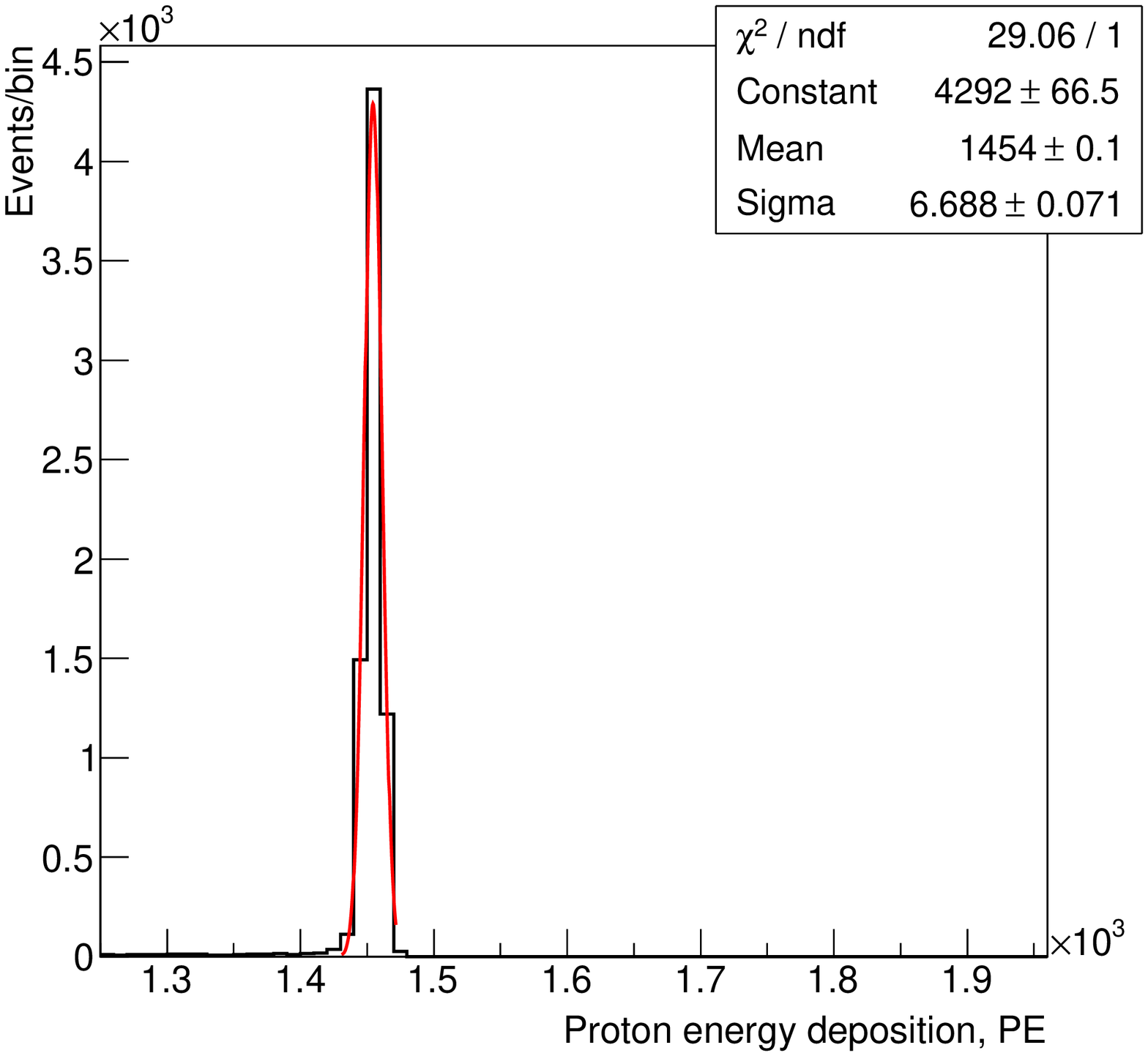}
 \includegraphics[scale=0.27]{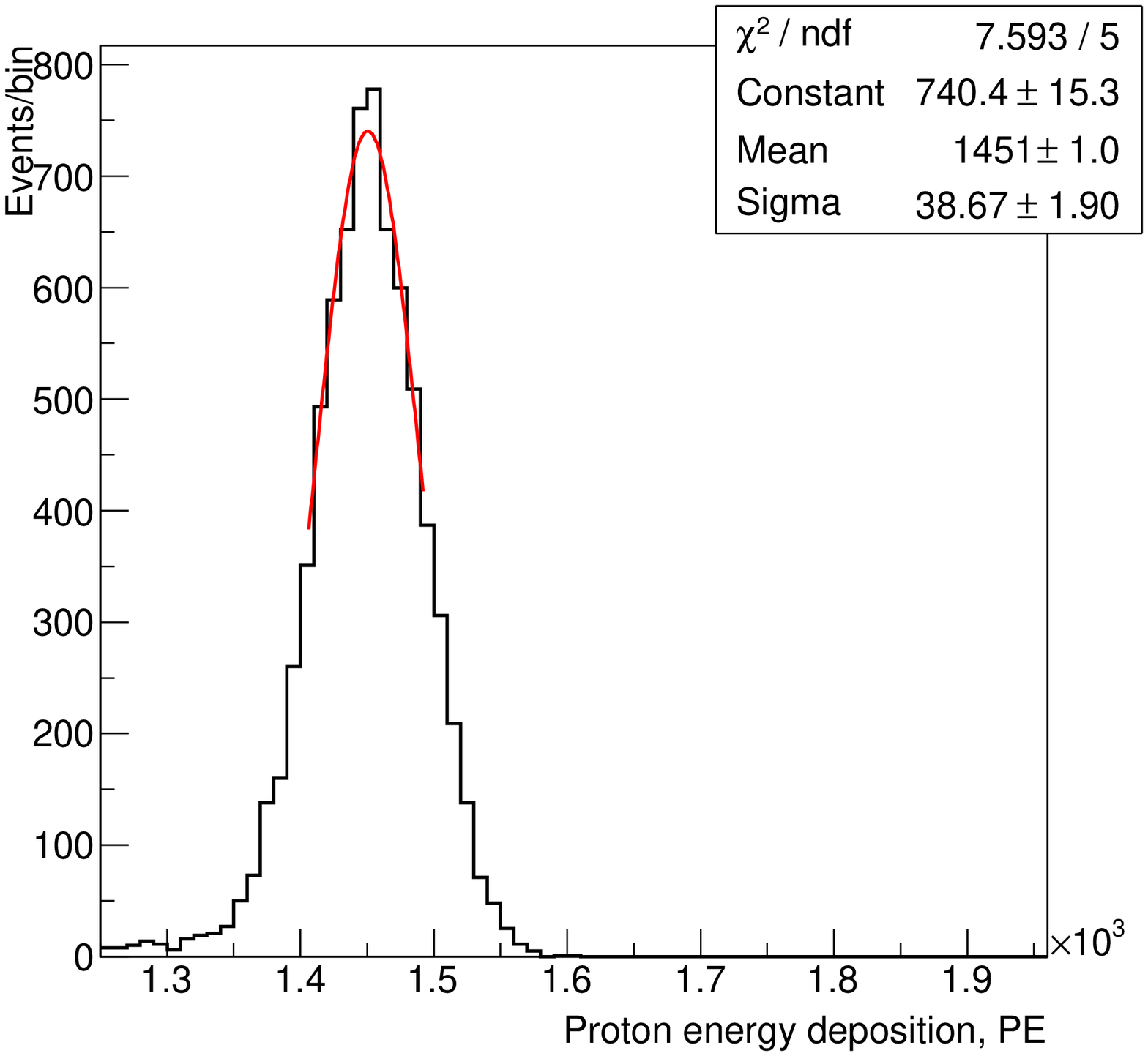}
 }
 \leftline{ \hspace{2.3cm}{\bf (a)} \hfill\hspace{3.2cm} {\bf (b)} \hfill \hspace{3.2cm} {\bf (c)} \hfill}
\caption{\label{fig:smearing} (a)  the 
total energy deposition in the range stack at beam energy of 26~cm (200~MeV ) in data; (b) the unsmeared total energy deposition in the range stack at beam energy of 26~cm (200~MeV) in GEANT;   (c) the smeared range stack energy measurement at beam energy of 26~cm (200~MeV) in GEANT.  }
\end{figure}

\begin{figure}[ht]
\centering
 {
 \includegraphics[scale=0.67]{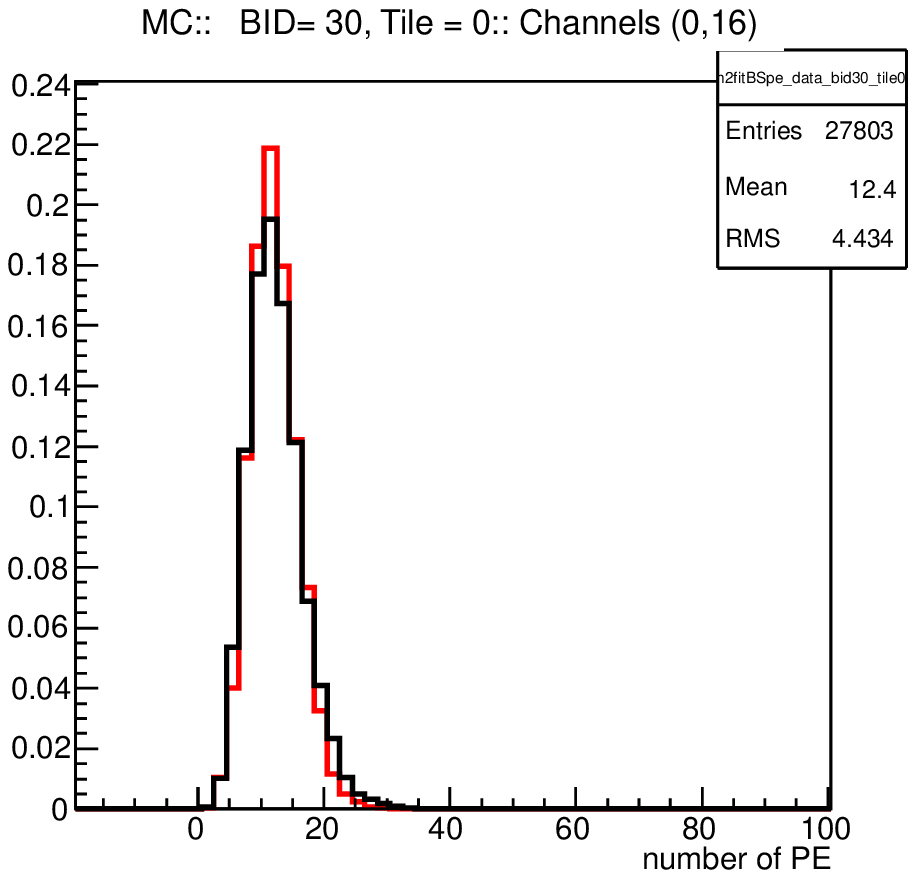}
 \includegraphics[scale=0.67]{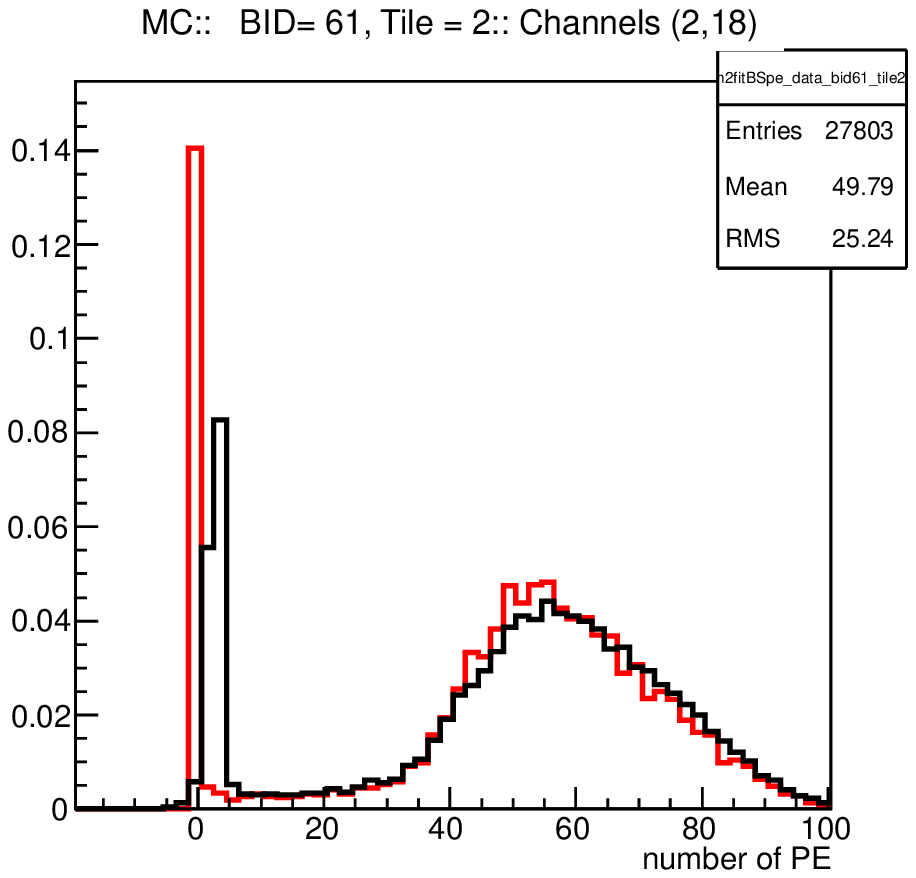}
 }
\leftline{ \hspace{4.5cm}{\bf (a)} \hfill\hspace{3.5cm} {\bf (b)} \hfill}
\caption{\label{fig:smearing_cmp} Comparison of data (blue histograms) and simulated signals (red histograms) from 200~MeV protons  in  (a) Tile0  and (b) Tile74 (the stopping tile, maximal signal).}
\end{figure}

To account for photo-statistics  and  SiPM readout,  smearing of the G4 signals in each tile was done as:\\
\centerline{ $S^{g4}_{tn} =  G(<S^{ped}_{tn}>) + P(E^{g4}_{tn}) - <S^{ped}_{tn}>$}, \\
where   $<S^{ped}_{tn}>$ is a mean sum of SiPM pedestals in tile $n$  from calibration runs,  $E^{g4}_{tn}$ is the energy deposition in
tile $n$ obtained from GEANT, and  $G(S^{ped}_{tn})$ and $P(E^{g4}_{tn})$  are the sum of SiPM pedestals smearing using Gaussian and $E^{g4}_{tn}$
smeared using Poisson distribution. The effect of smearing is shown in Fig.~\ref{fig:smearing}, where the left plot shows the 
total energy deposition in the range stack at a beam energy of 26~cm (or 200~MeV) in data; the center histogram
shows the unsmeared simulated signal, and the right histogram shows the smeared signal. Comparison of data and simulated signals from 200~MeV protons  in Tile0  and in Tile74 (the stopping tiles with the maximal signal for this energy) are shown in Fig.~\ref{fig:smearing_cmp}.

\subsection{\label{ecorr}Beam energy correction and smearing for the MC simulations}
The total stopping range of all material along the proton path, $R_{total}$,  before the proton stopping position 
is equal to the nominal beam energy of the  accelerator in $g/cm^2$,  $R_{total} \equiv  E^{beam}_{total}$.
In our test beam configuration, the total stopping range can be expressed as:
\[
          R_{total} =  R_{rs} + R_{beamline} + R_{tracker} + sft\_const,\\ 
 \]
where $R_{rs}$, $R_{beamline}$, $R_{tracker}$ are the proton ranges in the range stack, any material in 
the accelerator beam line, and the tracker, respectively.  
% New paragraph

\begin{figure}[ht]
\centering
 {
\includegraphics[scale=0.55]{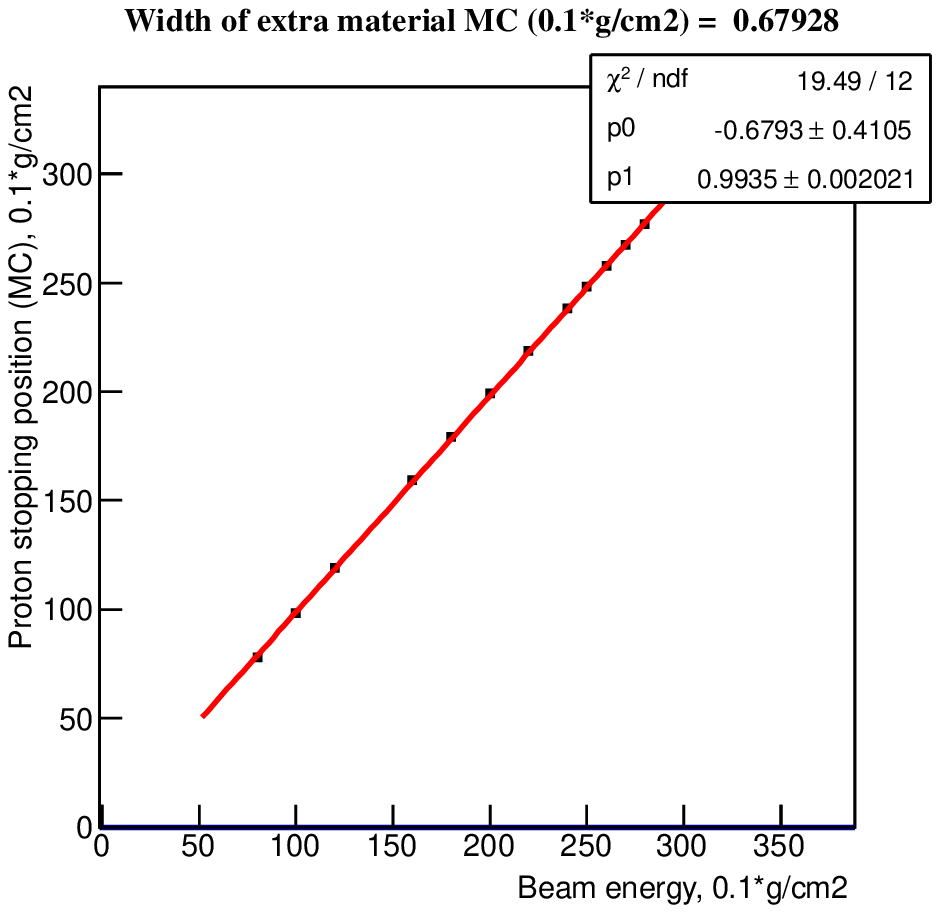}
\includegraphics[scale=0.55]{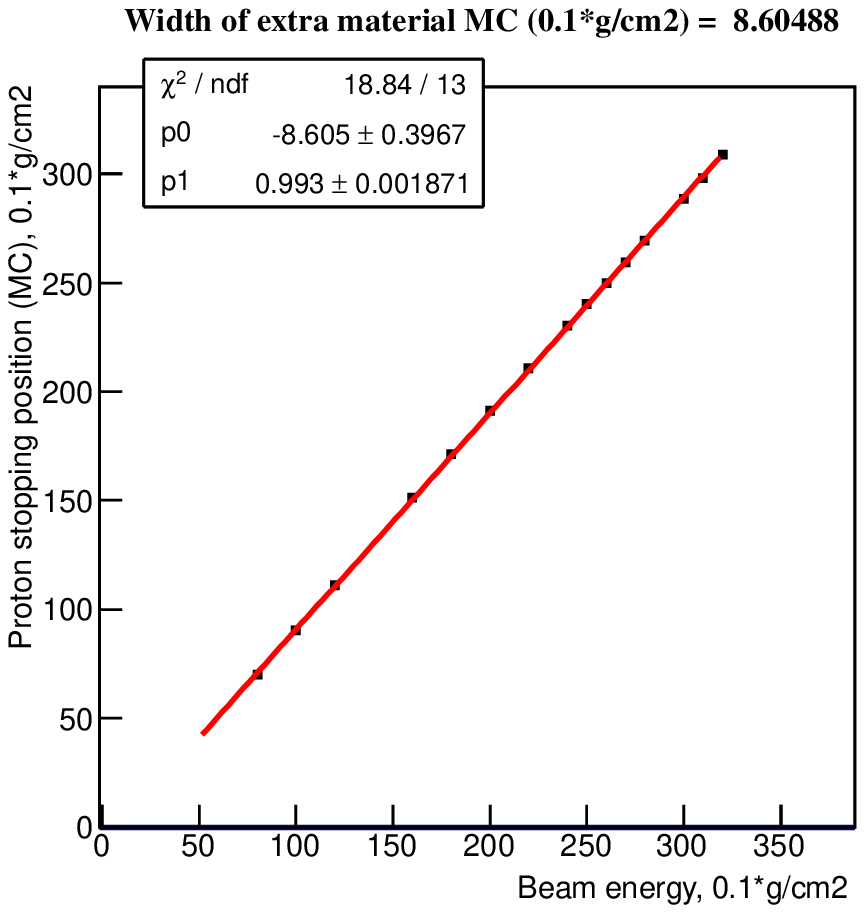}
\includegraphics[scale=0.55]{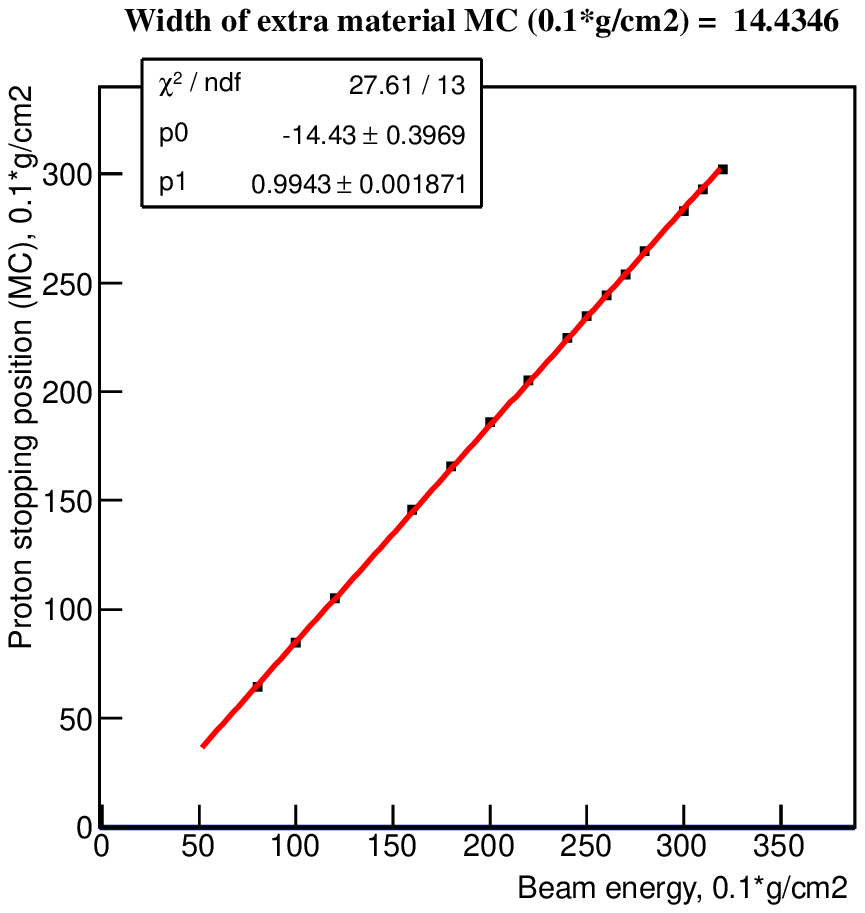}
 }
%\leftline{ \hspace{4.5cm}{\bf (a)} \hfill\hspace{3.5cm} {\bf (b)} \hfill}
\leftline{ \hspace{2.3cm}{\bf (a)} \hfill\hspace{3.2cm} {\bf (b)} \hfill \hspace{3.2cm} {\bf (c)} \hfill}
\caption{\label{fig:ebeam_correction} (a) the linearity of the proton stopping position measurement $R_{rs}$ obtained with GEANT using the nominal CDH beam energy points in a configuration with no tracker before the range stack; (b) the $R_{rs}$ linearity for nominal CDH beam energy points including the tracker; (c) the  $R_{rs}$ linearity after correcting beam energies by adding  ``extra material'' observed in data (5.7~mm).}
\end{figure}

The $sft\_const$ is the systematic shift of the range measurement due 
to initial and arbitrary origin of the range calculation. We extract the stopping range using the position of the tile with the maximum signal and
the total width of scintillator and wrapping layers including this stopping tile.  The definition of stopping position is arbitrary and for consistency estimated  with the MC.
We estimate the $sft\_const$ using simulations of the range stack response in configuration with no tracker.
In the GEANT model we  do not have any other material before the range stack, and the $sft\_const$ can be obtained 
from the fit of the proton stopping positions at different beam energies, as  shown in Fig.~\ref{fig:ebeam_correction}(a).
Evaluation of the fit function at zero beam energy results in $sft\_const$=$0.7\pm0.4$~mm in water (the $-p0$ parameter of the fit).
% Thus using the measured $wepl_{rs}$ we can find the width of material which the proton
% encountered before entering the range stack (the tracker and any other material), $wepl_{extram} + wepl_{tracker}$:\\
% \centerline{        $E^{beam}_{Acc} \approx wepl_{extram} + wepl_{tracker} + wepl_{rs}$\\}
From the fit of the measured proton stopping position $R_{rs}$ in Fig.~\ref{fig:wepl-lin-and-res}(a)  the $R_{beamline} + R_{tracker} + sft\_const$
is equal to $14.3\pm0.4$~mm.  This means that the proton energy at the range stack entry point, $Ep_{entry}$,
is lower than the nominal accelerator beam energy by $13.6\pm0.5$~mm (after subtracting the 0.7~mm $sft\_const$  parameter) for all test runs.  
To accurately compare the energy and range measurement with simulations, the $Ep_{entry}$ should be the same in data and in MC. 
Figure~\ref{fig:ebeam_correction}(b) shows the simulated proton stopping positiom $R_{rs}$ in a configuration with the tracker, and
here the $R_{tracker}+ sft\_const$ is equal to $7.9\pm0.5$~mm (again 0.7~mm is subtracted from the $-p0$ parameter of the fit).
Thus for simulations we subtract the difference of 5.7~mm (between the 13.6~mm observed in data and the 7.9~mm observed in MC)
from all nominal beam energy points in GEANT runs to compensate. 
The corrected results are shown in Fig.~\ref{fig:ebeam_correction}(c) and now the $Ep_{entry}$ in GEANT runs is equal to the $Ep_{entry}$ in data runs. 
%NP

Simulations also predict that this method of upstream material width estimation from the range stack measurements works
with an accuracy of about 0.5~mm in a configuration with a variable width of a rectangular water phantom installed before the range stack.
For the water phantom width of 2~mm,  5~mm, 10~mm, and 15~mm the simulated measurements 
are 2.1~mm, 5.4~mm, 10.3~mm and 14.3~mm, respectively. 

Additionally we smear $Ep_{entry}$ in GEANT in a range between 0.05\% at 100~MeV to 0.02\% at 200~MeV 
to account for the CDH beam energy spread.
% \begin{figure}[ht]
% \centering
%  {
% \includegraphics[scale=0.55]{figures/Range_linearity_mc_nominal.eps}
%  }
%\leftline{ \hspace{4.5cm}{\bf (a)} \hfill\hspace{3.5cm} {\bf (b)} \hfill}
% \caption{\label{fig:fit_vs_rs} The results of the phantom width  from fits of the proton stopping position in the range stack 
% as function of the actual with of the  rectangular water phantom installed before the range stack.}
% \end{figure}
%
\subsection{Comparison of proton stopping position measurements}

\begin{figure}[ht]
\centering
 {
 \includegraphics[scale=0.40]{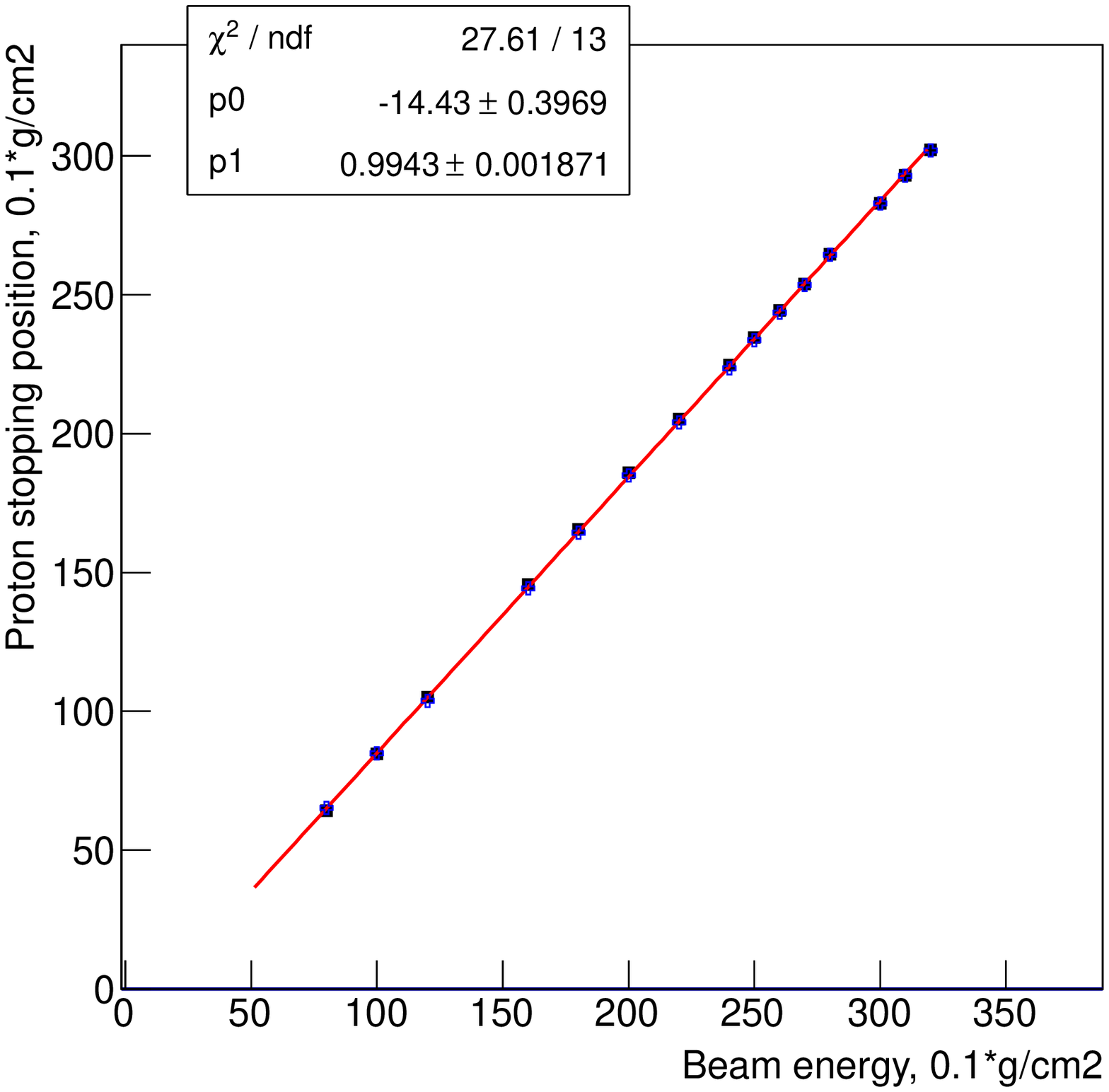}
 \includegraphics[scale=0.40]{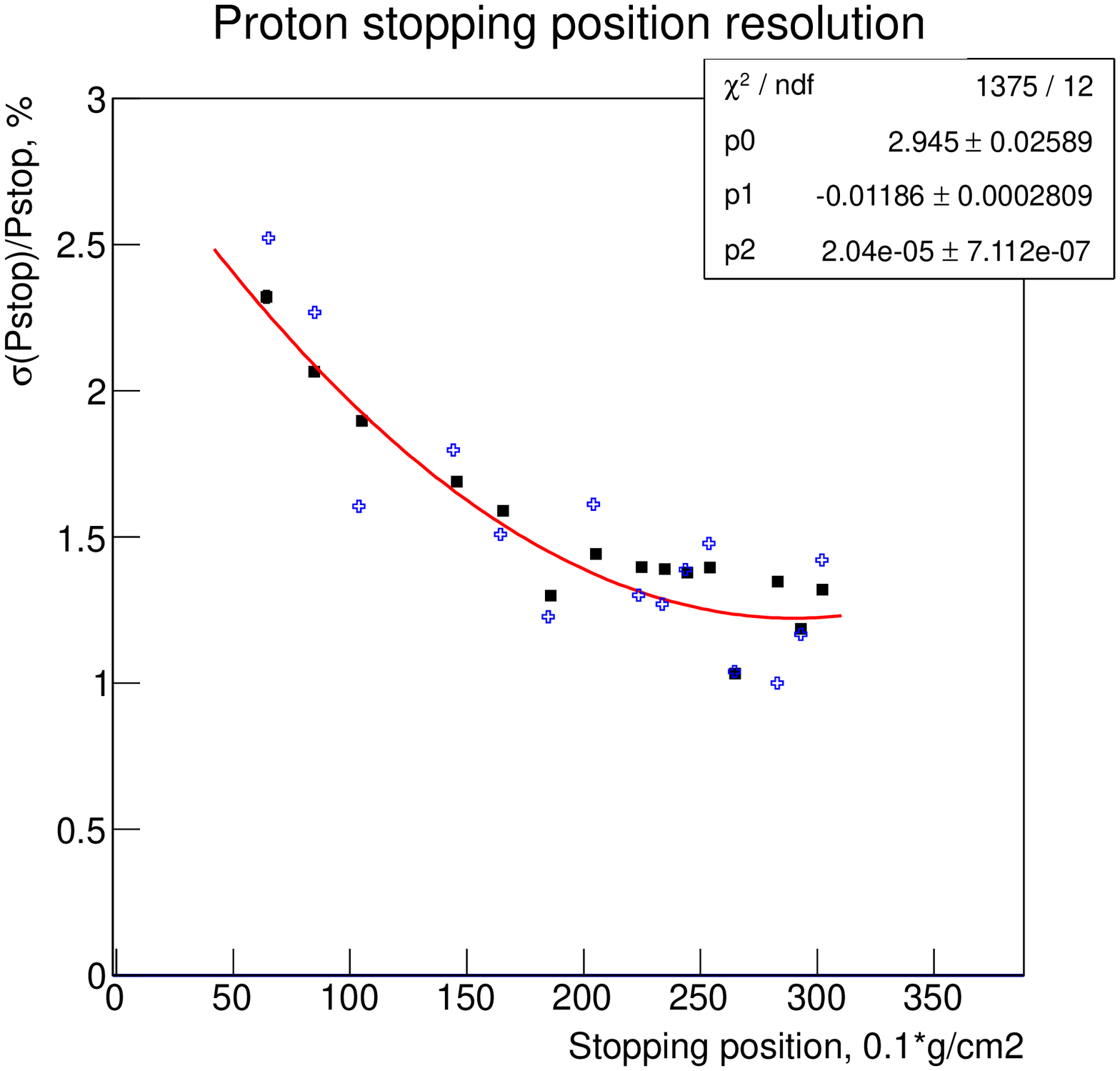}
 }
\leftline{ \hspace{4.5cm}{\bf (a)} \hfill\hspace{3.5cm} {\bf (b)} \hfill}
\caption{\label{fig:wepl-lin-and-res-cmp} Comparison of (a) the linearity and (b) resolution of the proton stopping position measurement $R_{rs}$ 
in data and GEANT.  Fits correspond to simulated results (black squares). Data shown as ``blue crosses``. }
\end{figure}

\begin{figure}[ht]
\centering
 {
\includegraphics[scale=0.27]{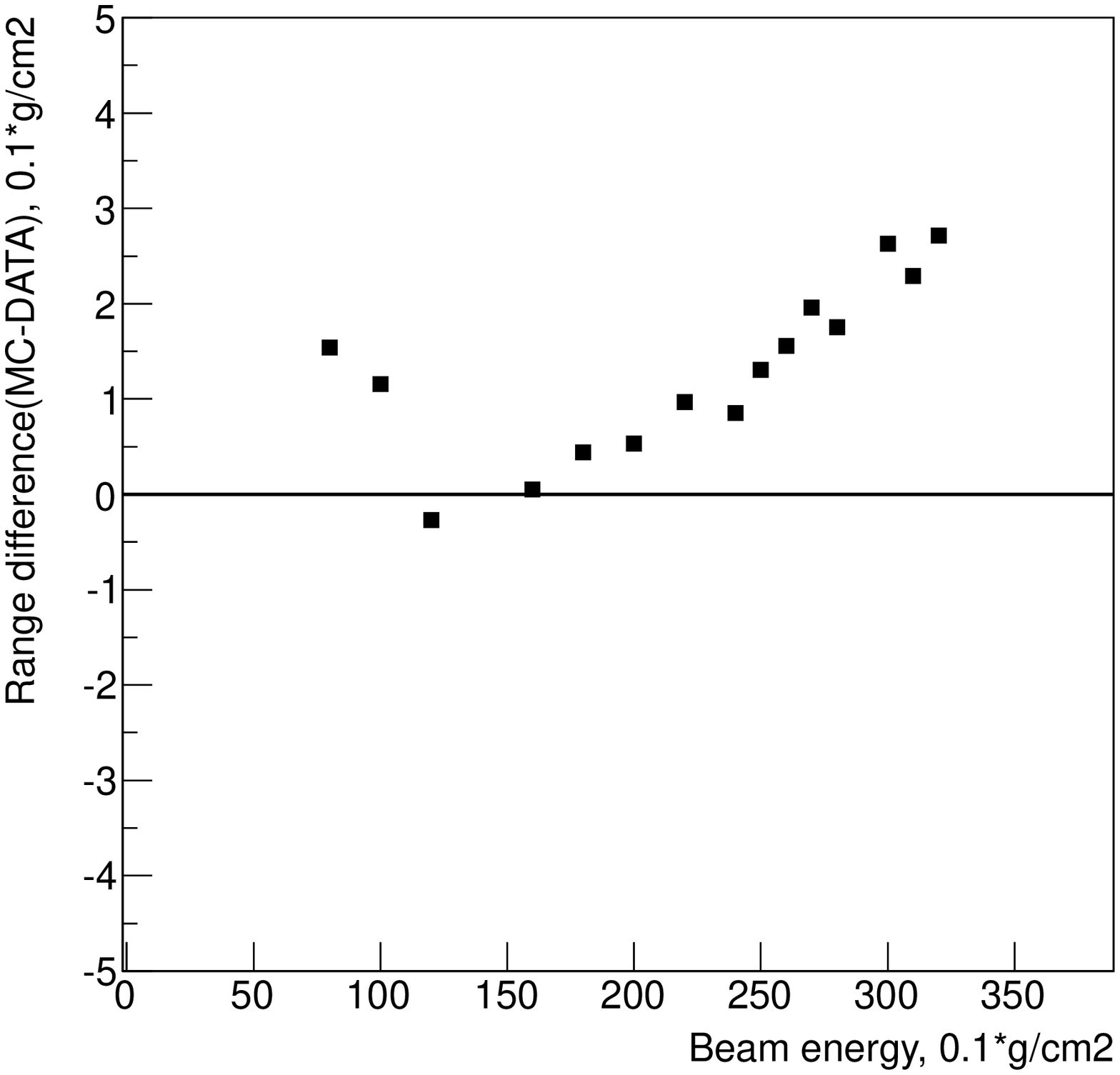}
\includegraphics[scale=0.27]{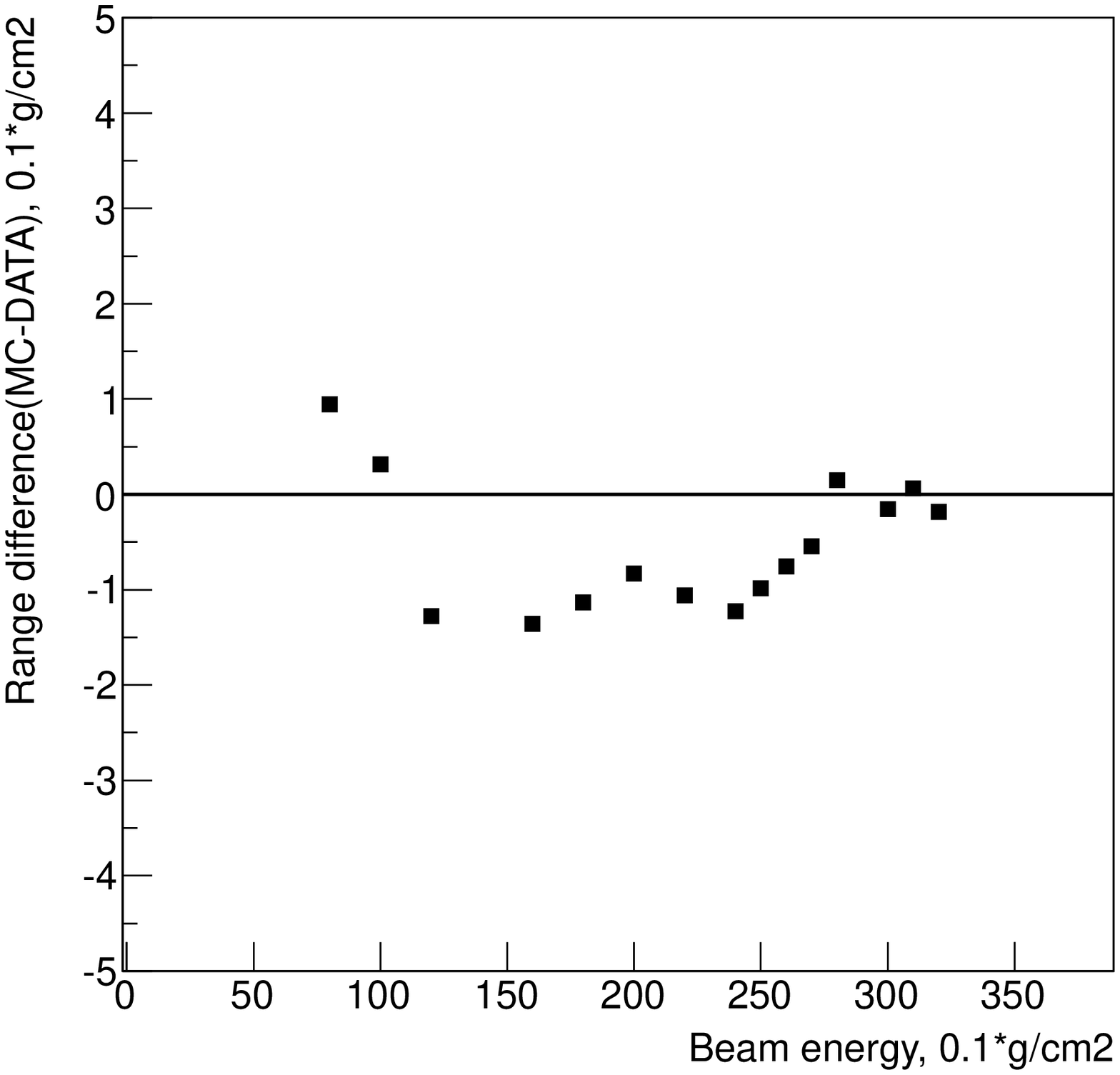}
\includegraphics[scale=0.27]{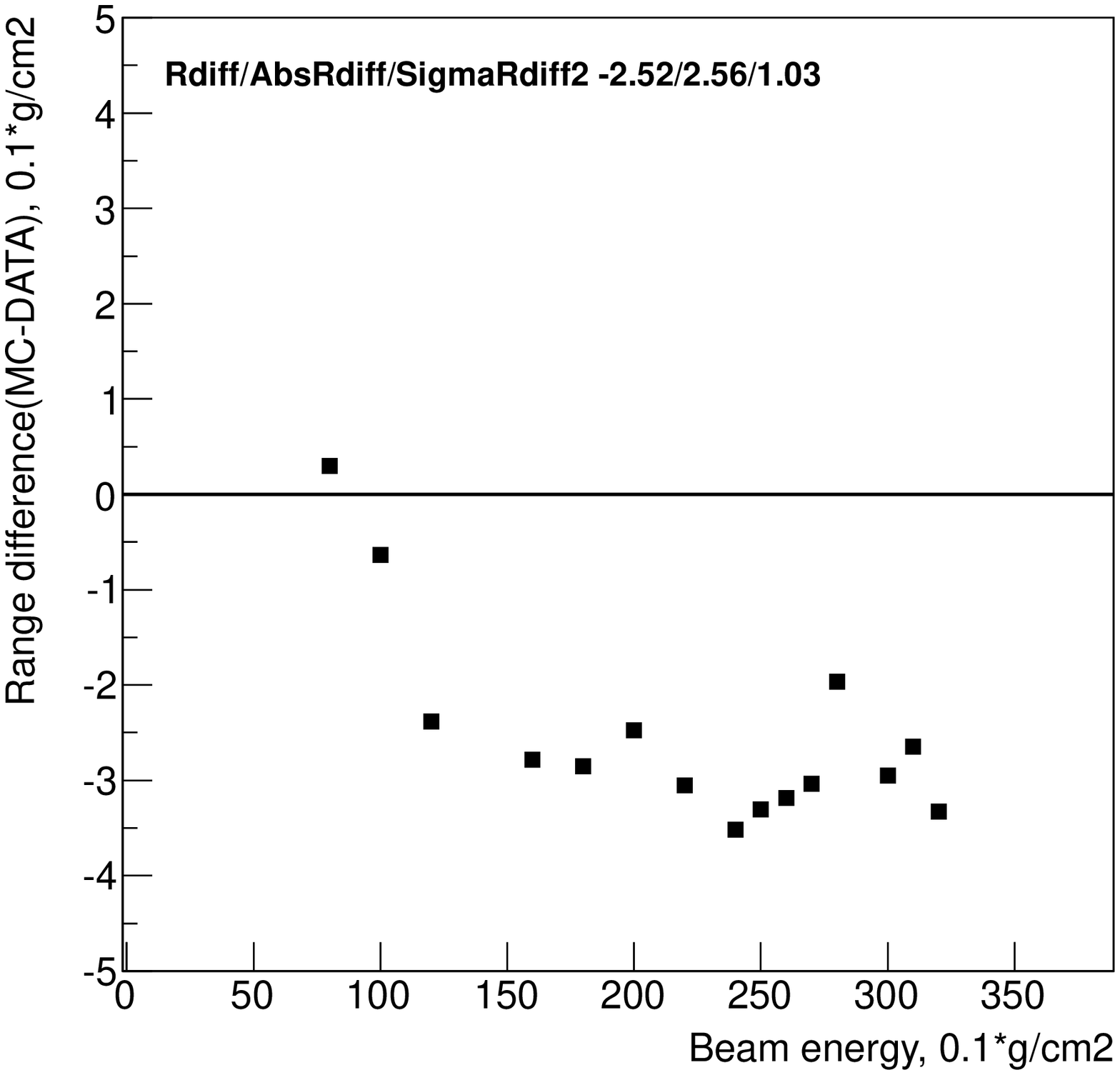}
 }
%\leftline{ \hspace{4.5cm}{\bf (a)} \hfill\hspace{3.5cm} {\bf (b)} \hfill}
\leftline{ \hspace{2.3cm}{\bf (a)} \hfill\hspace{3.2cm} {\bf (b)} \hfill \hspace{3.2cm} {\bf (c)} \hfill}
\caption{\label{fig:density-scan} The difference between measured and simulated proton stopping positions for a GEANT models with  
(a) nominal scintillator density of $1.025\pm 0.010$~$g/cm^3$; (b) nominal density decreased by 1\% (used in this Note); (c) nominal density decreased by 2\%.}
\end{figure}

The linearity and resolution plots for  the proton stopping position $R_{rs}$ are
shown in Fig.~\ref{fig:wepl-lin-and-res-cmp}.  Fits correspond to the simulated results.
We observe excellent agreement both in linearity and resolution. We used the detector model in which the density 
of the scintillator tiles is decreased by 1\% compare to the nominal value of $1.025\pm 0.010$~$g/cm^3$,
which provides the best agreement between measured and simulated proton stopping positions for different beam energies,
as shown in Figure~\ref{fig:density-scan}.
\subsection{Comparison of energy measurements}
\begin{figure}[ht]
\centering
 {
 \includegraphics[scale=0.40]{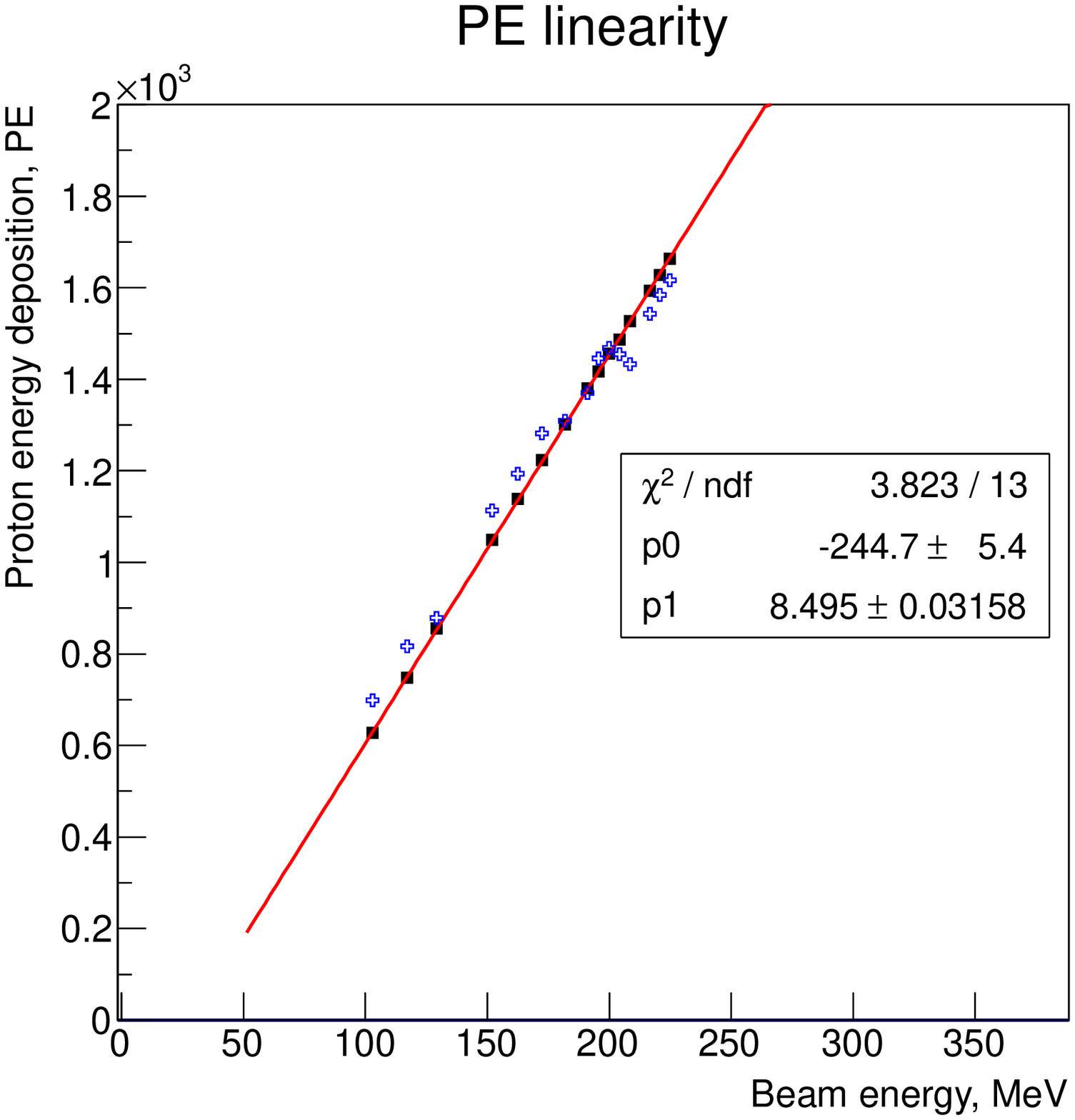}
 \includegraphics[scale=0.40]{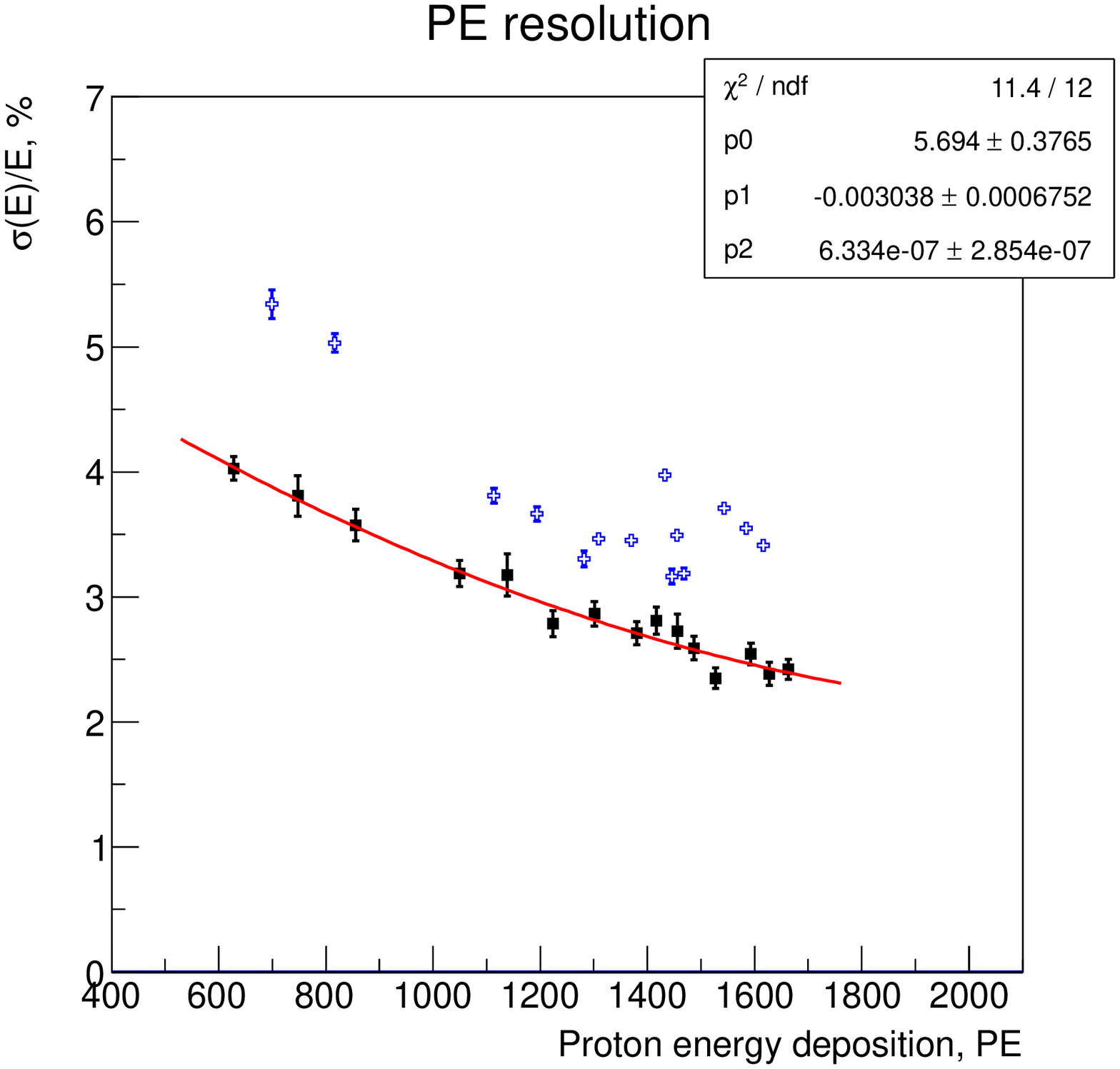}
 }
\leftline{ \hspace{4.5cm}{\bf (a)} \hfill\hspace{3.5cm} {\bf (b)} \hfill}
\caption{\label{fig:etot-lin-and-res-cmp} Comparison of (a) the linearity and (b) resolution of the energy measurements in the range stack  
in data and GEANT.  Fits correspond to simulated results (black squares). Data shown as ``blue crosses``. }
\end{figure}

\begin{figure}[ht]
\centering
{
 \includegraphics[scale=0.60]{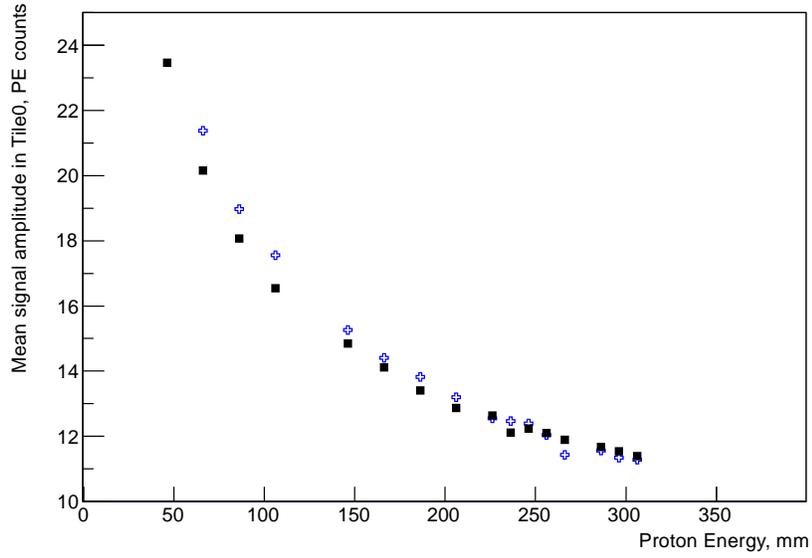}
}
%\leftline{ \hspace{4.5cm}{\bf (a)} \hfill\hspace{3.5cm} {\bf (b)} \hfill}
\caption{\label{fig:tile0_signal}Comparison of the measured (blue crosses) and simulated (black square) signal amplitudes in Tile0 for different proton energies. }
\end{figure}

The linearity and resolution plots for the energy measurements $E_{rs}= \sum E_{tn}$ in the range stack are
shown in Fig.~\ref{fig:etot-lin-and-res-cmp}.  Fits correspond to the simulated results.
The instrumental depression in the data linearity plot at beam energies of  27~cm and 28~cm is not present in simulations.     
The MC model response shows good linearity. The resolution is between 4\% and 2\% that is  higher than
observed in data (between 5\% and 3\%).  A comparison of measured (blue crosses) and simulated (black square)
energy measurements in Tile0 for different proton energies is shown in Fig.~\ref{fig:tile0_signal}.  
%Underestimation
%of the signal amplitude in the MC at low proton energies in this figure also can be seen in Figure~\ref{fig:stack-profiles-clb-8-16} 
%and in Figure~\ref{fig:stack-profiles-clb-18-24}. 

\begin{figure}[ht]
\centering
 {
 \includegraphics[scale=0.40]{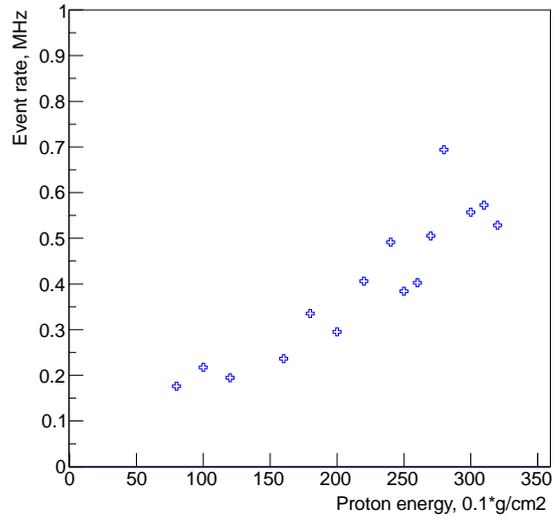}
 }
%\leftline{ \hspace{4.5cm}{\bf (a)} \hfill\hspace{3.5cm} {\bf (b)} \hfill}
\caption{\label{fig:event-rates} Event rate as function of beam energy.}
\end{figure}

\begin{figure}[ht]
\centering
 {
 \includegraphics[scale=0.27]{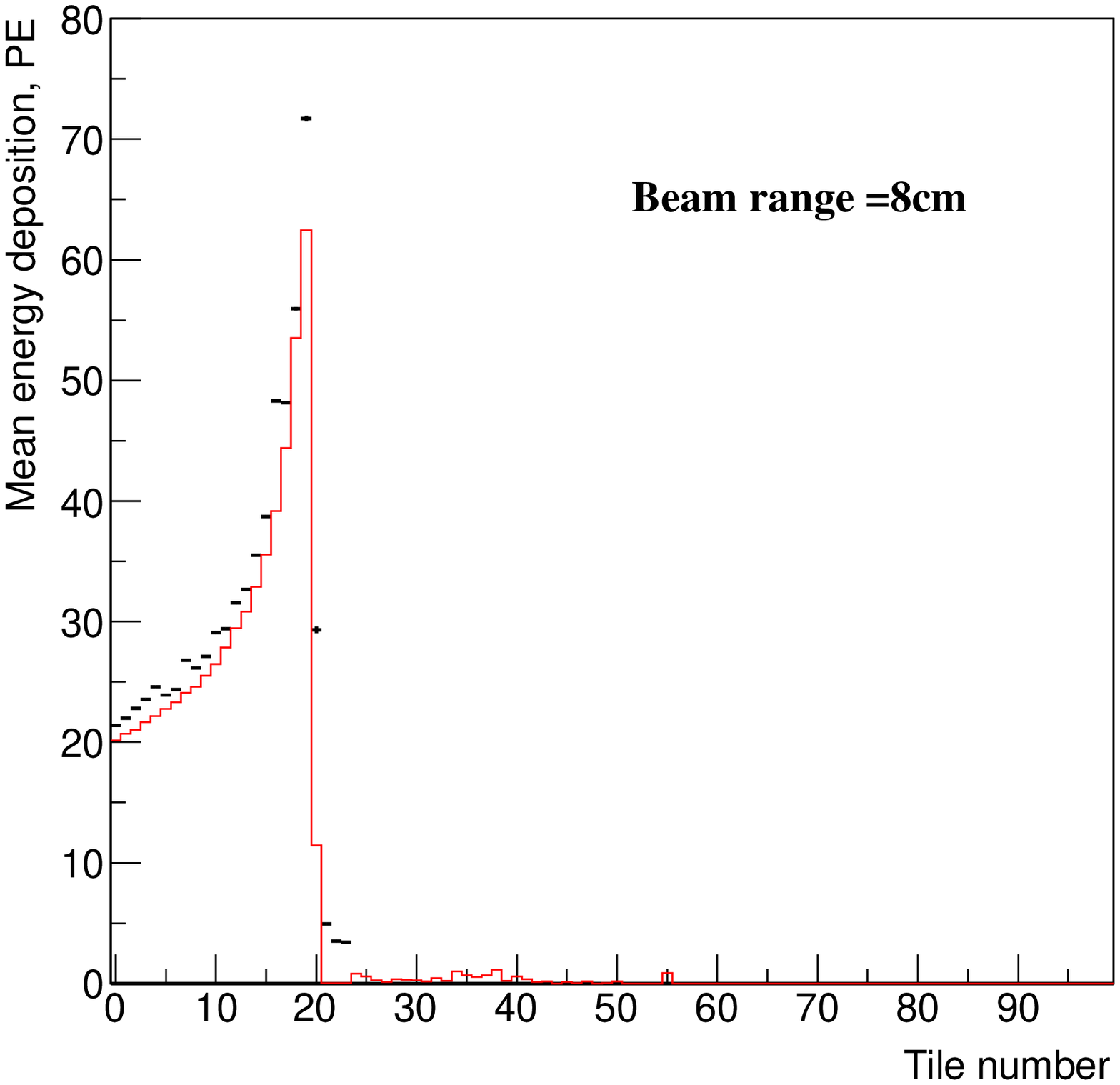}
 \includegraphics[scale=0.27]{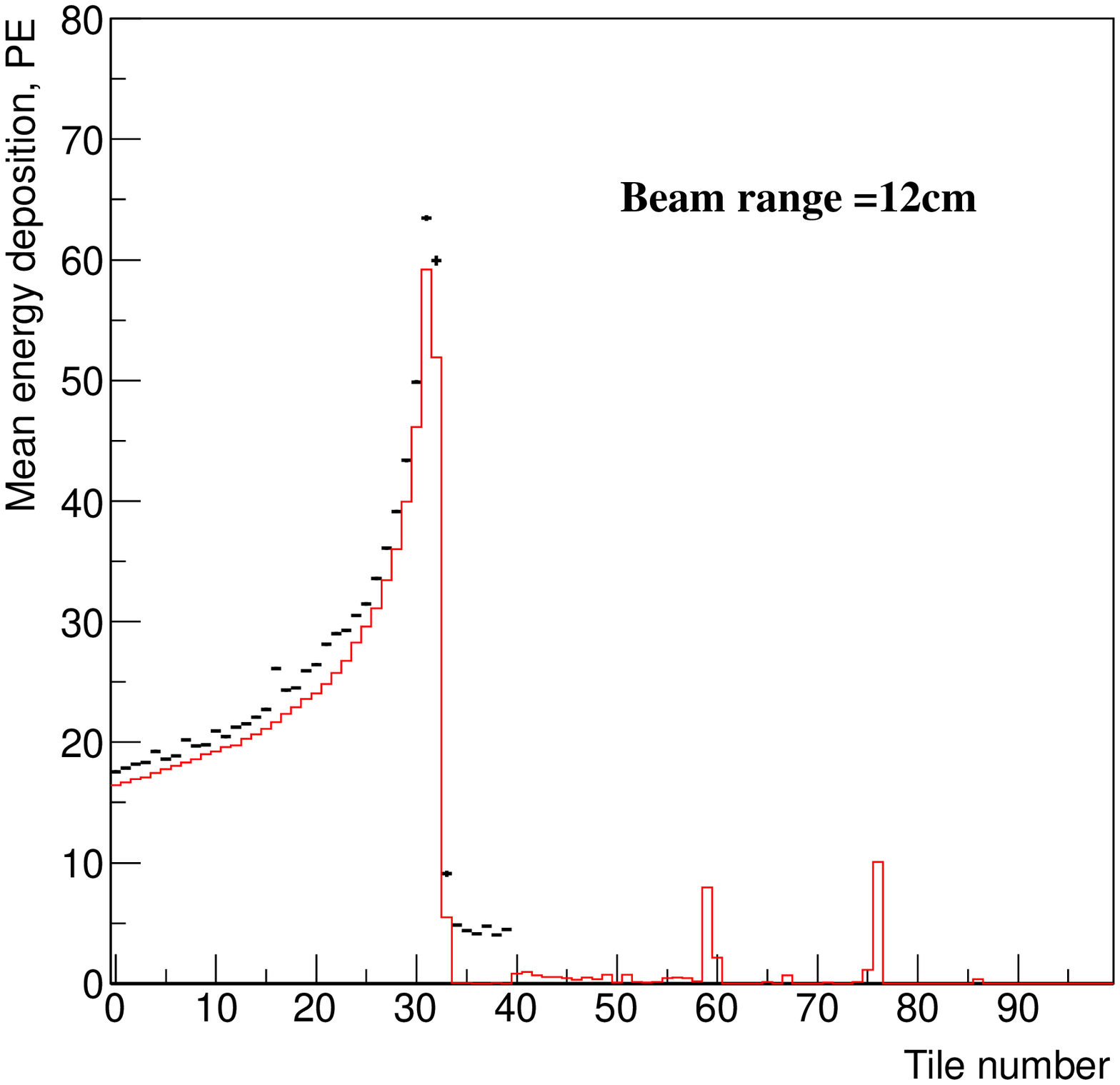}
 \includegraphics[scale=0.27]{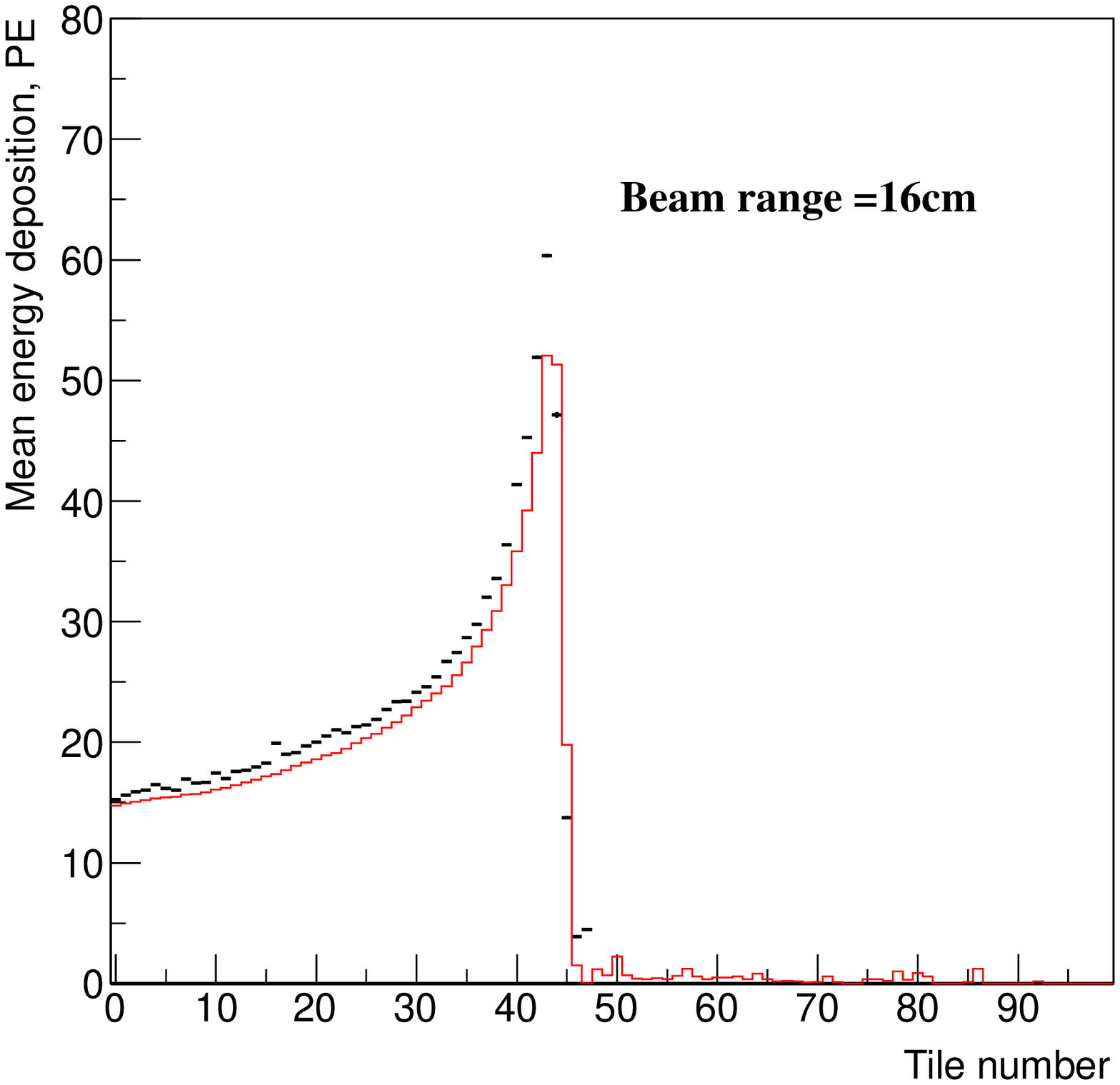}
 \leftline{ \footnotesize \hspace{2.3cm}{\bf  (a)} \hfill\hspace{3.2cm} {\bf (b)} \hfill \hspace{3.2cm} {\bf (c)} \hfill }
 \includegraphics[scale=0.27]{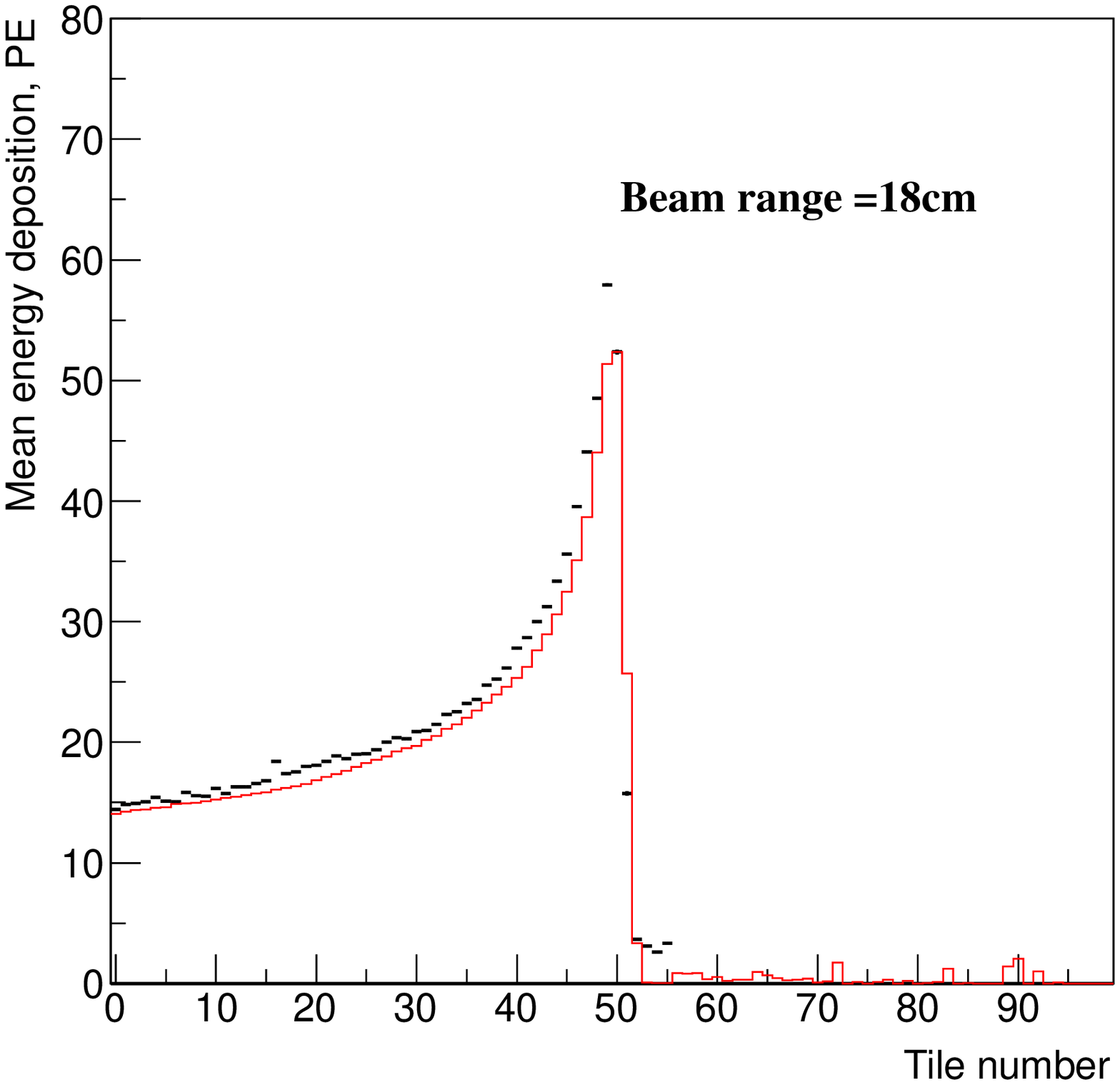}
 \includegraphics[scale=0.27]{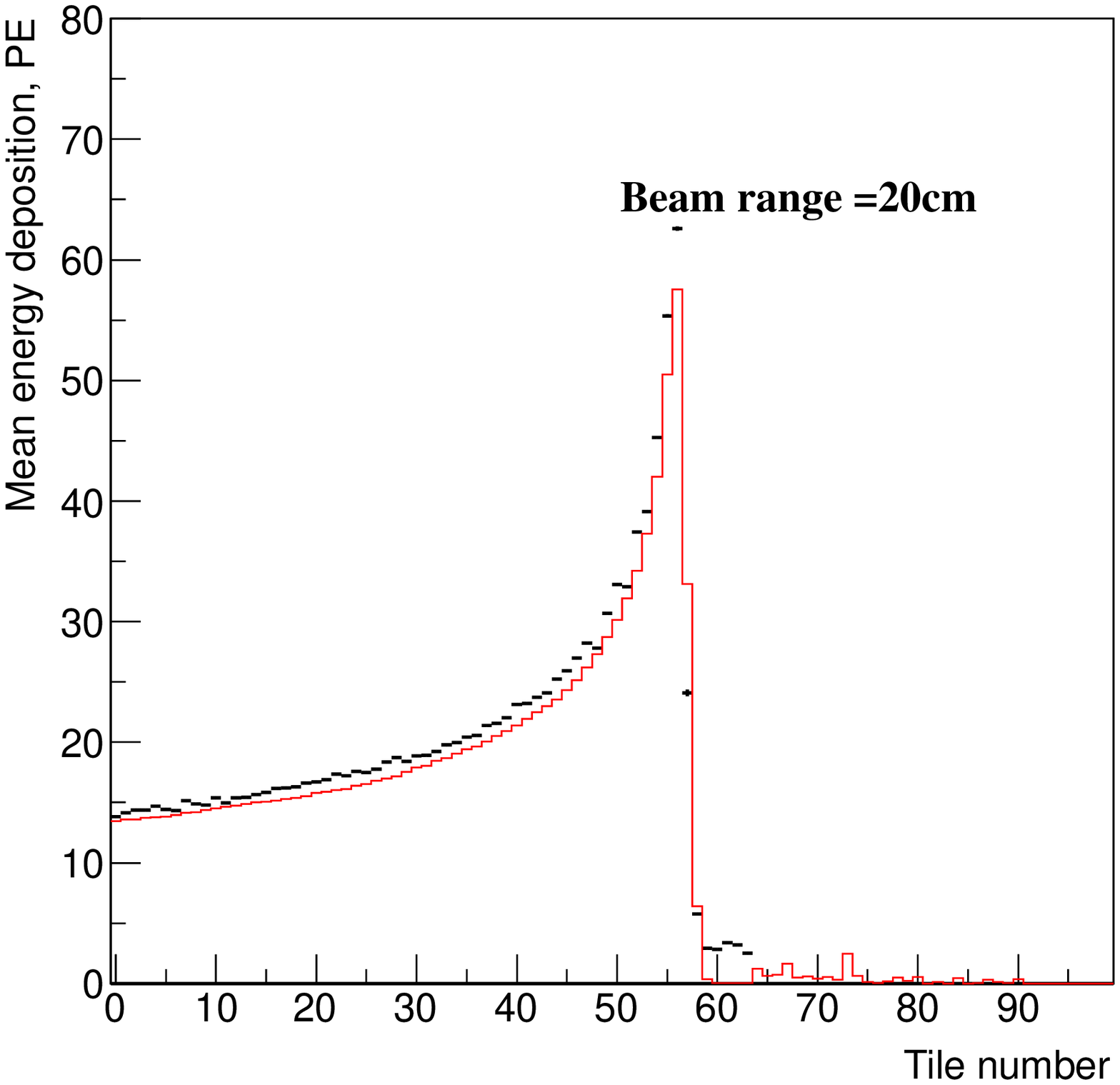}
 \includegraphics[scale=0.27]{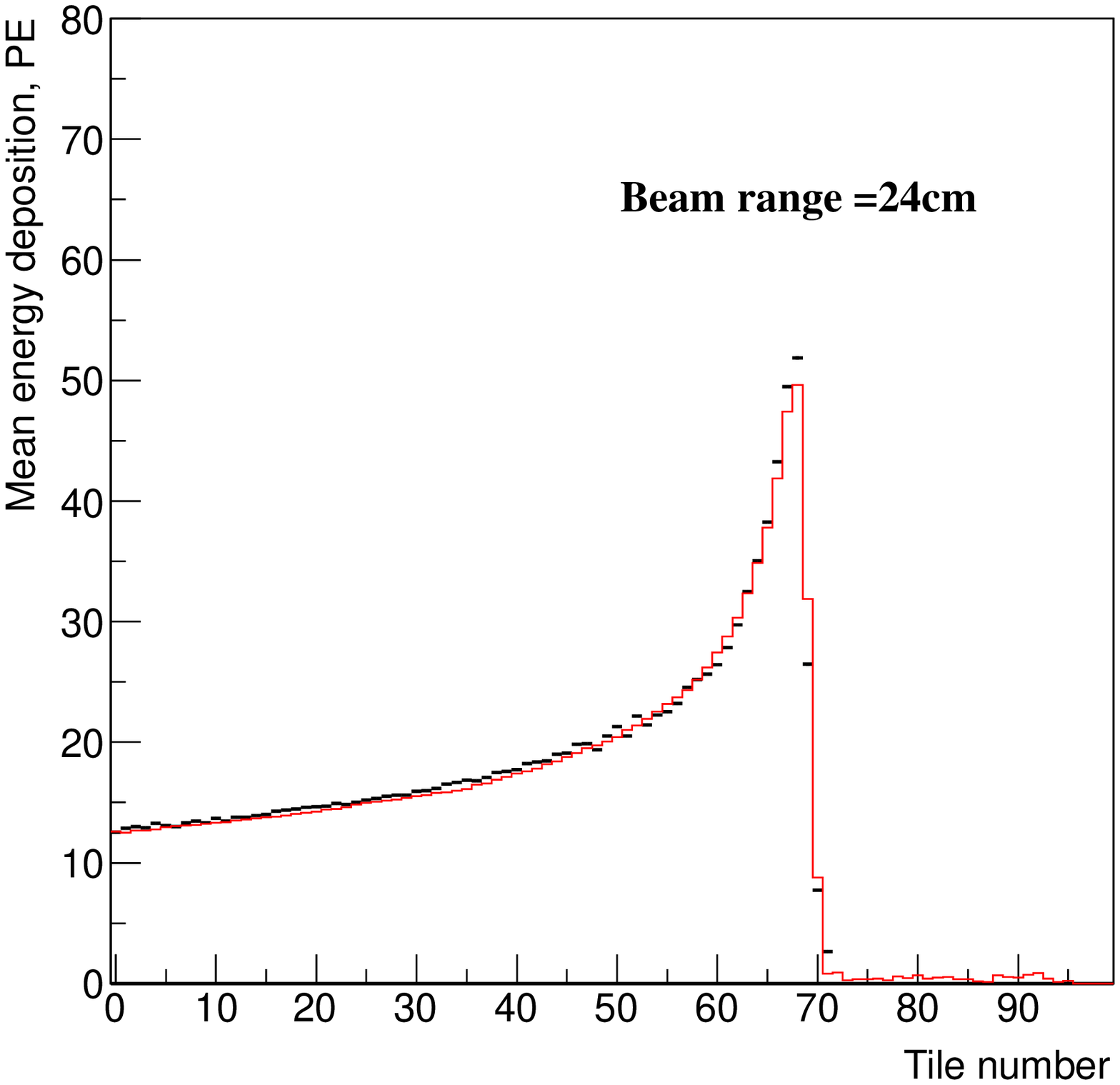}
 \leftline{  \footnotesize \hspace{2.3cm}{\bf (d)} \hfill\hspace{3.2cm} {\bf (e)} \hfill \hspace{3.2cm} {\bf (f)} \hfill}
 \includegraphics[scale=0.27]{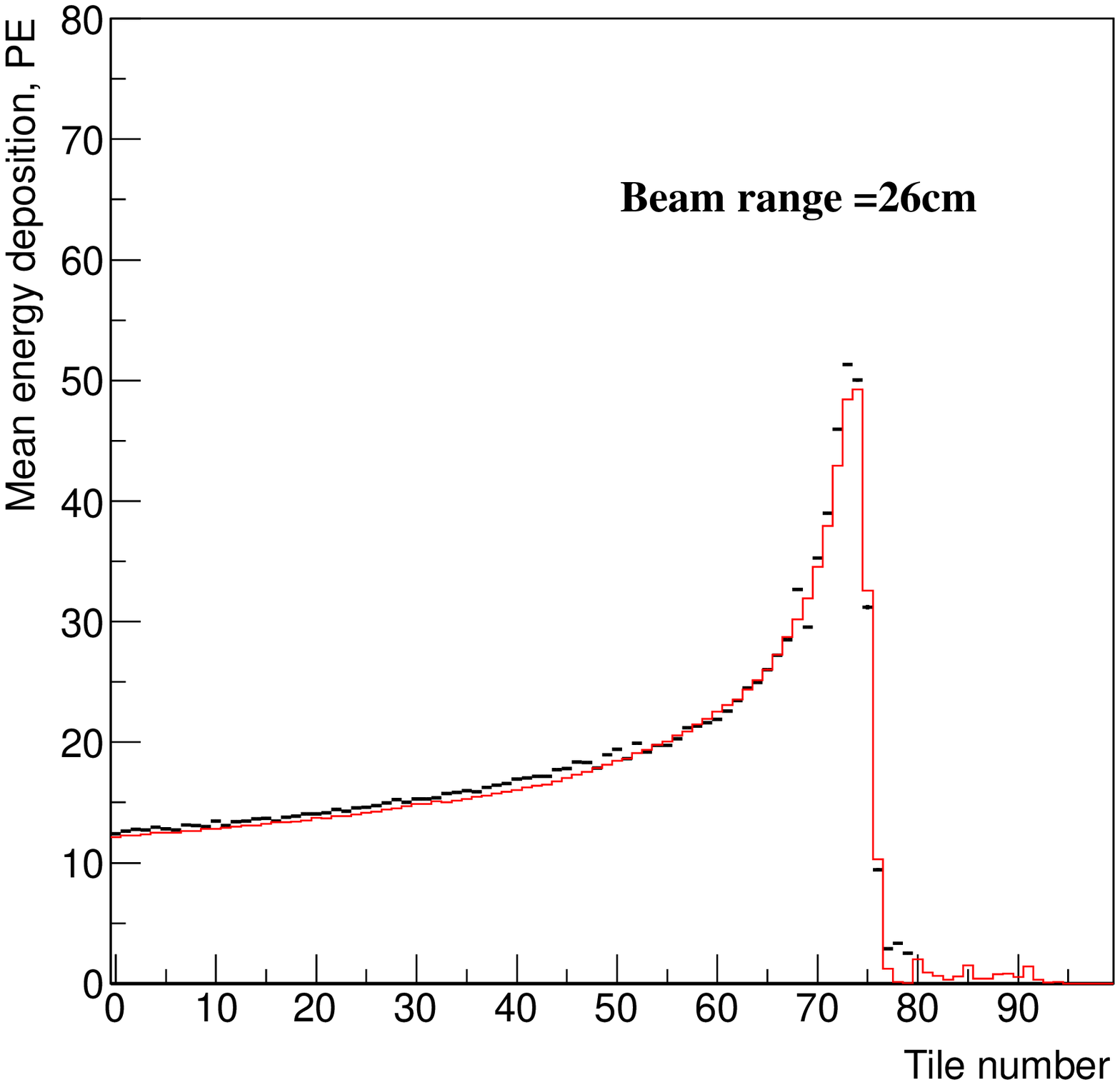}
 \includegraphics[scale=0.27]{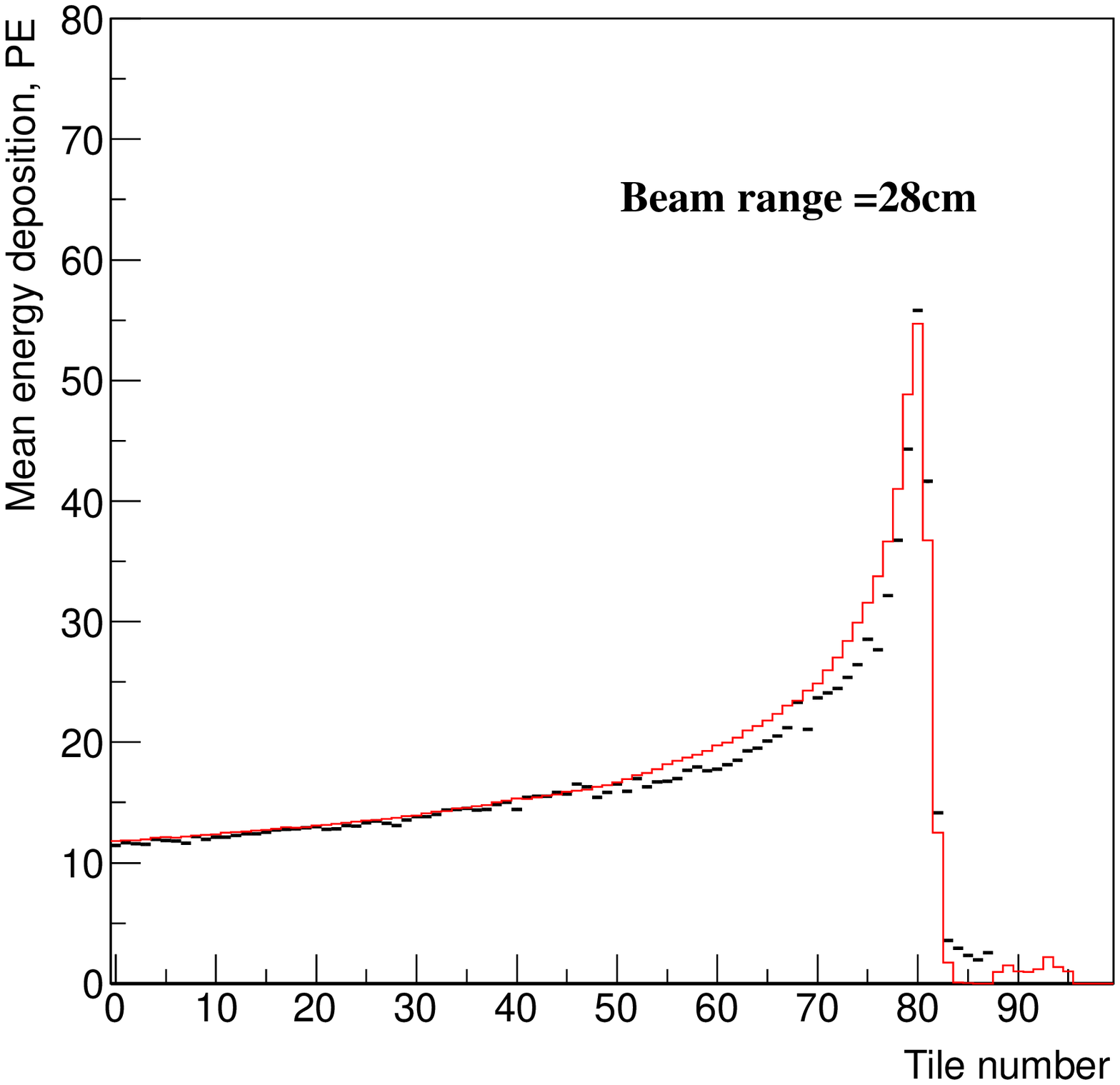}
 \includegraphics[scale=0.27]{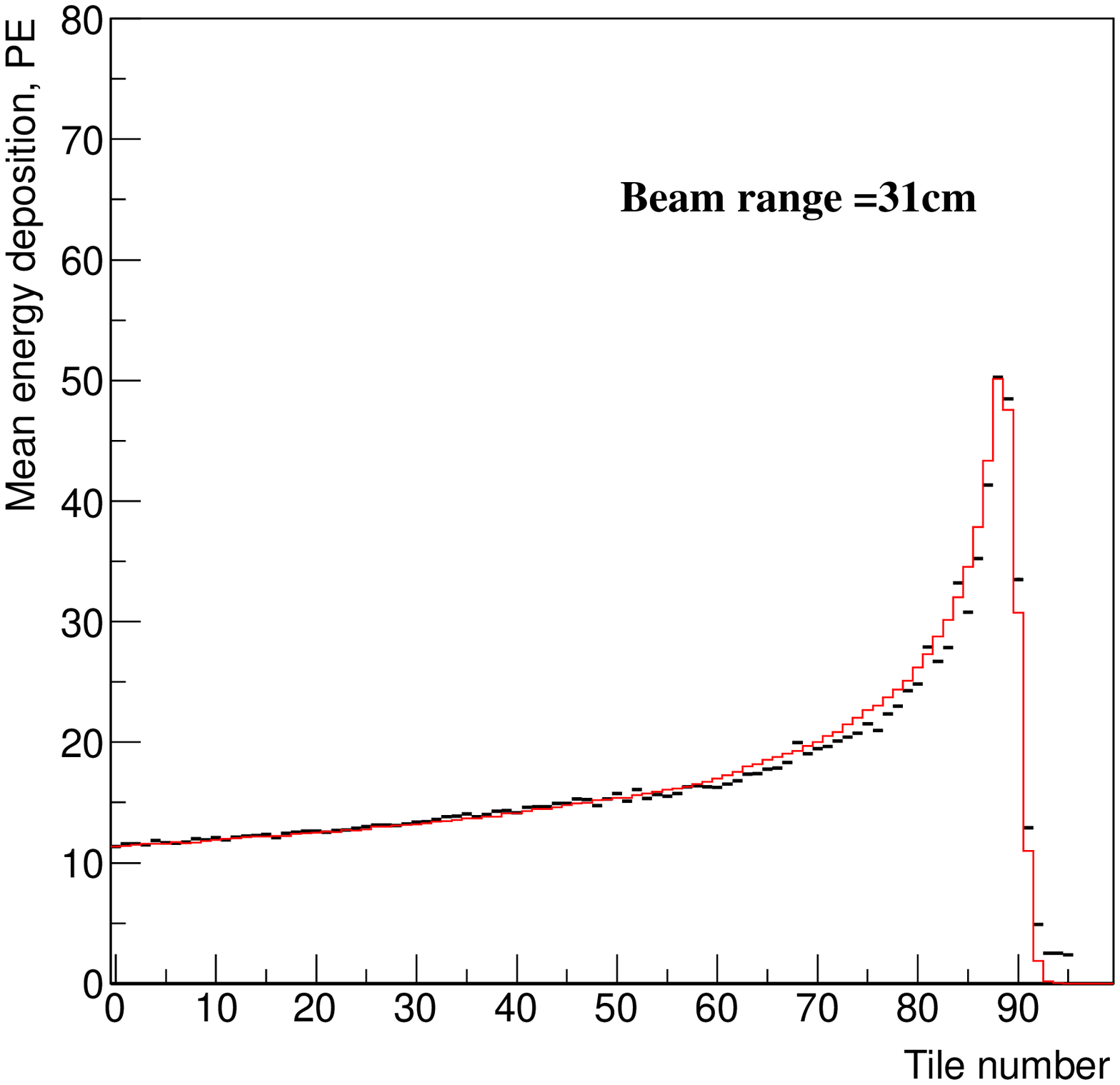}
 \leftline{ \footnotesize  \hspace{2.3cm}{\bf (g)} \hfill\hspace{3.2cm} {\bf (h)} \hfill \hspace{3.2cm} {\bf (i)} \hfill}
}
\caption{\label{fig:DATAvsMC-8-32} Comparison of the measured (black dots) and expected (red histograms) signal profiles in the range stack
from protons of incident energy of (a) 8~cm (103~MeV); (b) 12~cm (117~MeV);  (c) 16~cm (129~MeV); (d) 18~cm (162~MeV); (e) 20~cm (172~MeV);   (f) 24~cm (191~MeV); (g) 26~cm (200~MeV); (h) 28~cm (208~MeV);   (i) 31~cm (221~MeV).}
\end{figure}

The  normalized simulated energy amplitude profiles in the range stack in Fig.~\ref{fig:DATAvsMC-8-32} (red histograms) show fair agreement with 
the calibrated data (black dots) but diverge in amplitude at low proton energies (consistent with Fig.~\ref{fig:tile0_signal}).   The divergence could be
due to higher event rates in high proton energy runs, shown in  Fig.~\ref{fig:event-rates}.

%cfroot getTile0fit.C++
\section{Stopping range measurements in presence of a phantom}
\begin{figure}[ht]
\centering
 {
 \includegraphics[width=7.62 cm, height= 7.2 cm]{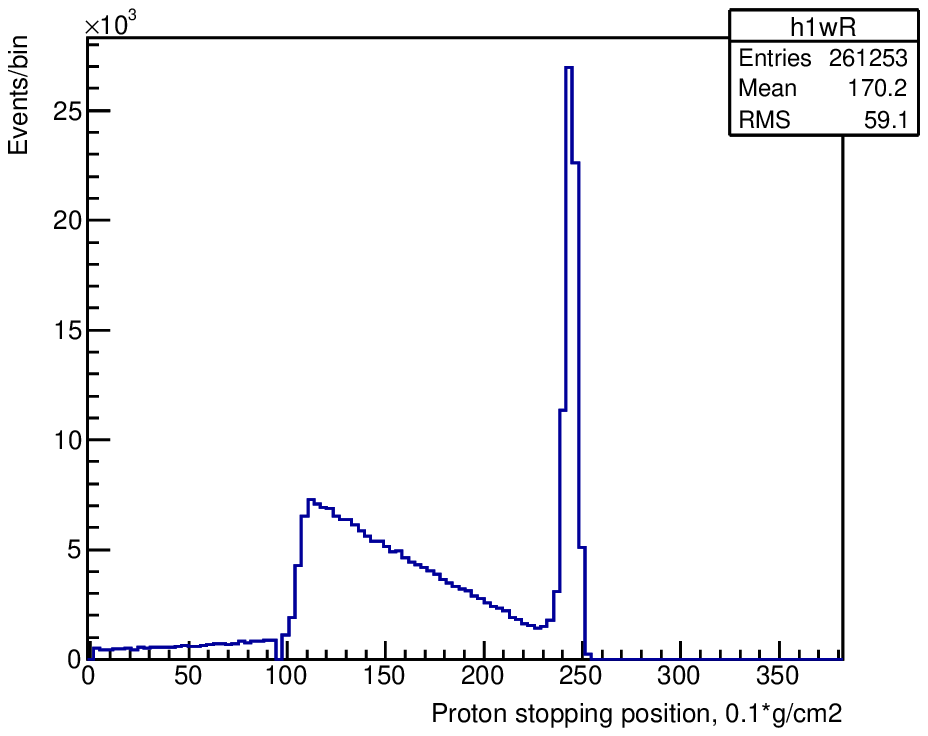}
 \includegraphics[width=7.62 cm, height= 7.2 cm]{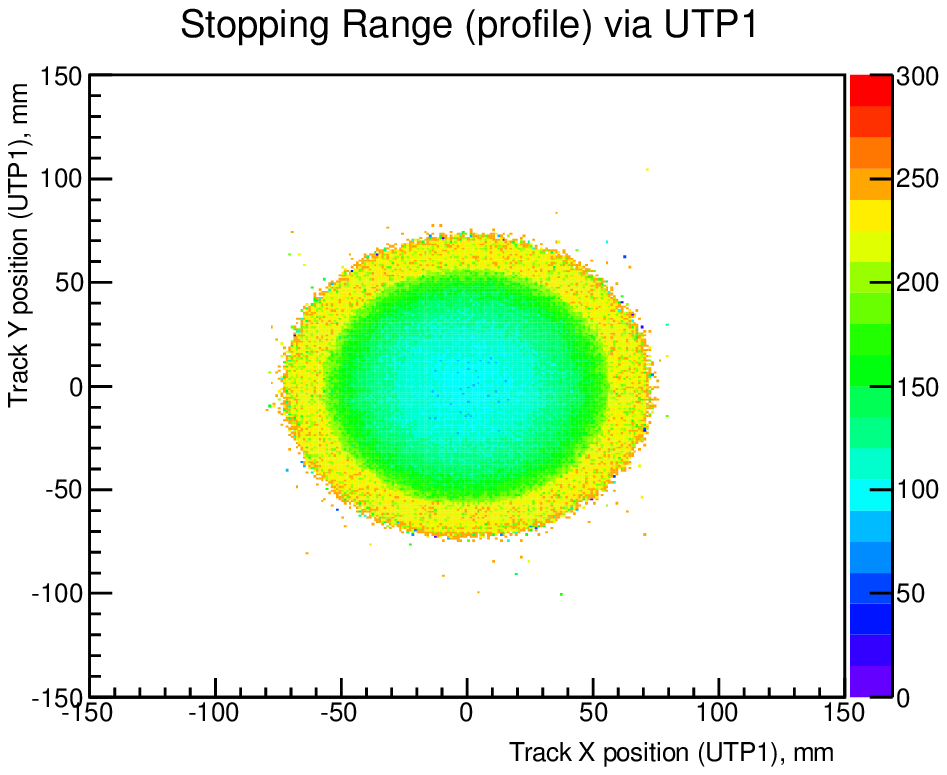}
 }
\leftline{ \hspace{4.5cm}{\bf (a)} \hfill\hspace{3.5cm} {\bf (b)} \hfill}
\caption{\label{fig:waterPh_profile} Water phantom (diameter of 14~cm) exposed to 300K protons of energy 200~MeV in GEANT simulations (a) the stopping range distribution 
(b) the stopping range profile as function of incident proton position at the first tracker station.}
\end{figure}

\begin{figure}[ht]
\centering
 {
 \includegraphics[width=7.62 cm, height= 7.2 cm]{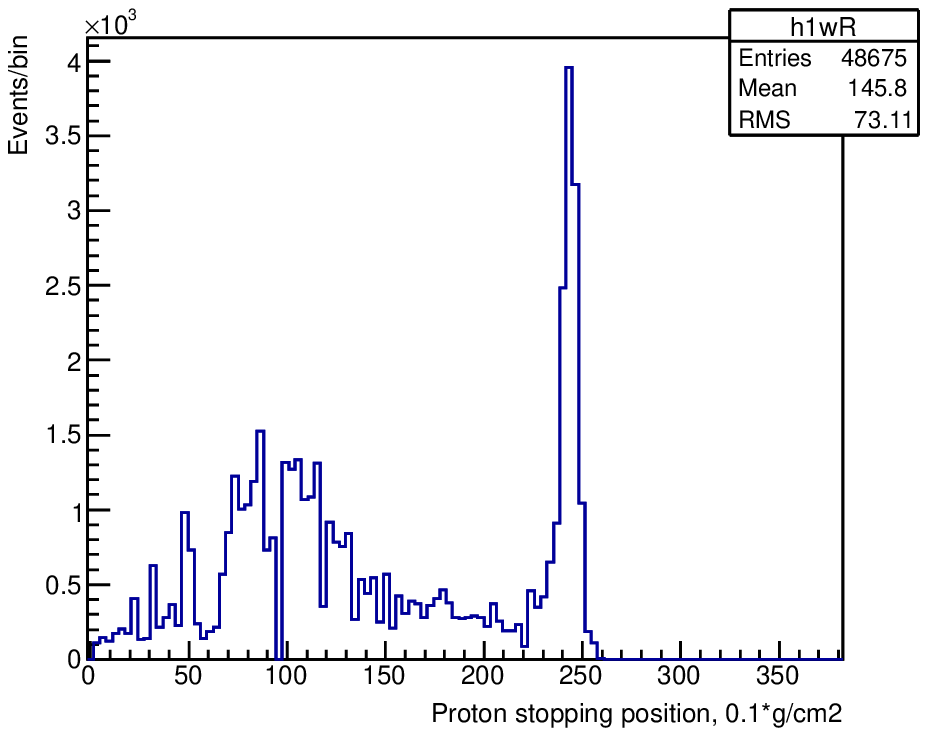}
 \includegraphics[width=7.62 cm, height= 7.2 cm]{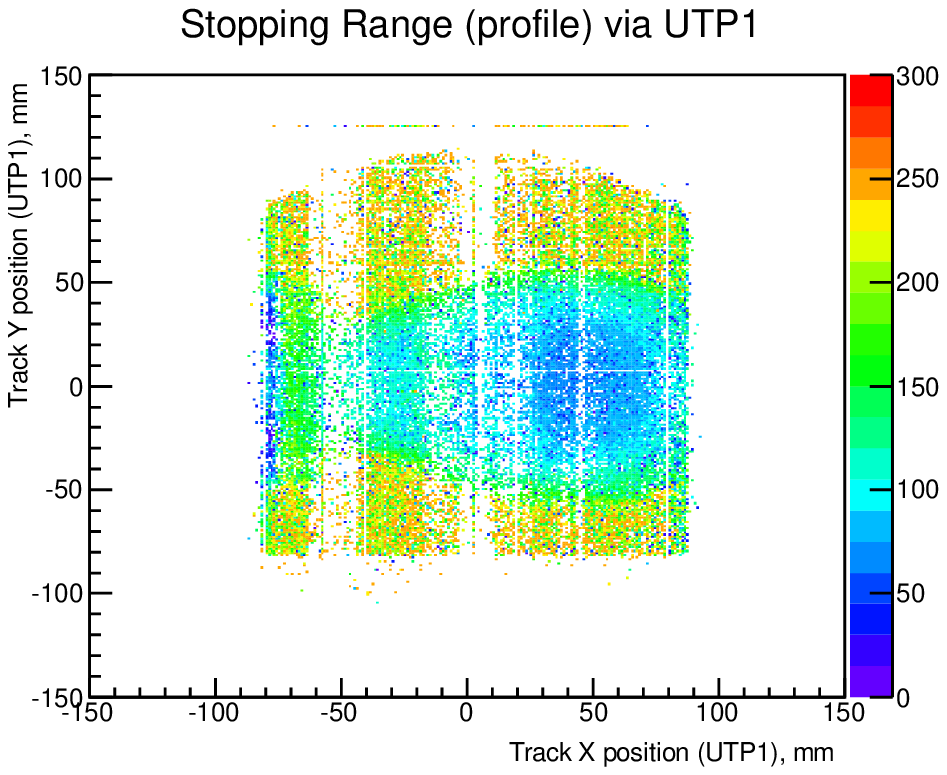}
 }
\leftline{ \hspace{4.5cm}{\bf (a)} \hfill\hspace{3.5cm} {\bf (b)} \hfill}
\caption{\label{fig:dataPhantopm_profile} Head phantom exposed to ~50K protons of energy 200 MeV at CDH (a) the stopping range distribution 
(b) the stopping range profile as a function of incident proton position at the first tracker station. }
\end{figure}

Figure~\ref{fig:waterPh_profile}(a) and Fig.~\ref{fig:waterPh_profile}(b), respectively, show the distribution of proton stopping range in the range stack and the simulated $(X,Y)$ distribution of protons  at the first tracker station of the GEANT~4 detector model obtained in the presence of a spherical ($D=14~cm$)  water phantom.  The first peak in the stopping range distribution corresponds to protons going through the center of the phantom, while the second peak corresponds to the stopping range of protons that missed the phantom.  Different colors for the reconstructed tracks correspond to the simulated proton stopping ranges. 
Figures~\ref{fig:dataPhantopm_profile}(a) and (b) show similar plots for the  head phantom obtained using ~50K reconstructed protons of energy 200~MeV at CDH. Contours corresponding to the different material width are
clearly visible. The missing bands correspond to  missing tracking channels.
\section{Summary}
% A calibration procedure for the pCT range stack detector has been developed.  The proton stopping position and  energy 
% measurements are obtained for fixed energies (8~cm to 32~cm)  beams at CDH facility.  The measurements
% are in good agreement with simulations.
%however both the calibration procedure and simulation model have space for improvement.
The stopping position measurements have better linearity and accuracy (2.2-1.2\%) than the energy measurements (5.5\% to 3.5\%), confirmed by simulations,
and thus are expected to provide more accurate WEPL calibration for the image reconstruction.  The behavior of range stack detector is well modelled by GEANT,
with a few dicrepancies in energy deposition at low energy and energy resolution.

\begin{thebibliography}{99}
\bibitem{brudge_paper}
G.~Coutrakon {\it et al.}, Proceedings AccApp 2013, Bruges, Belgium; S.~A.~Uzunyan {\it et al.}, arXiv:1409.0049 (2014).
%``A New Proton CT Detector'', 
% Proceedings Intl. Topical Meeting on Nuclear Applications of Accelerators, Bruges, Belgium, (2013).
\bibitem{su_daq}
S.~Uzunyan {\it et al.},  Proceedings of the New Trends in High-Energy Physics,  p. 152-157, Alushta, Crimea, Ukraine, Sep. 2013, ISBN 978-966-02-7015-2.
\bibitem{wepl_clb}
R.~F.~Hurley{\it et al.}, ``Water-equivalent path lengh claibration of a prototype proton CT scaner'',  Med. Phys. 39(5),  May 2012.
\bibitem{janni_tables}
J.~F.~Janni, "{Proton Range-Energy Tables, 1 keV-10 GeV, Energy Loss, Range, Path Length, Time-of-Flight, Straggling, Multiple Scattering, and Nuclear Interaction Probability. Part I. For 63 Compounds}", Atomic Data and Nuclear Data Tables, 27, 147, (1982).
%
\end{thebibliography}
\end{document}